\documentclass[prb,twocolumn,showpacs,amsmath,amssymb]{revtex4}
\usepackage{amssymb}
\usepackage{txfonts}
\usepackage{bbm}
\usepackage{graphicx}
\usepackage{appendix}
\usepackage{epsf}
\usepackage{epstopdf}

\begin{document}

\title{Quasiparticle states around a nonmagnetic impurity in electron-doped iron-based superconductors with spin-density-wave order}

\author
{Tao Zhou,$^{1,2}$  Huaixiang Huang,$^{1,3}$ Yi Gao,$^{1}$Jian-Xin
Zhu,$^{4}$ and C. S. Ting$^{1}$}

\affiliation{$^{1}$Texas Center for Superconductivity and Department of
Physics, University of Houston, Houston, Texas 77204, USA\\
$^{2}$Department of Physics, Nanjing University of Aeronautics and Astronautics, Nanjing 210016, China\\
$^{3}$Department of Physics, Shanghai University,
shanghai 200444, China\\
$^{4}$Theoretical Division, Los Alamos National Laboratory, Los
Alamos, New Mexico 87545, USA
}

\date{\today}

\begin{abstract}
The quasiparticle states around a nonmagnetic impurity in
electron-doped iron-based superconductors with spin-density-wave
(SDW) order are investigated as a function of doping and impurity scattering
strength. In the undoped sample, where a pure SDW state
exists, two impurity-induced resonance peaks are observed around the
impurity site and they are shifted to higher (lower) energies as the
strength of the positive (negative) scattering potential (SP) is
increased. For the doped samples where the SDW order and the
superconducting order coexist, the main feature is the existence of
sharp in-gap resonance peaks whose positions and intensity depend on
the strength of the SP and the doping
concentration. In all cases, the local density of states exhibits
clear $C_2$ symmetry. We also note that in the doped cases, the
impurity will divide the system into two sublattices with
distinct values of magnetic order. Here we use the band structure
of a two-orbital model, which considers the asymmetry of the As
atoms above and below the Fe-Fe plane. This model is suitable to
study the properties of the surface layers in the iron-pnictides and
should be more appropriate to describe the scanning tunneling
microscopy experiments.
\end{abstract}
 \maketitle

\section{introduction}
The new family of iron-based superconducting (SC) materials has
attracted much attention since their discovery.~\cite{kam} The
parent compounds exhibit a spin-density-wave (SDW) order at low
temperatures. Upon doping either electrons or holes into the system,
the SDW order is suppressed and superconductivity emerges,
suggesting the interplay and competition between these two
states.

The impurity effect is an important property in the studies of
superconductivity. One prominent feature of $d$-wave paring symmetry
in cuprates is the existence of bound states near the impurity,
which is revealed by both experiments and theoretical
calculations.~\cite{bal} For iron-based materials,  the impurity
effect in the SC state has also been theoretically studied
intensively.~\cite{chub,bang,park,seng,voro,zhou,zhang,tsai} It was
proposed that a single nonmagnetic impurity could be used to
distinguish the pairing symmetry and the in-gap bound states could
exist for the typical $s_{\pm}$ pairing
symmetry.~\cite{zhou,zhang,tsai} The in-gap bound states for
$s_{\pm}$ pairing should be different from those in cuprates due to
the absence of quasiparticle excitations at low energy. On the other
hand, in the SDW state, it was proposed experimentally that the
Fermi surface (FS) is only partially gapped and small ungapped Fermi
pockets exist at low
temperature.~\cite{dingh,hsieh,jong,myi,seba,ana} This feature was
recently reproduced based on a two-orbital model together with a
mean-field approach.~\cite{zho} The gap-like feature and the
existence of tiny ungapped regions along the diagonal direction of
the Brillouin zone (BZ) are quite similar to the case of the
$d$-wave SC gap in cuprates. Therefore, one would expect that the
low-energy bound states should also exist in the SDW state for the
iron-based materials. So far, the impurity effect in the SDW state
remains less explored and a systematic study for this topic is still
lacking. We believe it is timely and quite interesting to address this
issue theoretically and verify the aforementioned expectation
numerically. Furthermore, in some of the iron-based materials the SC
and SDW orders are proposed to coexist in the underdoped
regime.~\cite{pra,wan,lap,les,chr,jul,chen,zha,liu} The issue is
still a subject of discussion and we
 anticipate that the impurity effect could provide additional signatures
 for the coexistence of these two orders.

In this paper, we study theoretically the impurity effect on a
two-dimensional square lattice based on a two-orbital model and the
Bogoliubov-de Gennes (BdG) equations. By introducing a single
impurity into the system, the local density of states (LDOS) is
calculated and our results show that: (i) In the undoped sample,
there are two impurity-induced resonance peaks at and near the
impurity site and the LDOS spectra exhibit $C_2$ symmetry, with
one-dimensional modulation. (ii) The impurity effect in various
doped cases are also studied. Its effect on the LDOS is remarkable
only when the strength of the scattering potential (SP) is larger
than a certain value. For weak and moderate SPs, a distinct bound
state exists explicitly at the next-nearest-neighbor (nnn) sites of
the impurity, the energy of which depends on the strength and sign
of the SP, as well as on the doping concentration. For the unitary
impurity, there is a sharp in-gap peak at low doping; while at high
doping, the impurity induced bound state is close to the SC
coherence peaks. On the other hand, in a small range of moderate
doping there are two in-gap peaks only for positive SP. All the
above features could be used to detect the presence of the SDW order
as well as the coexistence of the SDW and SC orders.

The paper is organized as follows. In Sec.~\ref{SEC:model}, we
introduce the model and work out the formalism. In
Sec.~\ref{SEC:fs}, we study the FS. In Sec.~\ref{SEC:impforundoped},
the impurity effect in the parent compound is investigated. In
Secs.~\ref{SEC:positive} and ~\ref{SEC:negative}, we study the
impurity effect in doped regime for positive and negative SPs,
respectively. Finally, we give a summary in Sec. ~\ref{SEC:summary}.

\section{model and formalism}
\label{SEC:model} The iron-based superconducting materials have a
layered structure with the FeAs layers being the superconducting
planes. In the present work, following Refs.~[\onlinecite{zhang}]
and [\onlinecite{zho}] and taking into account the asymmetry of the
As ions located above and below the Fe-Fe plane, we start from a
two-orbital model with on-site interaction. This model is able to
qualitatively explain both the ARPES~\cite{arpes} and STM~\cite{stm}
experiments. For example, based on this model, the obtained phase
diagram,~\cite{zho} spin susceptibility,~\cite{gao1} as well as the
Andreev bound state inside the vortex core~\cite{gao2} are all
consistent with the experiments. Thus we also use it to investigate
the impurity effect. The Hamiltonian is written as,
\begin{equation}
H=H_{BCS}+H_{int}+H_{imp}\;.
\end{equation}
 Here $H_{BCS}$ is the BCS-like Hamiltonian, which includes the hopping term and the pairing term, expressed by,
\begin{eqnarray}
H_{BCS}=&-\sum_{{\bf i}\mu {\bf j}\nu\sigma}(t_{{\bf i}\mu
{\bf j}\nu}c^\dagger_{{\bf i}\mu\sigma}c_{{\bf j}\nu\sigma}+h.c.)-t_0\sum_{{\bf i}\mu\sigma}c^{\dagger}_{{\bf i}\mu\sigma}c_{{\bf i}\mu\sigma}
\nonumber\\&+\sum_{{\bf i}\mu {\bf j}\nu\sigma}(\Delta_{{\bf i}\mu
{\bf j}\nu}c^\dagger_{{\bf i}\mu\sigma}c^{\dagger}_{{\bf j}\nu\bar{\sigma}}+h.c.)\;,
\end{eqnarray}
where ${\bf i}=(i_x,i_y)$, ${\bf j}=(j_x,j_y)$ are the site indices,
$\mu,\nu=1,2$ are the orbital indices, and $t_0$ is the chemical
potential. $H_{int}$ is the on-site interaction term. At the
mean-field level, it can be written as:~\cite{zho,jiang,andr}
\begin{eqnarray}
H_{int}=&&U\sum_{{\bf i}\mu\sigma\neq\bar{\sigma}} \langle
n_{{\bf i}\mu\bar{\sigma}}\rangle
n_{{\bf i}\mu\sigma}+U^{\prime}\sum_{{\bf i},\mu\neq\nu,\sigma\neq\bar{\sigma}}\langle
n_{{\bf i}\mu\bar{\sigma}}\rangle n_{{\bf i}\nu{\sigma}}
\nonumber\\
&&+(U^{\prime}-J_H)\sum_{{\bf i},\mu\neq\nu,\sigma}\langle
n_{{\bf i}\mu\sigma}\rangle n_{{\bf i}\nu\sigma}\;,
\end{eqnarray}
where $n_{{\bf i}\mu\sigma}$ is the density operator at site ${\bf
i}$ and orbital $\mu$, with spin $\sigma$. The quantity $U^{\prime}$
is taken to be $U-2J_H$.~\cite{andr} The impurity part of the
Hamiltonian, $H_{imp}$, is given by:
\begin{equation}
H_{imp}=\sum_{{\bf i_m}\mu\sigma}V_s c^{\dagger}_{{\bf
i_m}\mu\sigma}c_{{\bf i_m}\mu\sigma}\;.
\end{equation}
Following Refs.~[\onlinecite{zho,zhang,tsai}], we here consider only
the intra-orbital scattering by a nonmagnetic impurity.

The mean-field Hamiltonian (1) can be diagonalized by solving the
BdG equations self-consistently,
\begin{equation}
\sum_{\bf j}\sum_\nu \left( \begin{array}{cc}
 H_{{\bf i}\mu {\bf j}\nu\sigma} & \Delta_{{\bf i}\mu {\bf j}\nu}  \\
 \Delta^{*}_{{\bf i}\mu {\bf j}\nu} & -H^{*}_{{\bf i}\mu {\bf j}\nu\bar{\sigma}}
\end{array}
\right) \left( \begin{array}{c}
u^{n}_{{\bf j}\nu\sigma}\\v^{n}_{{\bf j}\nu\bar{\sigma}}
\end{array}
\right) =E_n \left( \begin{array}{c}
u^{n}_{{\bf i}\mu\sigma}\\v^{n}_{{\bf i}\mu\bar{\sigma}}
\end{array}
\right),
\end{equation}
with
\begin{eqnarray}
H_{{\bf i}\mu {\bf j}\nu\sigma}=&&-t_{{\bf i}\mu {\bf
j}\nu}+[U\langle n_{{\bf i}\mu\bar{\sigma}}\rangle+(U-2J_H)\langle
n_{{\bf i}\bar{\mu}\bar{\sigma}}\rangle\nonumber
\\&&+(U-3J_H)\langle n_{{\bf i}\bar{\mu}\sigma}\rangle+v_s
\delta_{{\bf i},{\bf i_m}}-t_0]\delta_{\bf ij}\delta_{\mu\nu}\;,
\end{eqnarray}
and
\begin{eqnarray}
\Delta_{{\bf i}\mu {\bf j}\nu}=\frac{V_{{\bf i}\mu {\bf j}\nu}}{4}\sum_n
(u^{n}_{{\bf i}\mu\uparrow}v^{n*}_{{\bf j}\nu\downarrow}+u^{n}_{{\bf j}\nu\uparrow}v^{n*}_{{\bf i}\mu\downarrow})\tanh
(\frac{E_n}{2K_B T})\;,
\end{eqnarray}
\begin{eqnarray}
\langle n_{{\bf i}\mu}\rangle &=&\sum_n
|u^{n}_{{\bf i}\mu\uparrow}|^{2}f(E_n)+\sum_n
|v^{n}_{{\bf i}\mu\downarrow}|^{2}[1-f(E_n)]\;.
\end{eqnarray}
Here $V_{{\bf i}\mu {\bf j}\nu}$ is the pairing strength and $f(x)$
is the Fermi-Dirac distribution function. The SC order parameter at
site ${\bf i}$ is defined as

\begin{equation}
\Delta_{i}=\frac{\Delta_{i,i+\hat{x}+\hat{y}}+\Delta_{i,i-\hat{x}-\hat{y}}+
\Delta_{i,i+\hat{x}-\hat{y}}+\Delta_{i,i-\hat{x}+\hat{y}}}{4}\;,
\end{equation}
in accordance with the $s_{\pm}$ pairing symmetry.

The LDOS is calculated according to
\begin{equation}
\rho_{\bf i}(\omega)=\sum_{n\mu}[|u^{n}_{{\bf i}\mu\sigma}|^{2}\delta(E_n-\omega)+
|v^{n}_{{\bf i}\mu\bar{\sigma}}|^{2}\delta(E_n+\omega)]\;,
\end{equation}
where the delta function $\delta(x)$ is taken as
$\Gamma/\pi(x^2+\Gamma^2)$, with the quasiparticle damping
$\Gamma=0.01$.

Following Ref.~[\onlinecite{zhang}], we use the hopping constants,
\begin{eqnarray}
t_{{\bf i}\mu,{\bf i}\pm\hat{\alpha}\mu}&=&t_1 \qquad (\alpha=\hat{x},\hat{y})\;,\\
t_{{\bf i}\mu,{\bf i}\pm(\hat{x}+\hat{y})\mu}&=&\frac{1+(-1)^{\bf i}}{2}t_2+\frac{1-(-1)^{\bf i}}{2}t_3\;,\\
t_{{\bf i}\mu,{\bf i}\pm(\hat{x}-\hat{y})\mu}&=&\frac{1+(-1)^{\bf i}}{2}t_3+\frac{1-(-1)^{\bf i}}{2}t_2\;,\\
t_{i\mu,{\bf i}\pm\hat{x}\pm\hat{y}{\nu}}&=&t_4 \qquad (\mu\neq \nu)\;.
\end{eqnarray}

In the present work, we use $t_{1-4}=1, 0.4, -2, 0.04$.~\cite{zhang}
$t_0$ is determined by the electron filling per site $n$ $(n=2+x)$.
The on-site Coulombic interaction $U$ and Hund's coupling $J_H$ are
taken as $3.4$ and $1.3$, respectively. The pairing is chosen as nnn
intra-orbital pairing with the pairing strength $V=1.2$. This kind
of pairing is consistent with the $s_{\pm}$-pairing~\cite{mazin} and
has been widely used in previous theoretical studies based on the
BdG technique.~\cite{zho,zhou,jiang} The numerical calculation is
performed on a $32\times 32$ square lattice with the periodic
boundary conditions. A $30\times 30$ supercell is taken to calculate
the LDOS. Throughout the paper, the energy and length are measured
in units of $t_{1}$ and the Fe-Fe distance $a$, respectively. The
temperature is set to be $T=0$. In the following, all the results we
presented have been checked by using different initial values and
they remain qualitatively the same, suggesting the reliability of
our calculation.

\section{FS in the SDW state}
\label{SEC:fs}
\begin{figure}
      \includegraphics[width=1.65in]{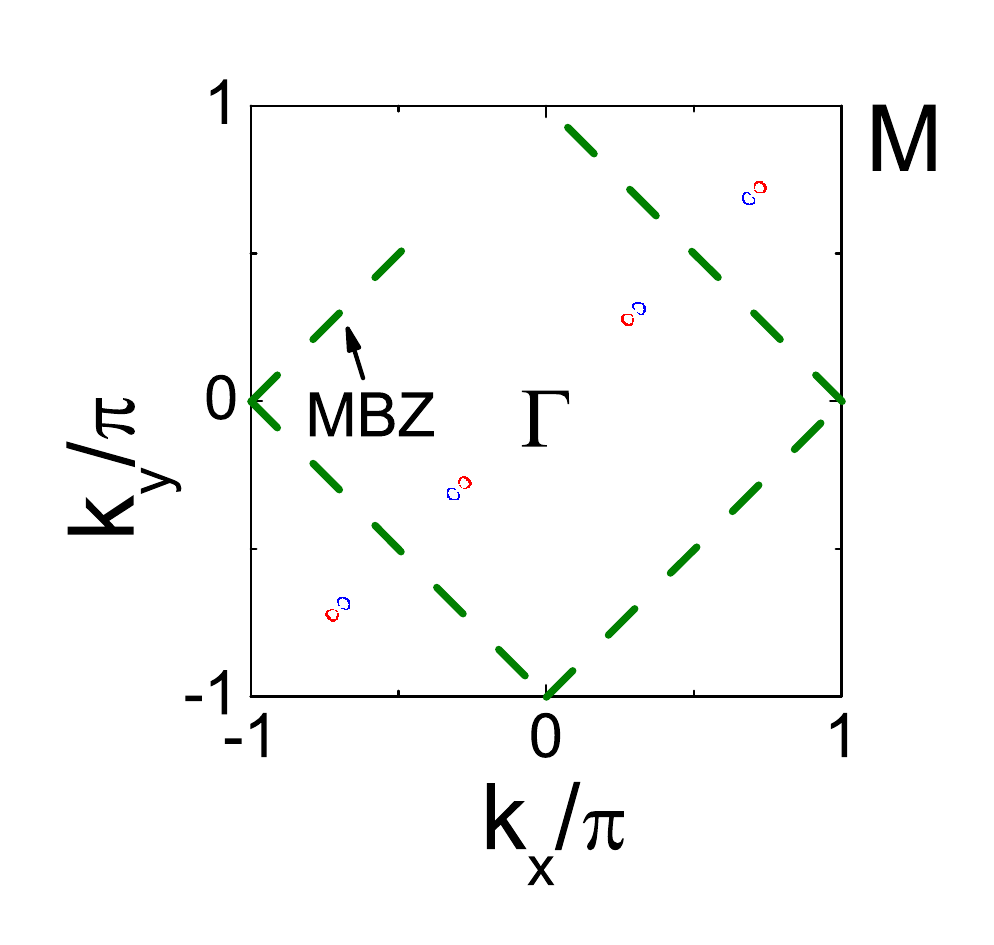}
      \includegraphics[width=1.7in]{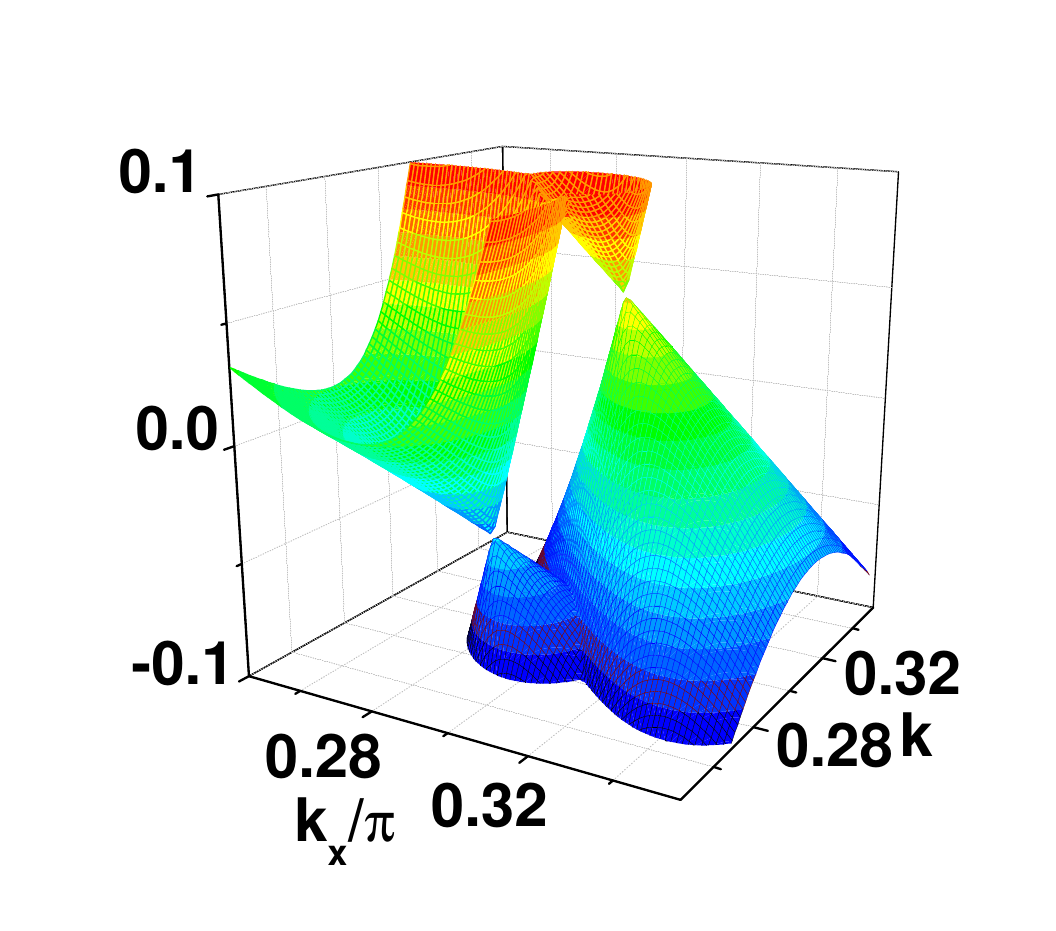}
      \includegraphics[width=1.8in]{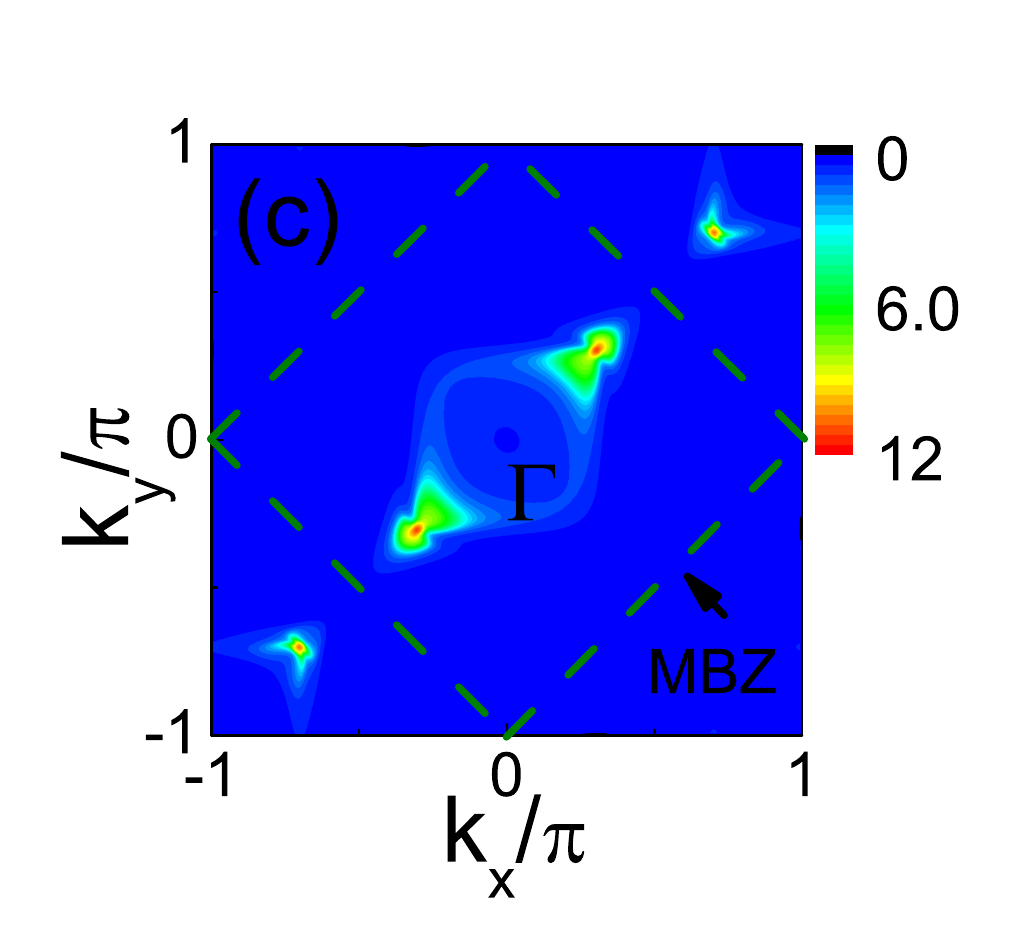}
      \includegraphics[width=1.55in]{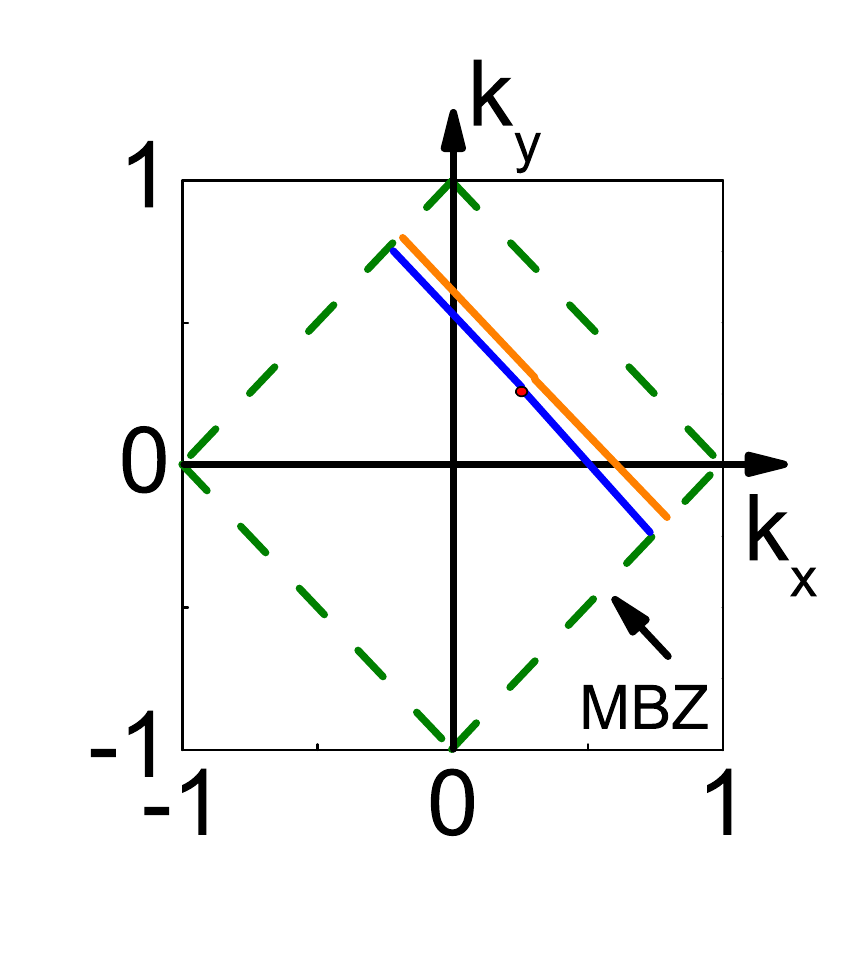}
      \includegraphics[width=1.65in]{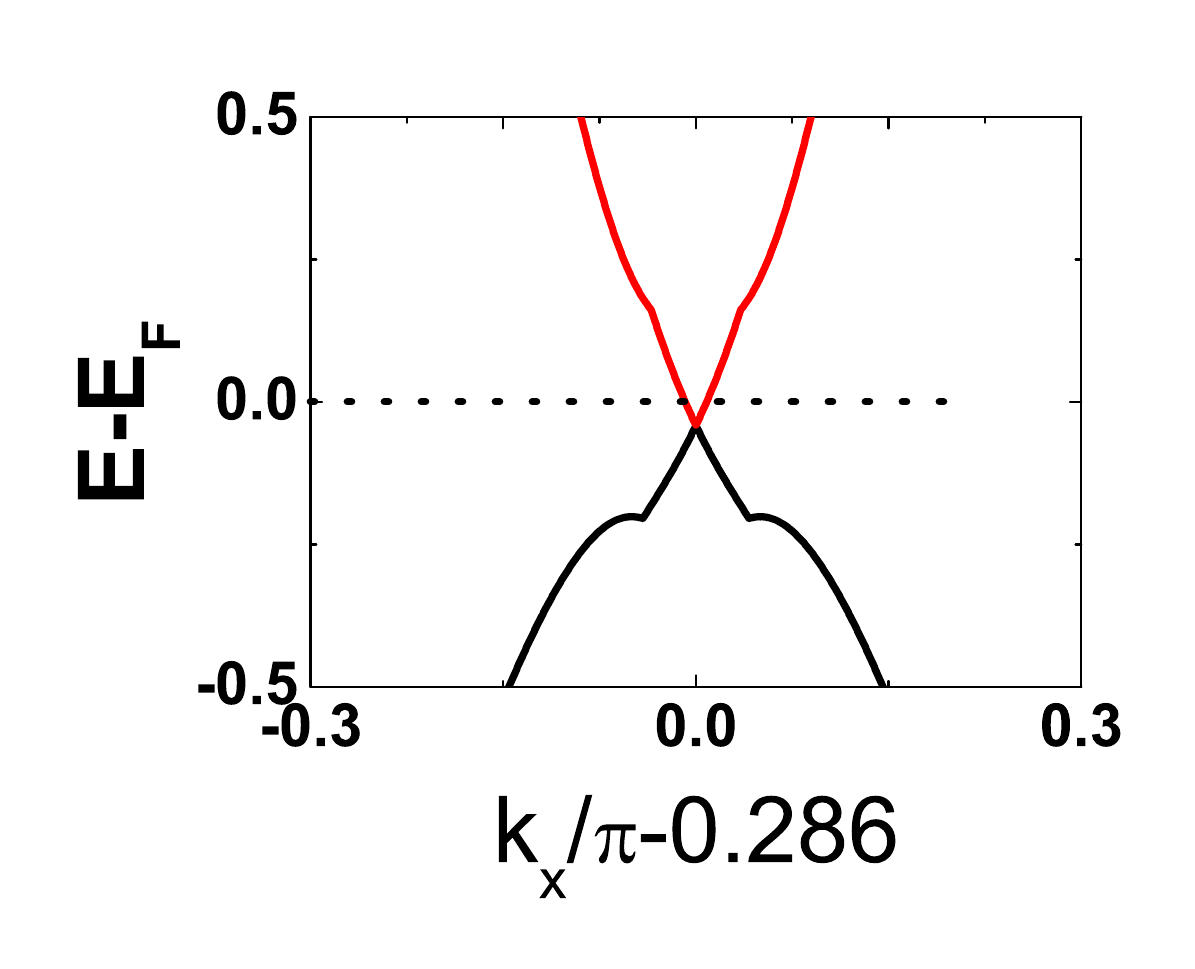}
      \includegraphics[width=1.65in]{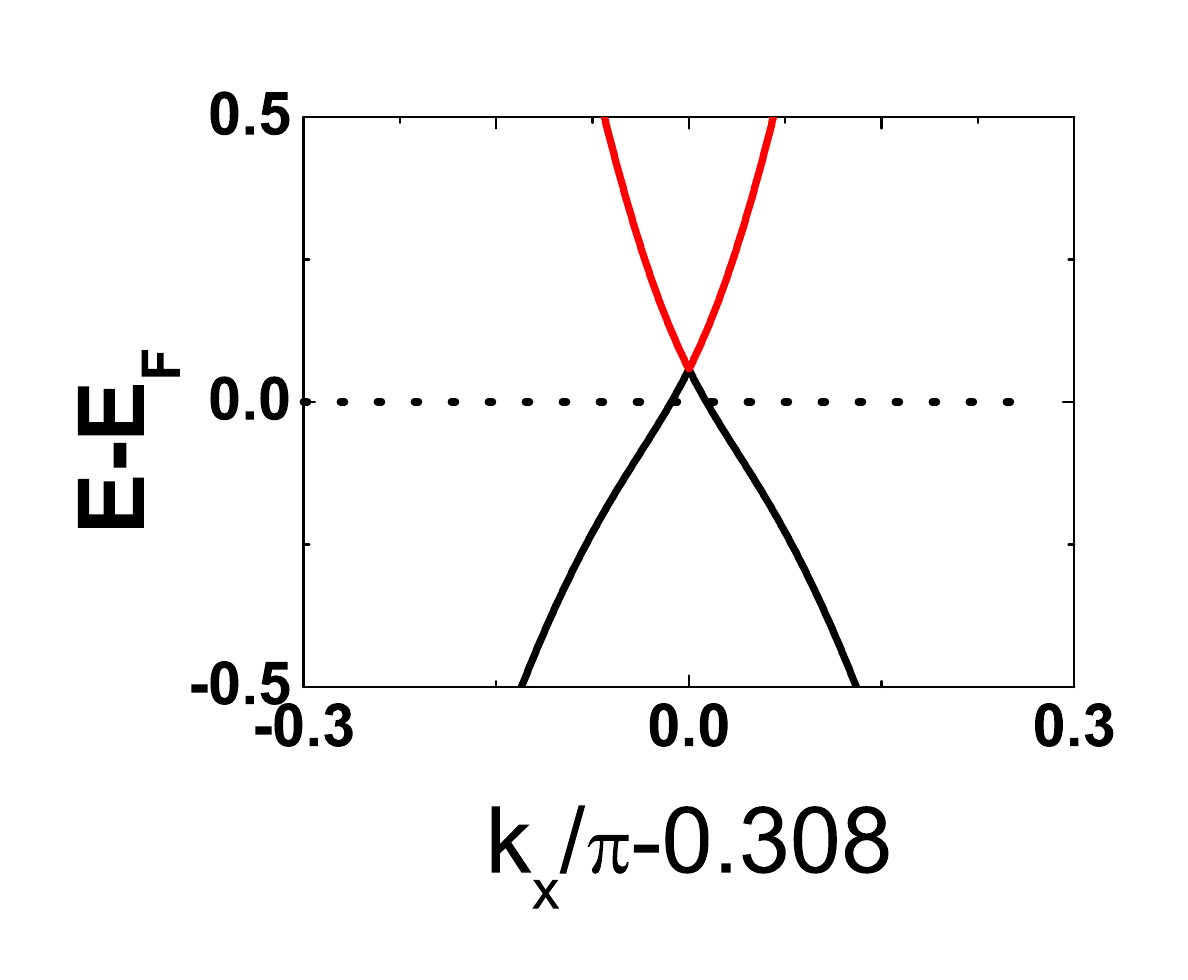}
\caption{(Color online) (a) The zero temperature SDW FS. (b) Two
Dirac cones at $(k_{x},k_{y})=(0.286\pi,0.286\pi)$ and
$(k_{x},k_{y})=(0.308\pi,0.308\pi)$, respectively. (c) The spectral
function $A(\mathbf{k},\omega)$ integrated from $\omega=-0.1$ to
$\omega=0.1$. (e) and (f) are the band structures near the Fermi
energy along the blue [goes through
$(k_{x},k_{y})=(0.286\pi,0.286\pi)$] and orange [goes through
$(k_{x},k_{y})=(0.308\pi,0.308\pi)$] lines in (d), respectively. The
BZ is defined in the 2Fe/cell representation and the green dashed
line in (a), (c) and (d) represents the MBZ.} \label{FIG:ferm}
\end{figure}

\begin{figure}
\centering
\includegraphics[width=1.6in]{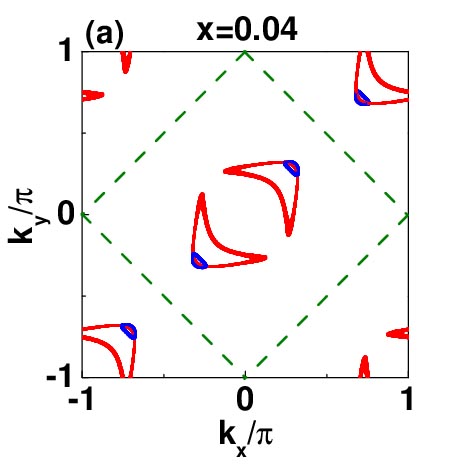}
 \includegraphics[width=1.6in]{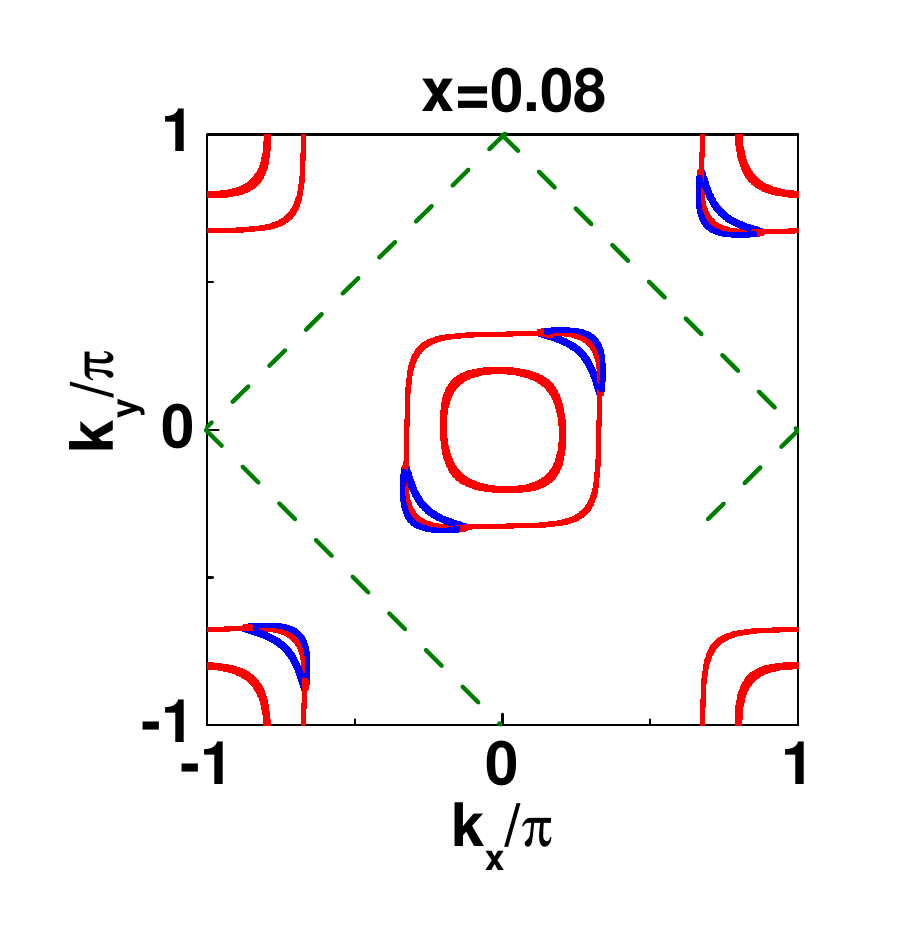}
  \includegraphics[width=1.6in]{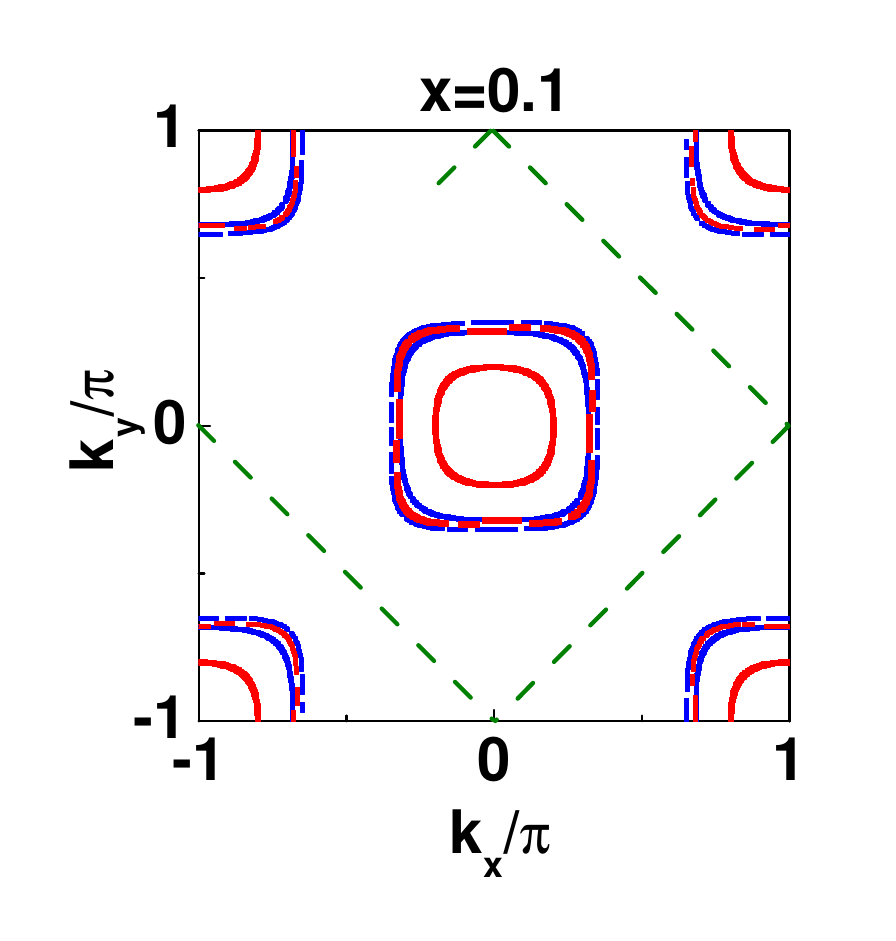}
   \includegraphics[width=1.6in]{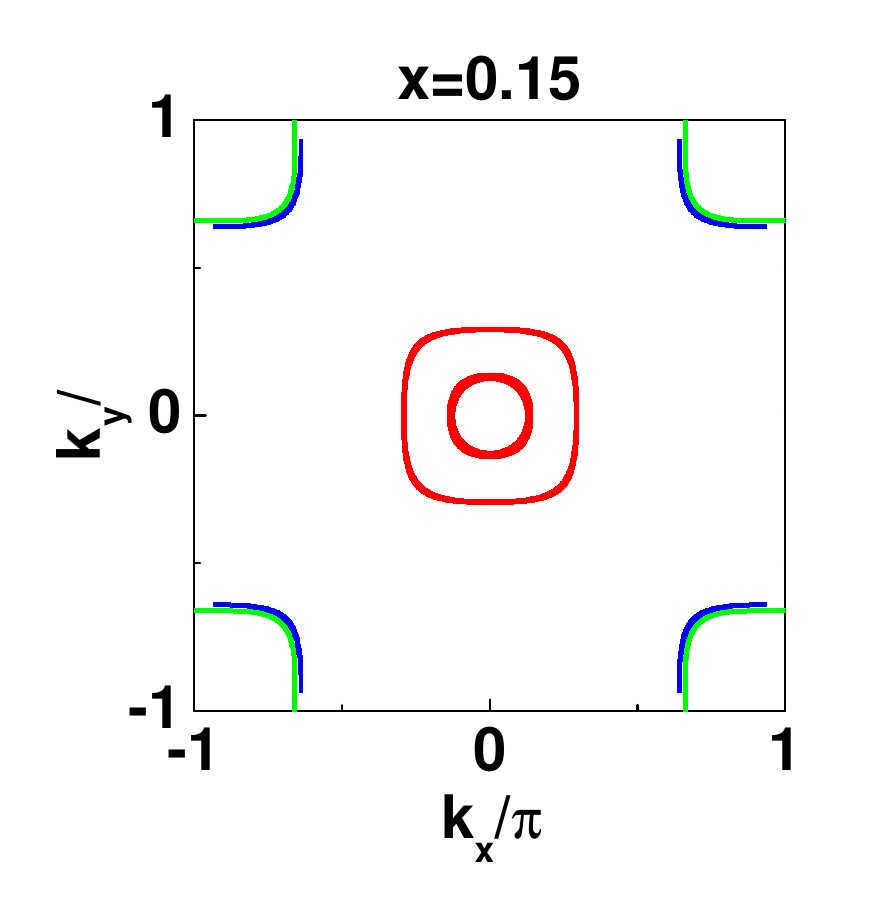}
\caption{(Color online) The zero temperature SDW FS at various
doping levels. The SC order $\Delta$ is artificially set to zero in
order to illustrate the effect of SDW on the evolution of the FS.
The blue and red pockets in the $x=0.04$, $0.08$ and $0.1$ cases are
both electron pockets. The green dashed line is the same as that in
Fig.~\ref{FIG:ferm}} \label{FIG:ferm2}
\end{figure}

In order to investigate the impurity effect in the iron-based
superconductors with SDW order, we first study the FS in the SDW
state.

At zero doping, below the SDW temperature, it was proposed
experimentally that there exist small FSs along the $\Gamma-M$ line
of the BZ and Dirac cones in the electronic structure form inside
these FSs, with their apices being located close to the Fermi
energy. However, whether these FSs and Dirac cones are electron- or
hole-like is within uncertainties of the experiment.~\cite{dingh} On
the other hand, theoretically it was shown that in both a two-band
model and a five-band model, nodes in the SDW gap function must
exist due to the symmetry-enforced degeneracy at the $\Gamma$ and
$M$ high-symmetry points, even in the presence of perfect nesting,
but the number and locations of these nodes are model
dependant.~\cite{theorydiraccone} Therefore, whether they correspond
to the experimentally observed Dirac cones is still unclear. In
Fig.~\ref{FIG:ferm}, we plot the zero temperature SDW FS and the
corresponding band structure near the Fermi energy obtained by our
self-consistent calculation. As we can see from Fig.~\ref{FIG:ferm}
(a), in the SDW state, there remain four small FS pockets in the
magnetic Brillouin zone (MBZ), two of which are electron-like (red)
located around $(k_{x},k_{y})=\pm(0.286\pi,0.286\pi)$, while the
other two are hole-like (blue) located around
$(k_{x},k_{y})=\pm(0.308\pi,0.308\pi)$. The pockets outside the MBZ
are just replica of those inside it due to band-folding in the SDW
state and they can be connected by the SDW wave vector
$\mathbf{Q}=(\pi,\pi)$. The areas enclosed by these pockets are
equal, thus keeping the doping level at $x=0$. Inside these four
pockets, there are four Dirac cones. As shown in Fig.~\ref{FIG:ferm}
(b), the apex of the Dirac cone is $0.026$ below the Fermi energy at
$(k_{x},k_{y})=\pm(0.286\pi,0.286\pi)$ while it is $0.046$ above the
Fermi energy at $(k_{x},k_{y})=\pm(0.308\pi,0.308\pi)$, suggesting
that they are electron- and hole-like Dirac cones, respectively. The
spectral function $A(\mathbf{k},\omega)$, which is proportional to
the photoemission intensity measured in ARPES experiments, is
integrated from $\omega=-0.1$ to $\omega=0.1$ and shown in
Fig.~\ref{FIG:ferm} (c). As we can see, the locations of the bright
spots are around $(k_{x},k_{y})=\pm(0.3\pi,0.3\pi)$ and the
equivalent symmetry points outside the MBZ, on the $\Gamma-M$ line,
in qualitative agreement with experiment.~\cite{dingh} In addition,
although most parts of the original FSs around $\Gamma$ are gapped
by the SDW order, the gap value is extremely small on these FSs.
Thus, around $\Gamma$, the low-energy spectral function has moderate
intensity and this can be seen from the ring structure around
$\Gamma$ with lower intensity, as compared to those bright spots. We
also notice that the system has only two-fold symmetry when entering
the SDW state while the experimentally observed four-fold symmetry
is due to the superposition of twin domains or domain averaging, as
suggested in Refs.~\onlinecite{dingh} and ~\onlinecite{hsieh},
respectively. The band structures near the Fermi energy scanned
along the blue and orange cuts in Fig.~\ref{FIG:ferm} (d) are
plotted in Figs.~\ref{FIG:ferm} (e) and ~\ref{FIG:ferm} (f),
respectively. It clearly shows the X-like structure of Dirac cones
and again suggests that the Dirac cone is electron-like at
$(k_{x},k_{y})=(0.286\pi,0.286\pi)$ and hole-like at
$(k_{x},k_{y})=(0.308\pi,0.308\pi)$. The locations of the FS pockets
and the bright spots in the spectral function are consistent with
the experimental observation, but in our calculation, the electron-
and hole-like Dirac cones appear in-pairs and are located very close
to each other along the $\Gamma-M$ line of the BZ, the apices of
which are both in the vicinity of the Fermi energy, thus we propose
this to be directly verified by future ARPES experiments with higher
resolution. In addition, the existence of electron and hole Dirac
cone states in-pairs has already been confirmed indirectly by
measuring the magnetoresistance.~\cite{diracconepair}

Fig.~\ref{FIG:ferm2} shows the evolution of the FS with doping.
Here, we set the SC order $\Delta$ to zero to illustrate the effect
of SDW on the evolution of the FS. In the MBZ, as doping increases,
the size of the electron pockets [the red pockets shown in
Fig.~\ref{FIG:ferm} (a)] is enlarged while that of the hole pockets
[the blue pockets shown in Fig.~\ref{FIG:ferm} (a)] is reduced. When
doping increases to about $x=0.02$, the hole pockets vanish
completely. By further increasing doping, another two electron
pockets appear in the MBZ [the blue pockets in the $x=0.04$, $0.08$
and $0.1$ cases shown in Fig.~\ref{FIG:ferm2}], exactly at the same
locations where the hole pockets vanish and overlap with the
original electron pockets. The size of all these electron pockets is
enlarged with doping. If we define the areas enclosed by the inner
and outer red lines to be $S_{1}$ and those enclosed between the
inner and outer blue lines to be $S_{2}$, then we have
$x=2N_{x}N_{y}(S_{1}+S_{2})$, with $N_{x}$, $N_{y}$ being the linear
dimensions of the square lattice. Finally, when $x=0.15$, the SDW
order disappears and there is no more band-folding due to it, in
this case, there are two electron pockets and two hole pockets
around the $M$ and $\Gamma$ points of the BZ, respectively.

\section{Impurity scattering effect in undoped sample}
\label{SEC:impforundoped}

Based on a toy model and phenomenological calculation,~\cite{han} it
was proposed that the impurity induced bound state should appear
near the impurity site for the undoped sample. While actually the
band structure and FS should be important issues for the features of
the order parameters and LDOS. Thus we will reexamine this issue
based on the two-orbital model and present a detailed investigation
of the nonmagnetic impurity effect in the parent compound. Here we
consider both positive and negative impurity SPs. A single impurity
is put at site $(16,16)$. We define the on-site magnetic order
parameter, $M_{\bf i}=(-1)^{i_x}\frac{1}{4}\sum_{\mu}(n_{{\bf
i}\mu\uparrow}-n_{{\bf i}\mu\downarrow})$. This definition is
suitable for the typical $(\pi,0)$ SDW order, consistent with
previous experiments~\cite{cruz} and theoretical
calculations.~\cite{zho,jiang}

\begin{figure}
\centering
      \includegraphics[width=1.6in]{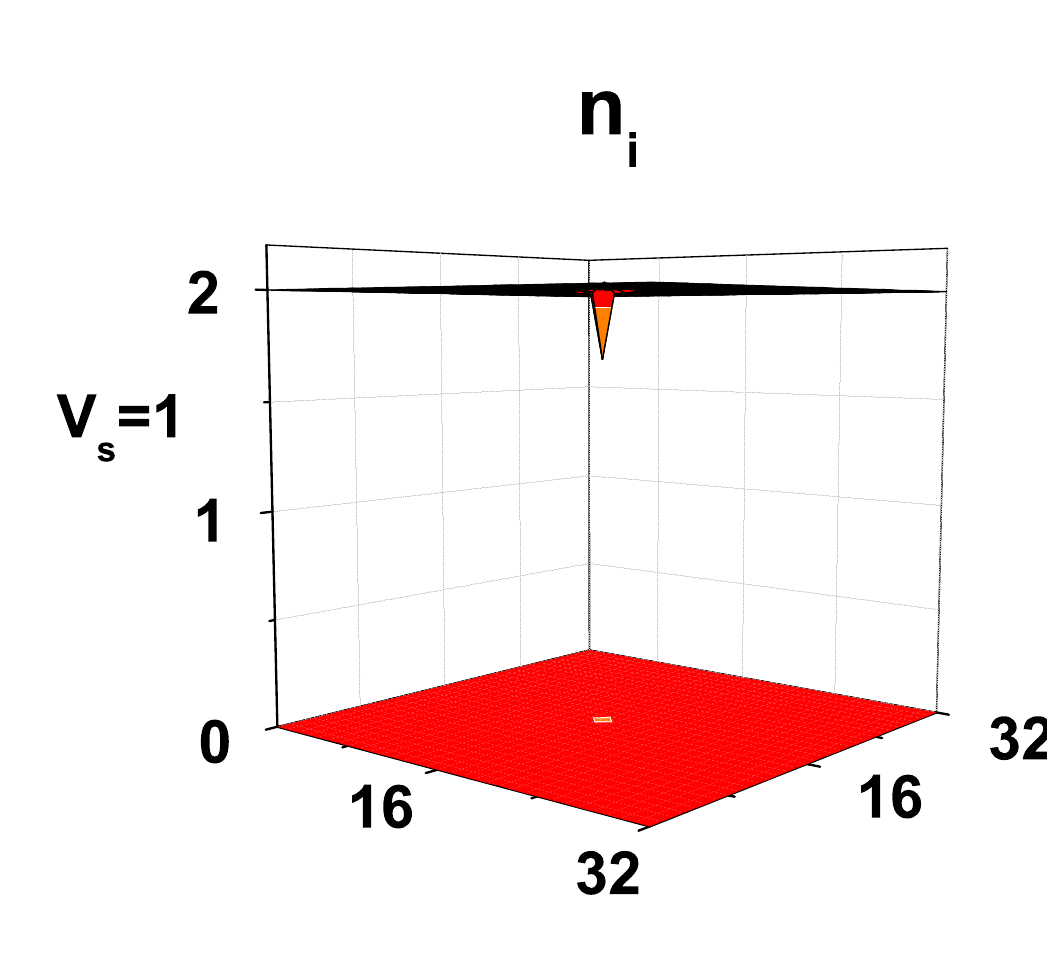}
      \includegraphics[width=1.6in]{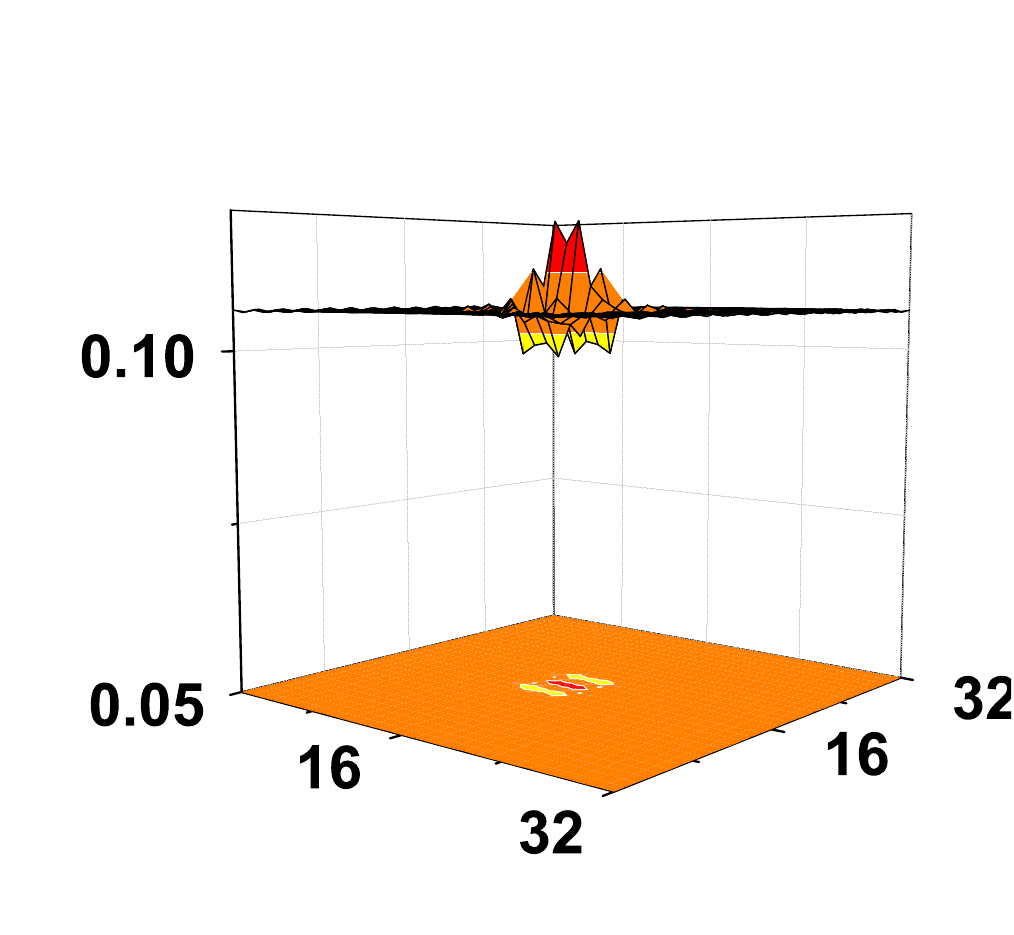}
      \includegraphics[width=1.6in]{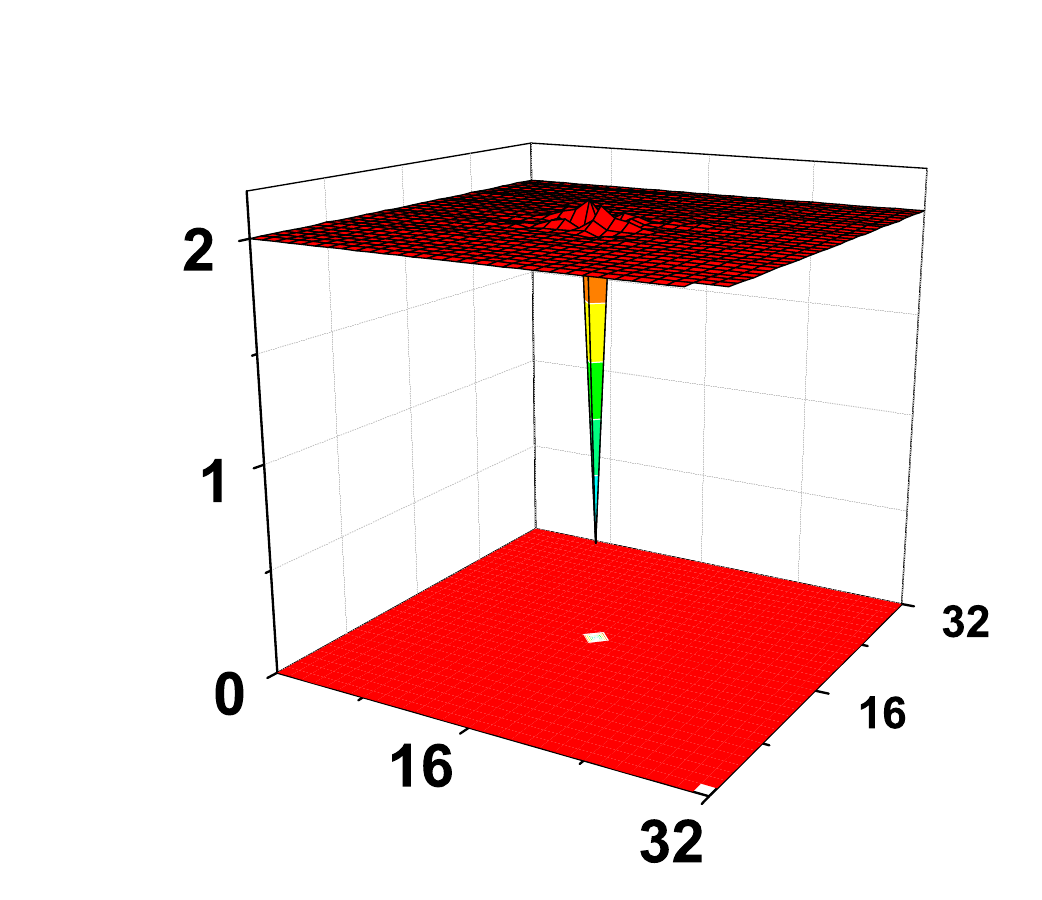}
      \includegraphics[width=1.6in]{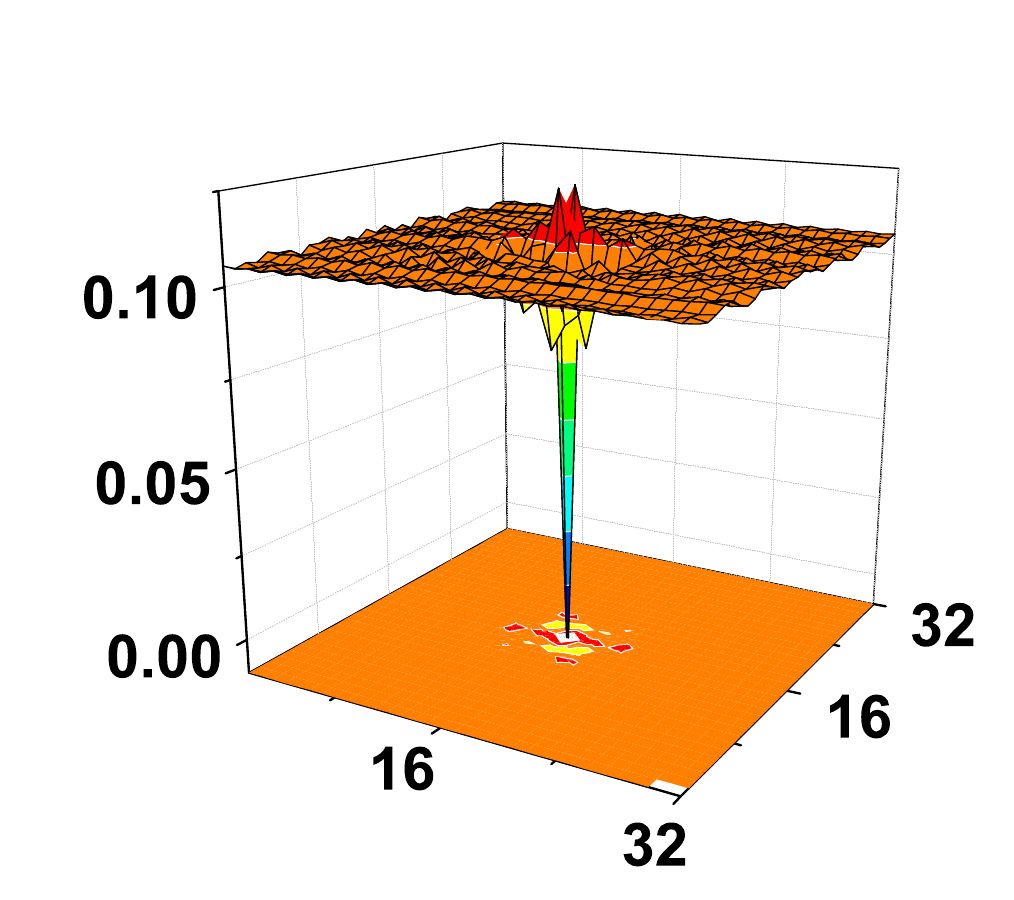}
      \includegraphics[width=1.6in]{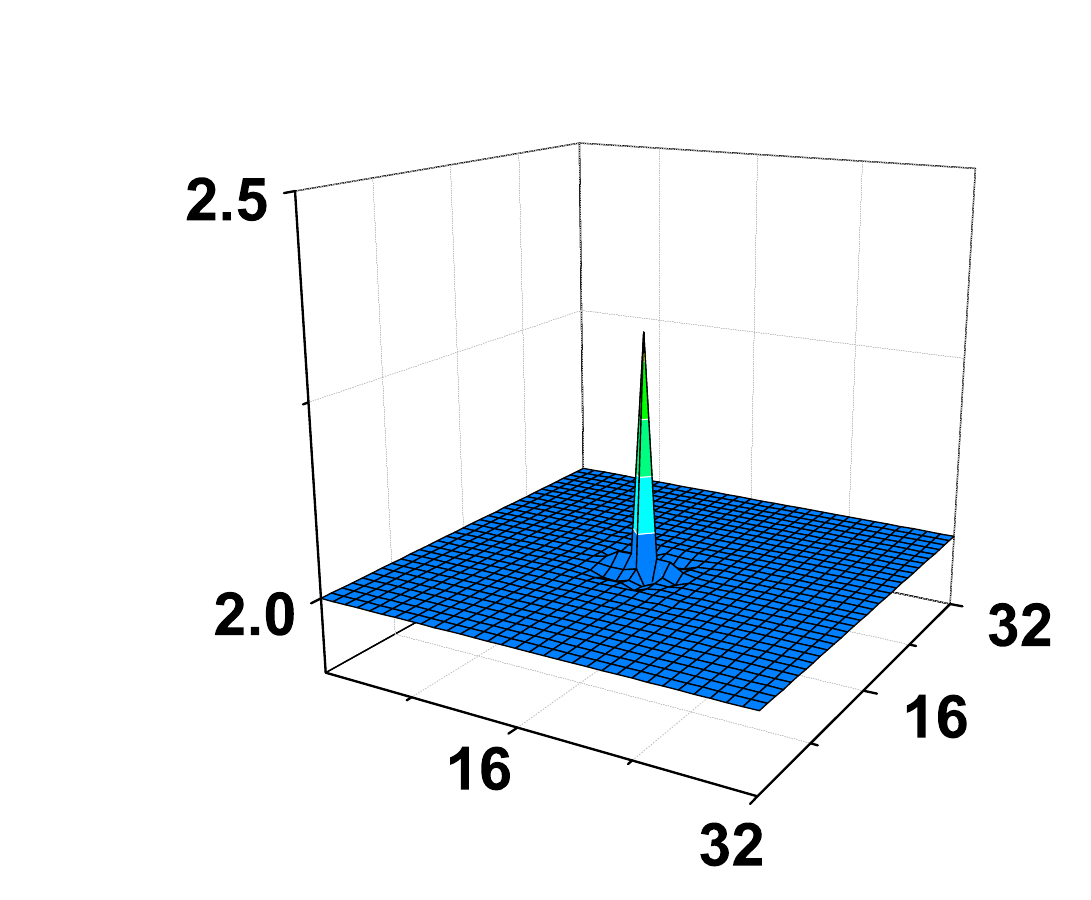}
      \includegraphics[width=1.6in]{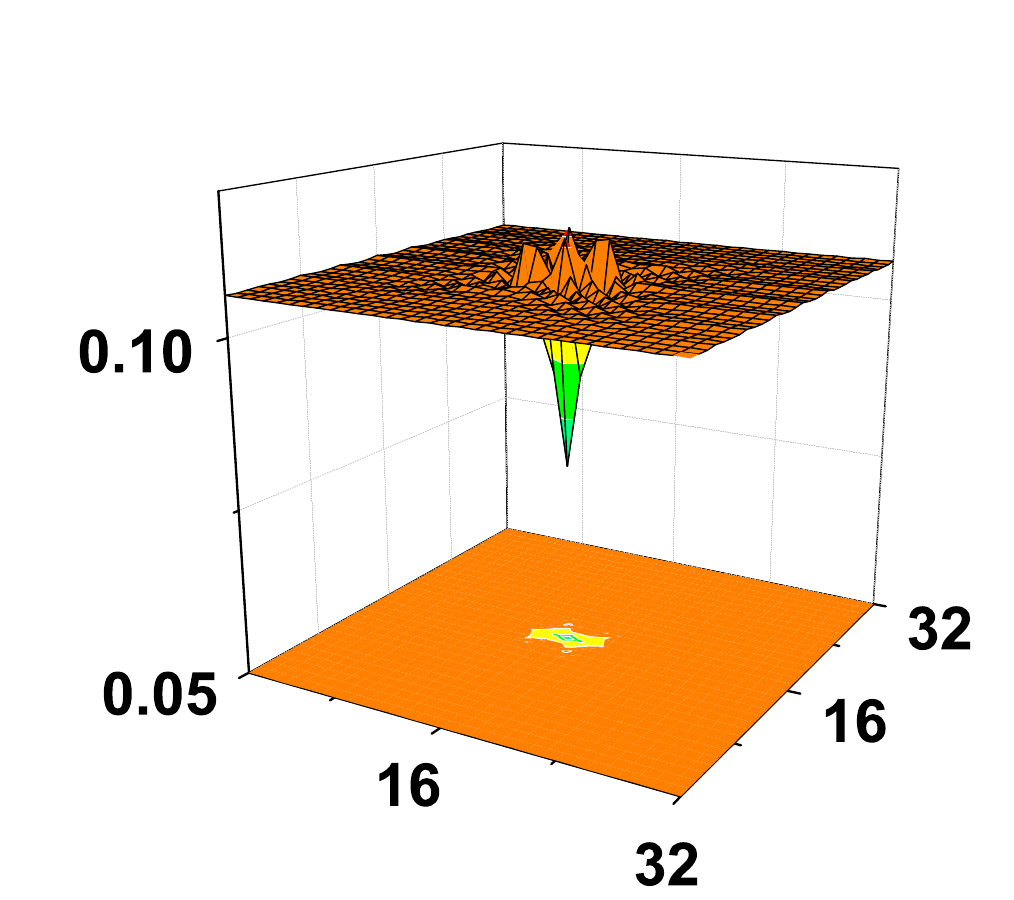}
      \includegraphics[width=1.6in]{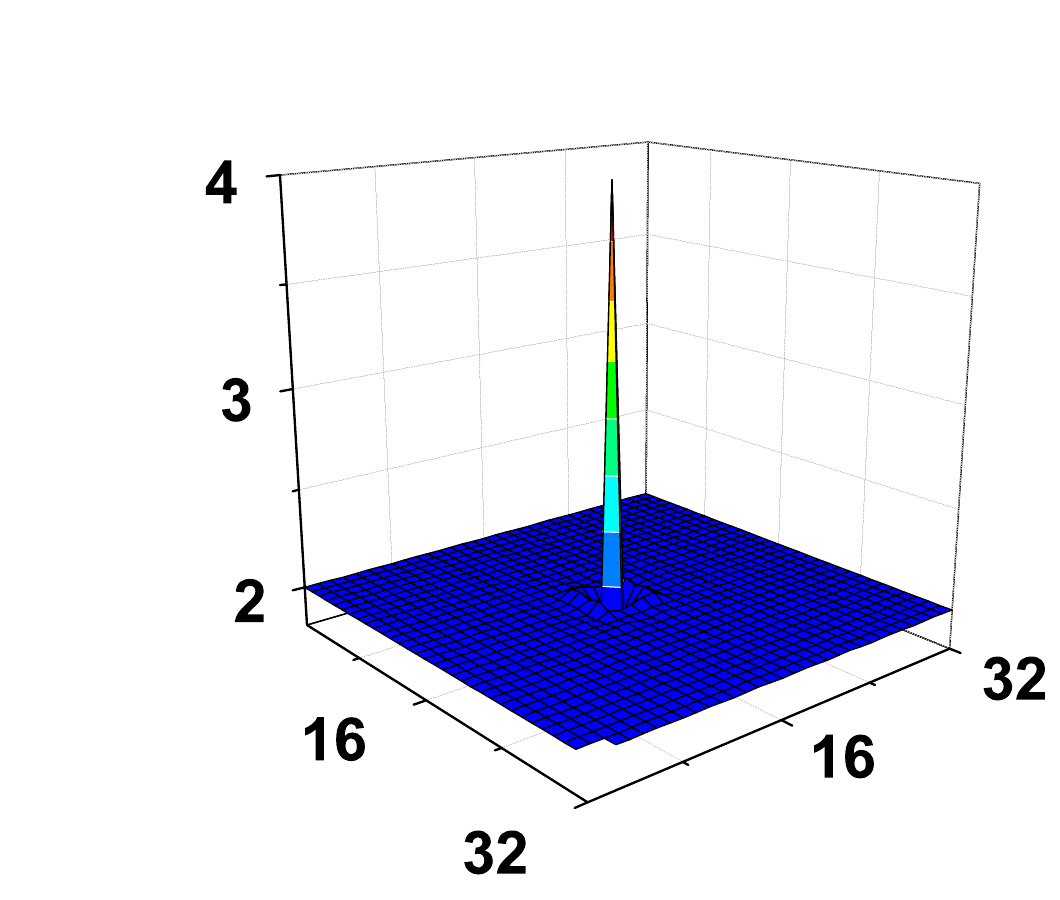}
      \includegraphics[width=1.6in]{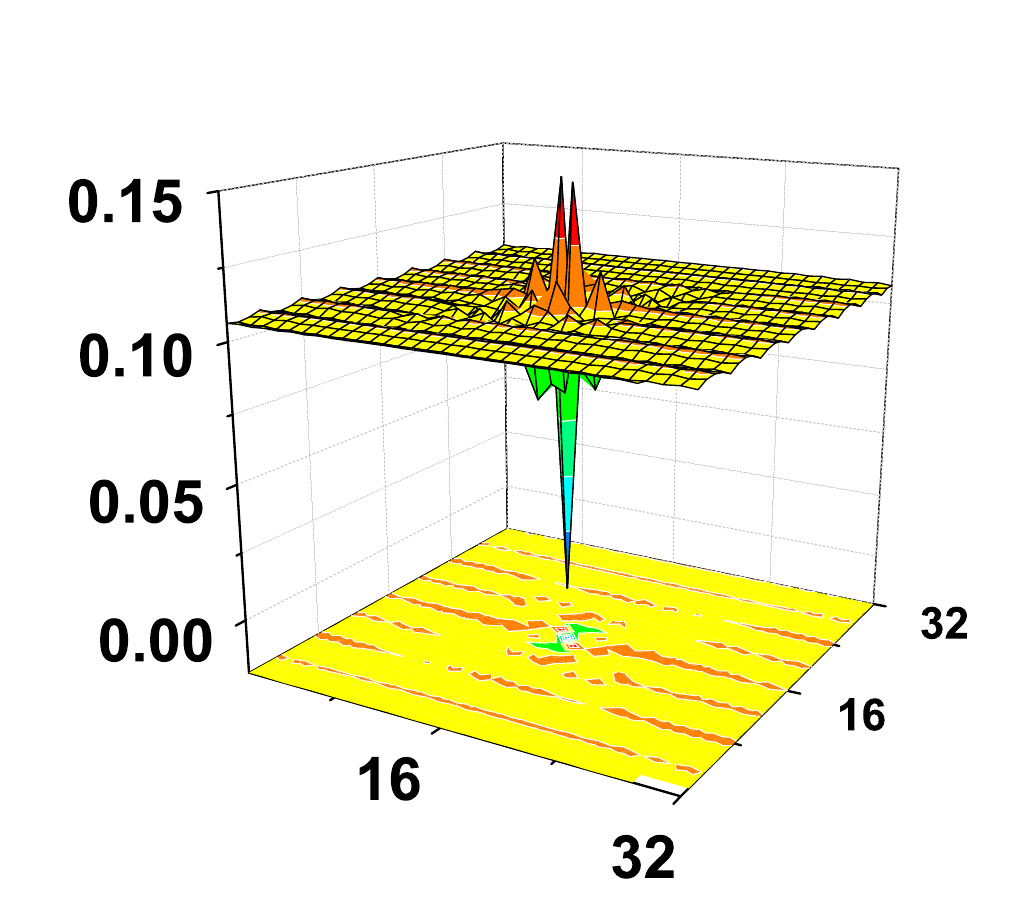}
\caption{(Color online) The intensity plots of the particle number
(left panels) and magnetic order (right panels) at zero doping and
zero temperature for different SPs $V_s=1,5,-1,-100$.}
\label{FIG:dop0}
\end{figure}

The intensity plots of the site-dependent particle number $n_i=
\sum_{\mu}(n_{i\mu\uparrow}+n_{i\mu\downarrow})$  and magnetic order
$M_i$ in real space are shown in Fig.~\ref{FIG:dop0}. The left
panels of Fig.~\ref{FIG:dop0} plot the spatial distribution of the
particle number. For positive SP, electrons are repelled by the
impurity, therefore at the impurity site the value of particle
number is reduced. Increasing the positive SP will lead to smaller
values of $n_i$ at the impurity site, which, when $V_s>6$, will
vanish. For negative SP, on the contrary, electrons are attracted to
the impurity and larger $|V_s|$ will lead to a higher particle
number at the impurity site. The particle numbers will recover to
the bulk value $2.0$ at about $2$ lattice constants away from the
impurity site for both positive and negative SPs. In doped samples,
these characteristics of $n_i$ do not change except that the value
of $n_i$ far away from the impurity site will be $2+x$, where $x$ is
the electron doping concentration.

The right panels of Fig.~\ref{FIG:dop0} show the real space
modulation of the magnetic order $M_i$. For small positive SP $V_s=1$, the values of
magnetic order oscillate near the impurity site with the maximum
$M_i=0.115$ at the impurity site, slightly higher than the bulk
value $0.105$. For moderate SP $V_s=5$, the magnitude of $M_i$ drops
down almost to zero at the impurity site, with the maximum
$M_i\sim0.12$ appearing in the vicinity of the impurity site. We
thus expect that stronger SP will lead to stronger oscillation of
$M_i$ around the impurity and this is proved by setting $V_s=\pm100$
which we take $V_s=-100$ as an example as shown in
Fig.~\ref{FIG:dop0}. From the corresponding plot we can see that the
modulation of $M_i$ is very strong around the impurity, similar to
Friedel oscillation and it will recover to the impurity-free value
at about $4\backsim5$ lattice constants away from the impurity site.
Order parameters for negative SP $V_s=-1$ are also shown, unlike the
enhanced $M_i$ in the $V_s=1$ case, we find that the magnitude of
$M_i$ is suppressed at the impurity site and the oscillation of
$M_i$ is stronger than the $V_s=1$ case.

\begin{figure}
\centering
  \includegraphics[width=4.2cm]{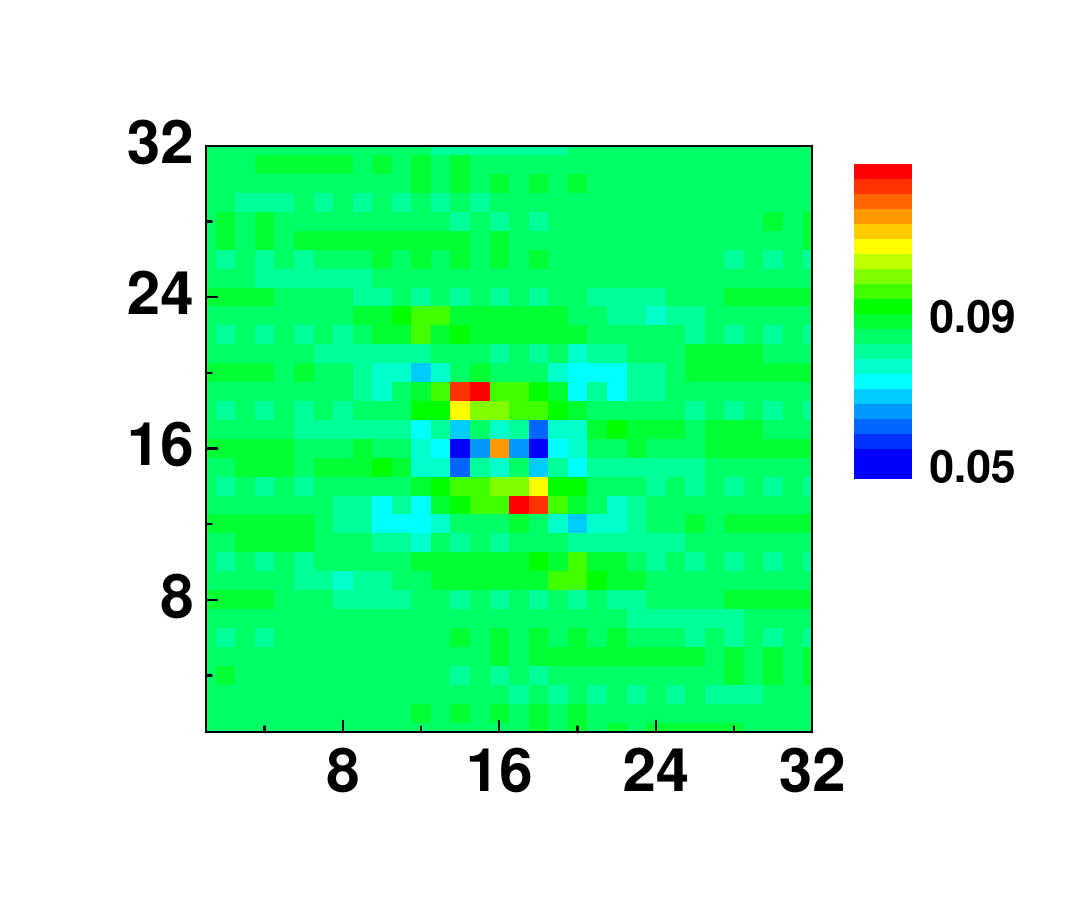}
  \includegraphics[width=4.2cm]{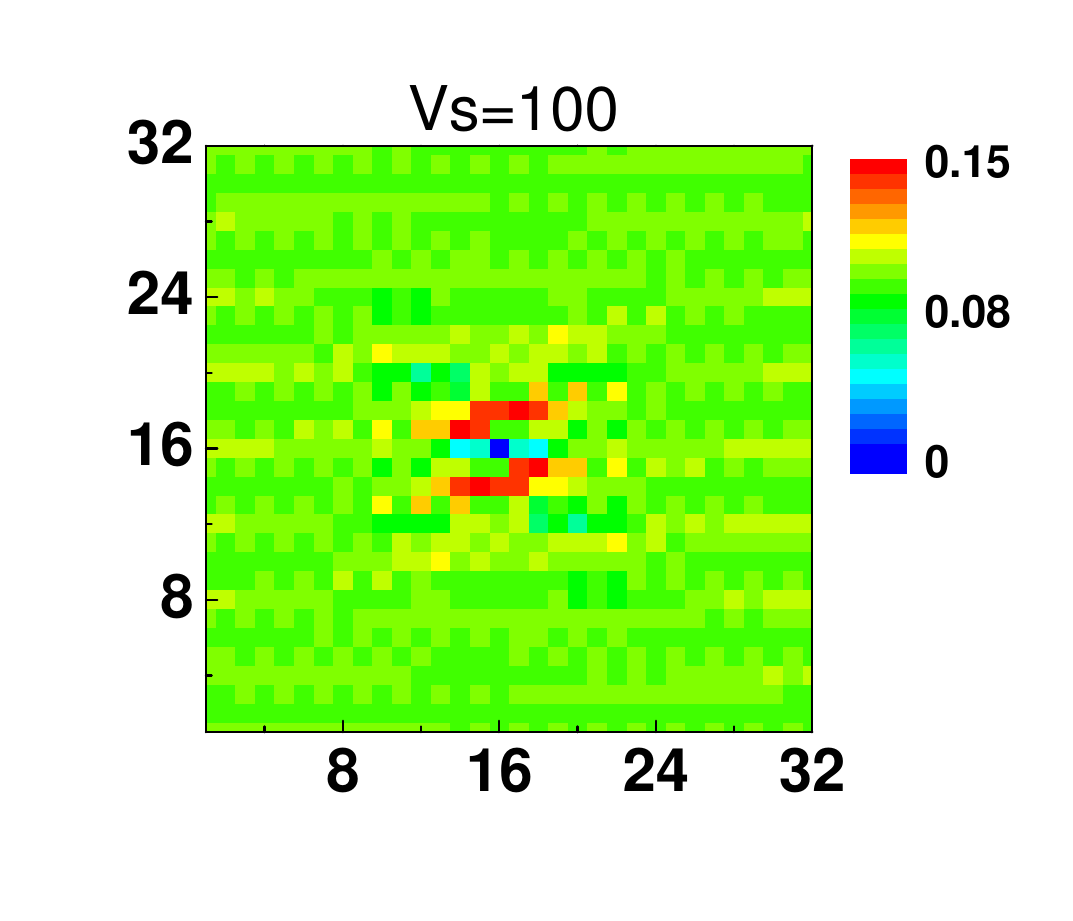}
\caption{(Color online) The intensity plots of the LDOS at zero
energy for different SPs $V_s=3$ and $V_s=100$. } \label{FIG:c2}
\end{figure}

We now study the low-energy impurity-induced bound states. The
intensity plots of the LDOS in real space at zero energy are shown
in Fig.~\ref{FIG:c2} for $V_s=3$ and $V_s=100$, respectively. As seen,
for $V_s=3$ the LDOS at the impurity site $(16,16)$ is
finite while for nearly unitary SP $V_s=100$ it vanishes. The LDOS modulates
near the impurity site and some bright spots can be seen clearly
around the impurity, indicating the existence of bound states at low
energy. Another prominent feature revealed from the LDOS map in
Fig.~\ref{FIG:c2} is the four-fold symmetry breaking which is more
obvious near the impurity. The symmetry of the system reduces to
$C_2$ and it survives for various SPs no matter they are positive or
negative. Furthermore, there also exists one-dimensional modulation
of the LDOS along the $y-$axis even when it is far away from the
impurity. This feature is similar to the experimentally observed
nematic electronic structure,~\cite{nematic} thus supporting the
impurity effect as a possible candidate for the formation of nematic
order. Since in the two-orbital model we use, each unit cell
contains two inequivalent Fe atoms, the existence of an impurity on
either site of the unit cell will naturally break the four-fold
symmetry of the system, thus we conclude that the breaking of the
four-fold symmetry in the LDOS is induced not only by the SDW order,
but also by the intrinsic asymmetry pinned by the impurity.

\begin{figure}
\centering
  \includegraphics[width=7cm]{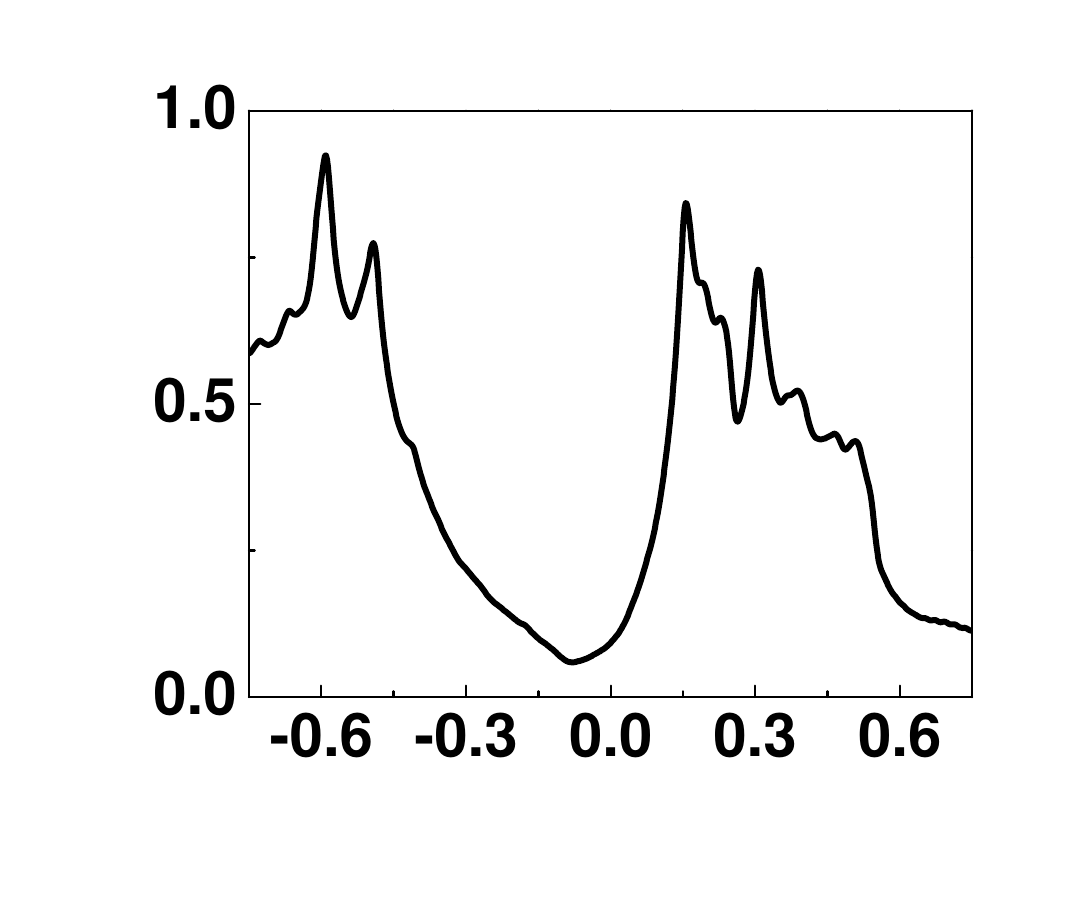}
\caption{The bulk LDOS in the SDW state of undoped sample.}
\label{FIG:bulkdp0}
\end{figure}

\begin{figure}
\centering
  \includegraphics[width=4.2cm]{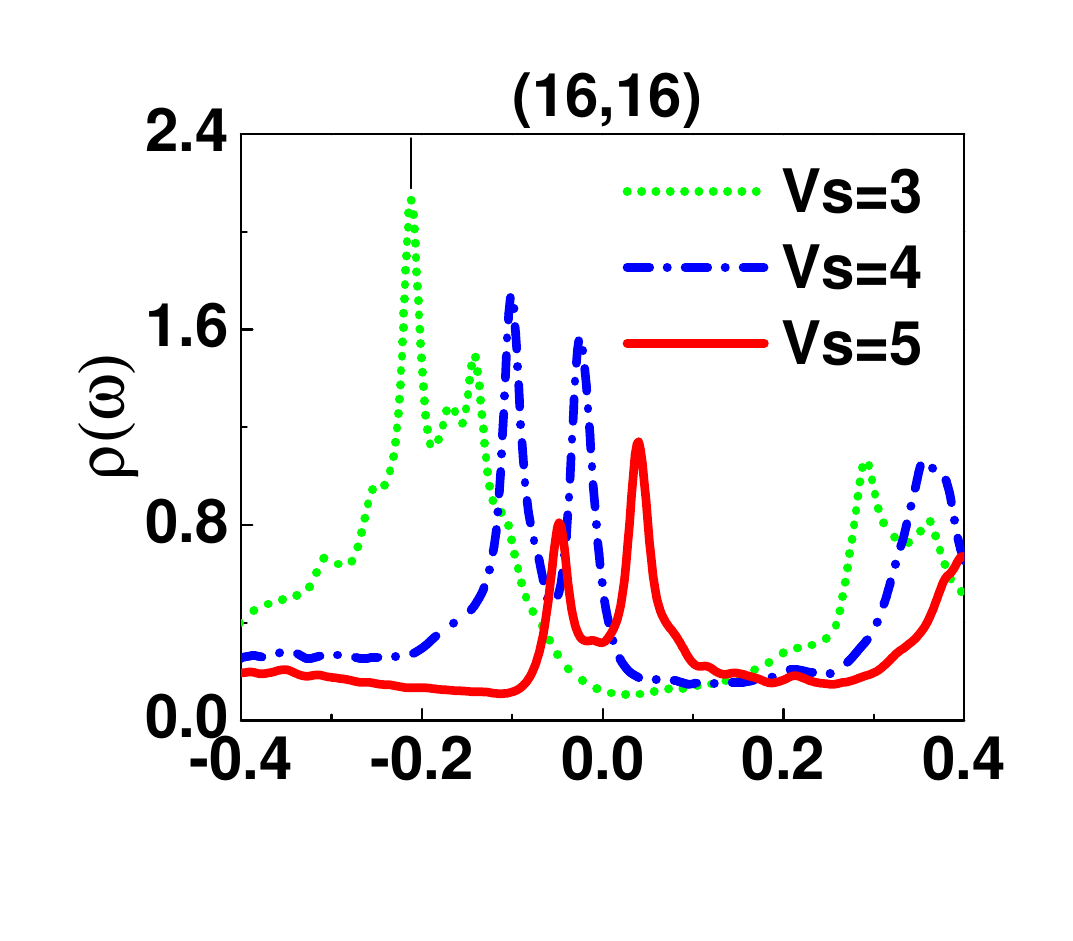}
  \includegraphics[width=4.2cm]{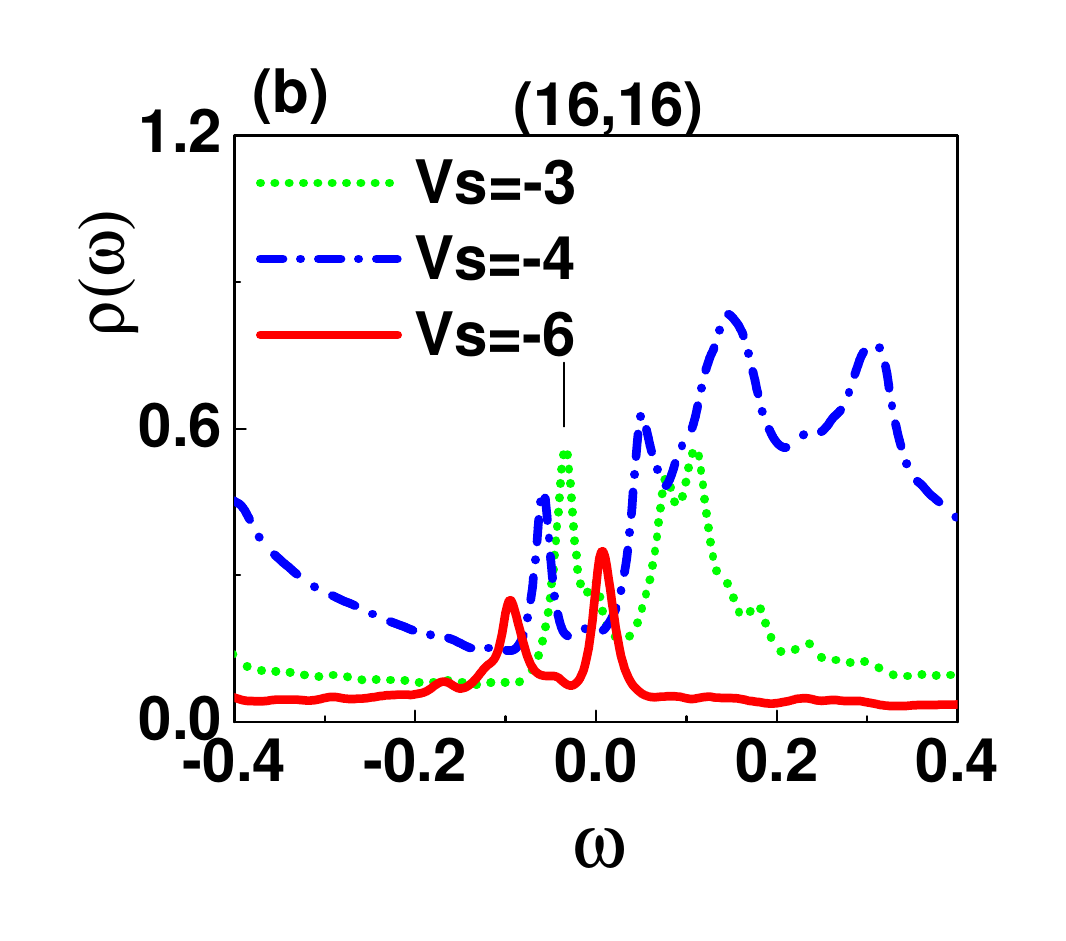}
  \includegraphics[width=4.2cm]{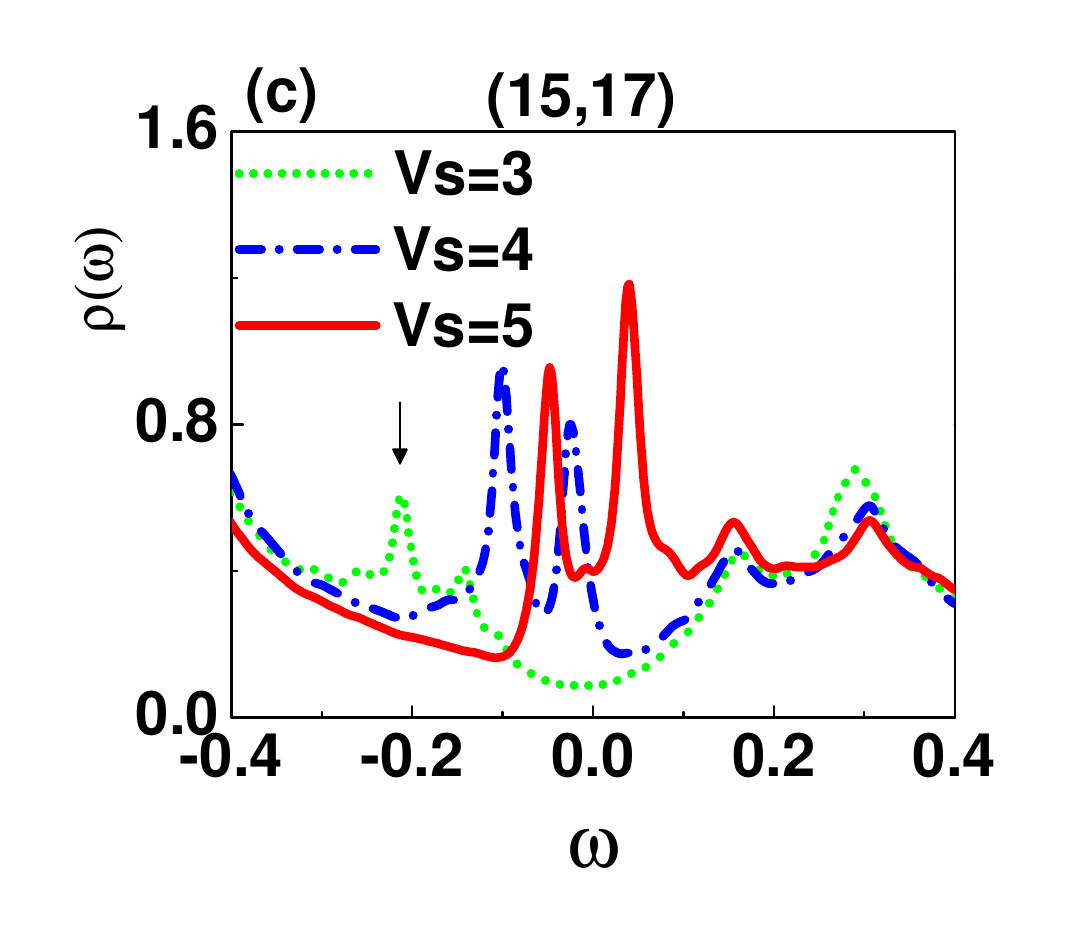}
  \includegraphics[width=4.2cm]{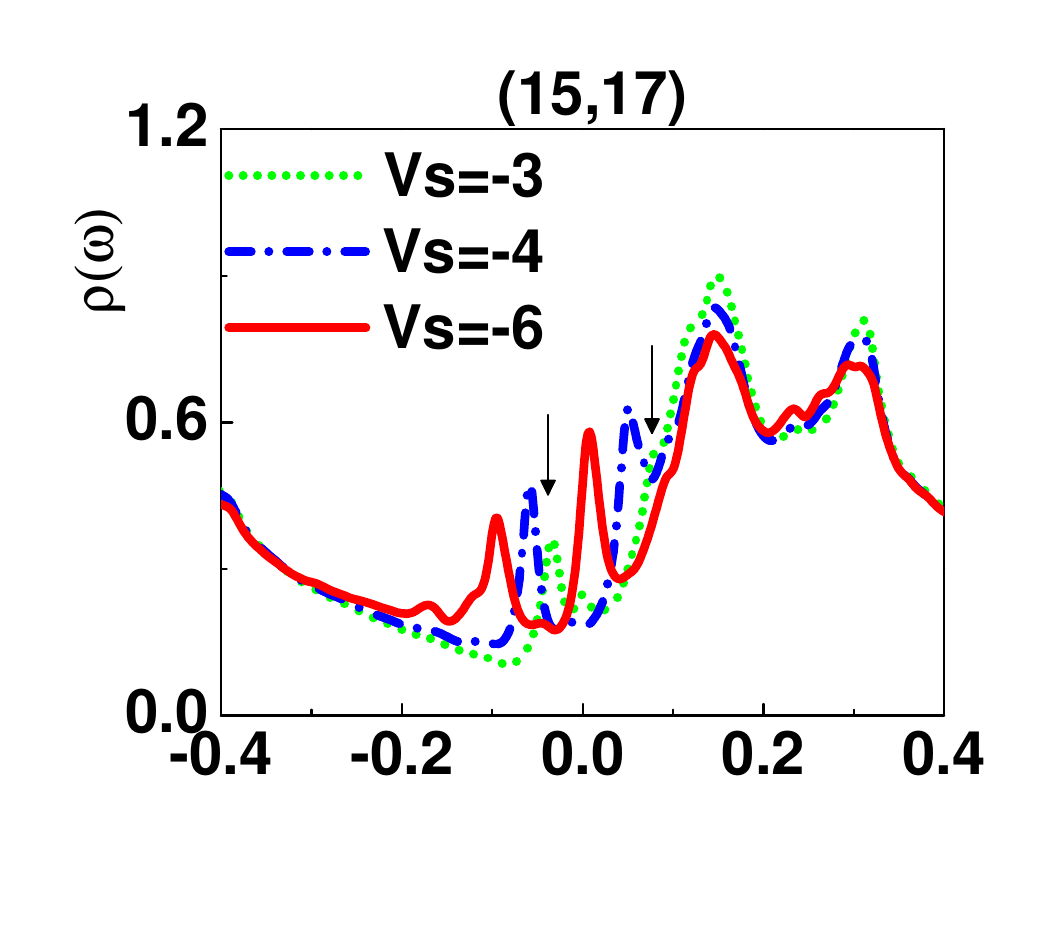}
  \includegraphics[width=4.2cm]{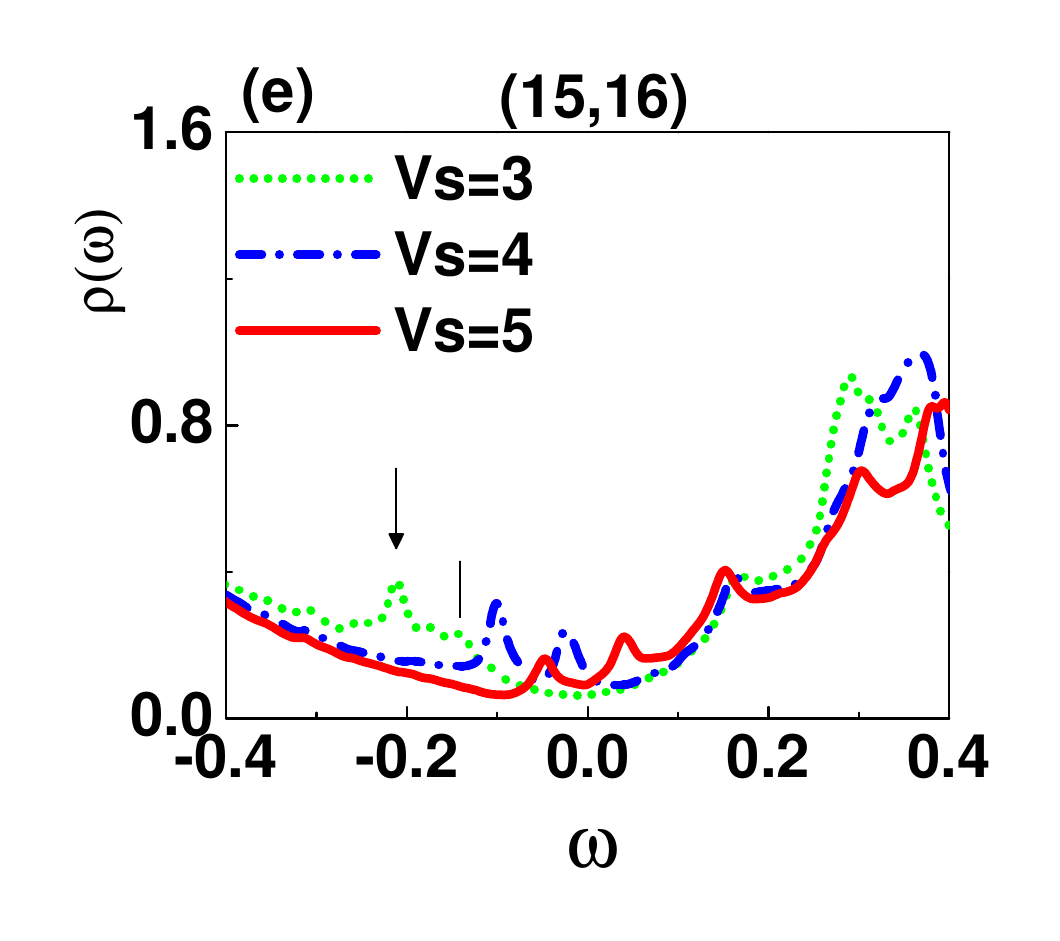}
  \includegraphics[width=4.2cm]{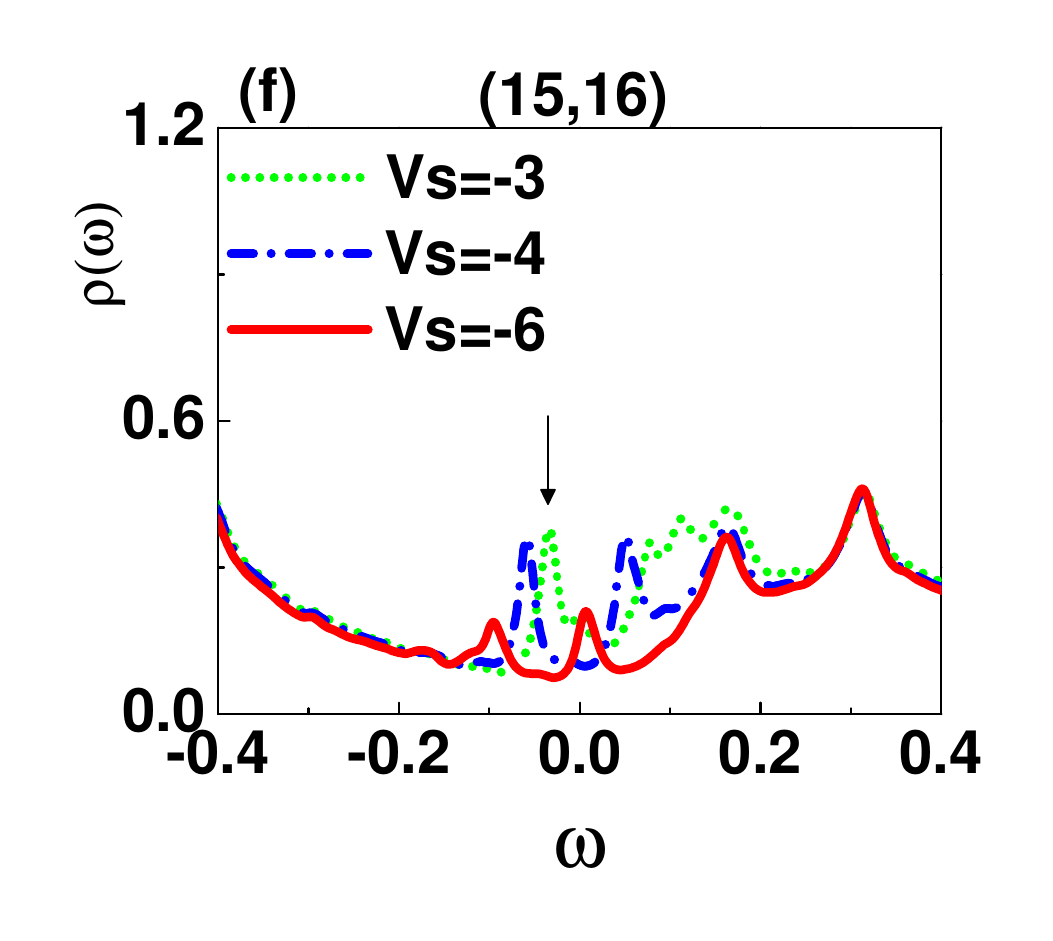}
\caption{(Color online) (a) The LDOS at the impurity site as a
function of $\omega$, for different positive SPs $V_s=3,4,5$. (b) is
similar to (a), but for negative SPs $V_S=-3,-4,-6$. (c) and (d)
[(e) and (f)] are similar to (a) and (b), respectively, but are
plotted at site $(15,17)$ [$(15,16)$].} \label{FIG:den0}
\end{figure}
Then we study the energy and site dependance of the LDOS. Without
impurity, the LDOS is uniform and site independent, with its minimum
located at negative energy, consistent with STM
experiment.~\cite{panli} There are two coherence peaks at negative
energies and two at positive energies, as shown in
Fig.~\ref{FIG:bulkdp0}. Since the LDOS has finite value at the
impurity site for weak SP, we plot it exactly at the impurity site
for both positive and negative SPs in Fig.~\ref{FIG:den0}. For
positive impurity potential $V_s=3$ the spectrum displays two
distinct resonance peaks at negative energies which are denoted by
two arrows. The intensity of the left peak is higher than that of
the right one. The splitting of the resonance peaks is due to the
presence of the inter-orbital coupling $t_4$ and the resonance peaks
are related to the opening of SDW gap. As $V_s$ is increased the two
resonance peaks shift to higher energies and the intensities of the
peaks decrease as shown in Fig.~\ref{FIG:den0} (a). At last, the
intensity of the right peak becomes higher than that of the left
one. For $V_s=6$, the LDOS at the impurity site nearly vanishes. The
feature of the LDOS for negative SPs shown in Fig.~\ref{FIG:den0}
(b) is different from that in the positive SP case, for example, the
intensities of the peaks are much lower. For $V_s=-3$, double peaks
show up at both sides of the Fermi energy, which we also denote by
two arrows. These peaks shift to lower energies with increased value
of $|V_s|$. As $V_s$ reaches to $V_s=-8$, the LDOS at the impurity
site also vanishes.

In Figs.~\ref{FIG:den0} (c) and ~\ref{FIG:den0} (d) we plot the LDOS
at nnn site $(15,17)$ of the impurity for positive and negative SPs,
respectively. One can see that with increased strength of SP, the
double peaks move to higher (lower) energies for positive (negative)
SP. For positive SP, increasing impurity strength will lead to
increased peak intensities and this is in contrast to that at the
impurity site. However, for negative SP, the situation is similar to
that at the impurity site. Figs.~\ref{FIG:den0} (e) and
~\ref{FIG:den0} (f) plot the LDOS at nn site $(15,16)$. It is shown
that the intensities of the impurity-induced resonance peaks are
much lower than those at the impurity site $(16,16)$, although the
characteristics are similar. Since the impurity has four nnn and nn
sites, and the system has only $C_2$ symmetry, there are two
inequivalent nnn and nn sites, respectively. The LDOS at the other
nnn site $(15,15)$ and nn site $(16,15)$ does not show the
impurity-induced resonance peaks at low energies (not shown here)
and resembles the bulk LDOS, again suggesting the four-fold symmetry
breaking.

The properties of the low-energy bound states shown in
Fig.~\ref{FIG:den0} are significantly different from those in the
pure SC state.~\cite{zho,zhang,tsai} In the pure SC state, the
positions of the bound states are close to the SC coherence peaks so
that they may be masked by the SC coherence peaks and may be hard to
detect by experiments. However, in the SDW state, the energies of
the bound states are close to the Fermi energy so that they can be
easily detected by experiments.

\section{Positive impurity scattering in doped samples}
\label{SEC:positive}
\begin{figure}
      \includegraphics[width=1.65in]{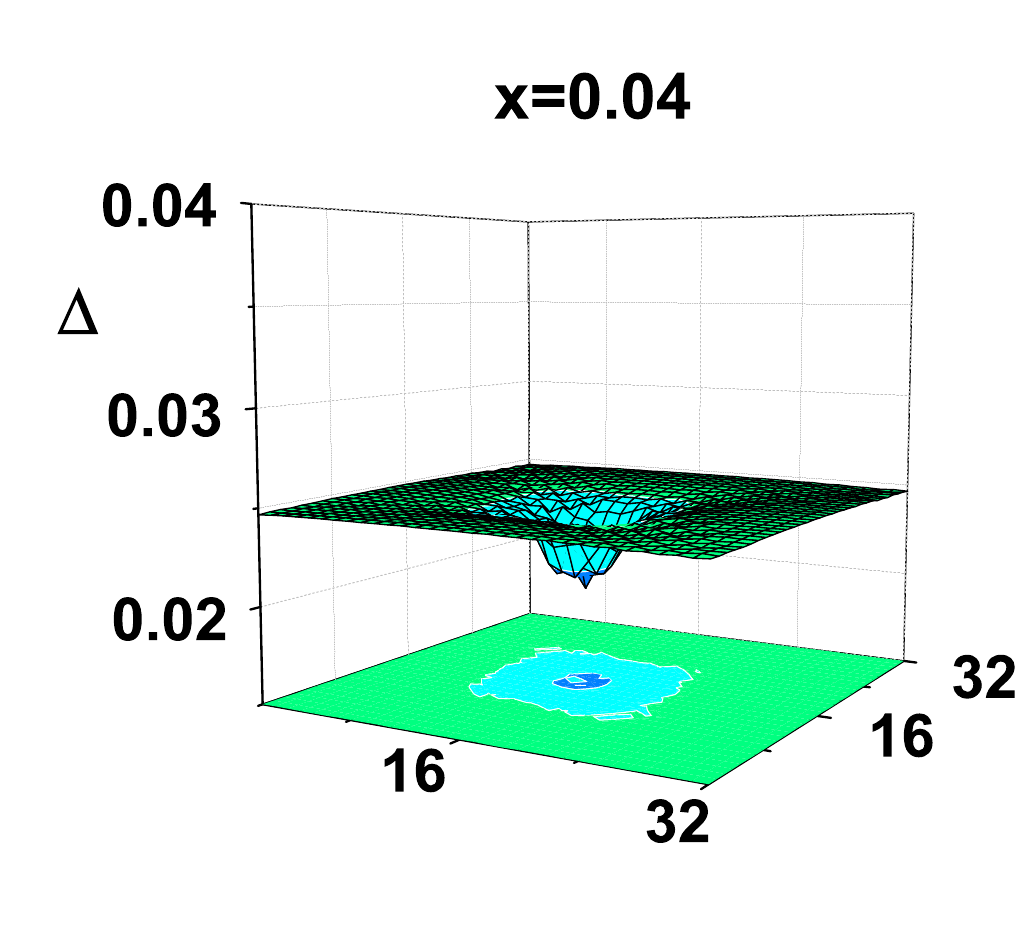}
        \includegraphics[width=1.65in]{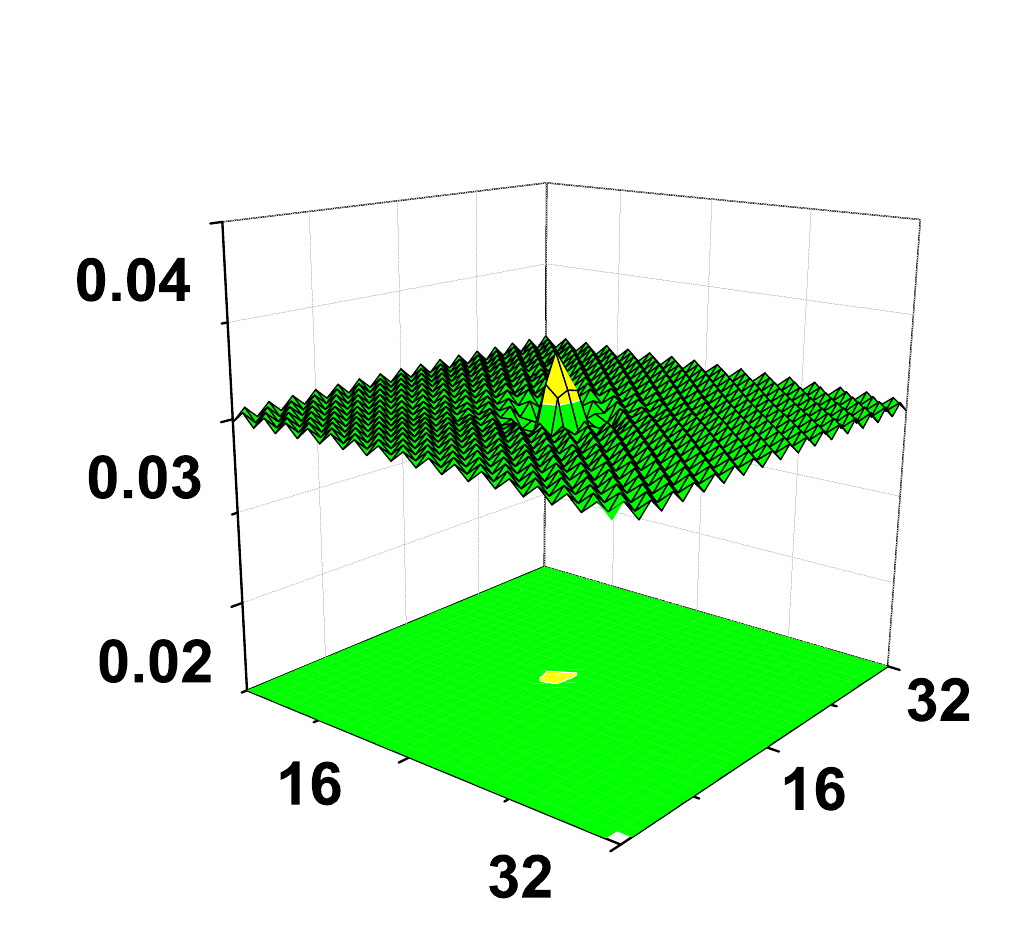}
         \includegraphics[width=1.65in]{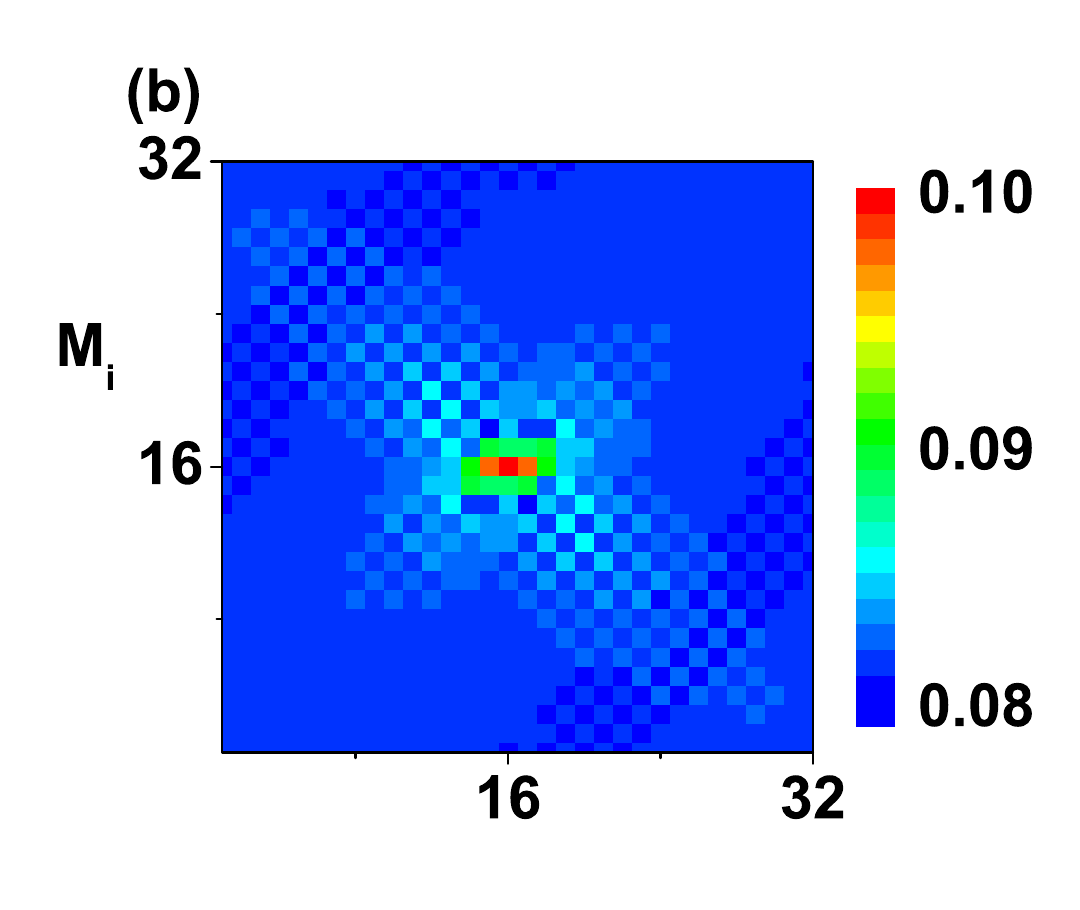}
          \includegraphics[width=1.65in]{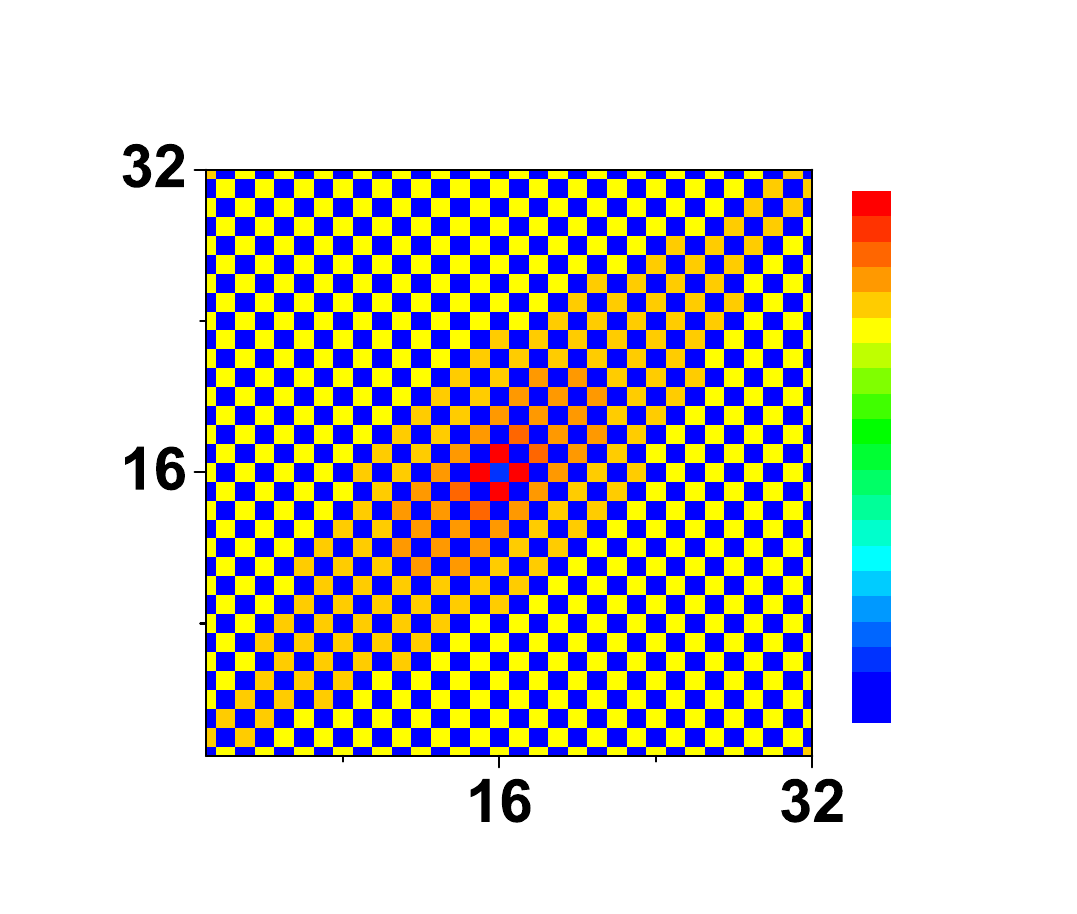}
\caption{(Color online) The intensity plots of the SC [(a) and (c)]
and magnetic [(b) and (d)] orders for weak SP $V_s=1$, at doping
$x=0.04$ and $x=0.08$, respectively.} \label{v1:OP1}
\end{figure}

As we have discussed above, in both pure SDW state and pure SC
state, the bound states are induced by a single nonmagnetic
impurity. Since the detailed features of the resonance peaks are
quite different between these two cases, the impurity effect in the
underdoped regime where the SC and SDW orders coexist is an intriguing question. In particular, both
theoretical analyses~\cite{zho} and experimental
observations~\cite{pra,wan,lap,les,chr,jul,chen,zha,liu} do suggest
the coexistence of these two orders in this regime. In the
following, we will not plot the real space particle number since it
is similar to the undoped case. We want to mention that around the
moderate doping $x=0.08$, the impurity could induce a weak charge
density wave for various SPs. However, the $\delta n/(2+x)$ is less
than $0.5/100$, so we neglect it.

For small SP $V_s=1$, we can see from Fig.~\ref{v1:OP1} (a) that at
low doping $x=0.04$, the amplitude of the SC order $\Delta_i$ is
reduced at and around the impurity site, which will recover to the
impurity-free value at about $6$ lattice constants away from the
impurity. But the SC order is not always suppressed at the impurity
site. As doping is increased to $x\geq0.08$, at the impurity site the magnitude of the SC order is
enhanced [see Fig.~\ref{v1:OP1} (c)], which means the impurity is
not a pair breaker in this case. At $x=0.04$, the magnetic order
$M_i$ at the impurity site is enhanced, similar to that in the
undoped case and we notice that there exist modulations along the
diagonal directions as can be seen from Fig.~\ref{v1:OP1} (b). At
$x=0.08$, the pattern of magnetic order changes, the system
separates into two sublattices explicitly. The value of $M_i$ in one
sublattice is about $0.05$, while in the other one is $\sim0.007$,
with $M_i\sim0.01$ at the impurity site. This impurity-induced
two-sublattice pattern of magnetic order survives until the doping
level is beyond the region where the SDW and SC orders coexist. At
higher doping $x=0.12$, $\Delta_i$ is enhanced just like the
$x=0.08$ case, but with a vanishingly small value of $M_i$.

For larger SP $V_s=3.0$, the order parameters are similar to those
for $V_s=1.0$, except that $M_i$ is reduced at the impurity site at
all doping levels. As doping increases to $x=0.08$, the system also
separates into two sublattices. An enhanced SC order will appear at
the impurity site when $x$ reaches to $0.12$ (not shown here).

\begin{figure}
\centering
\includegraphics[width=9.5cm]{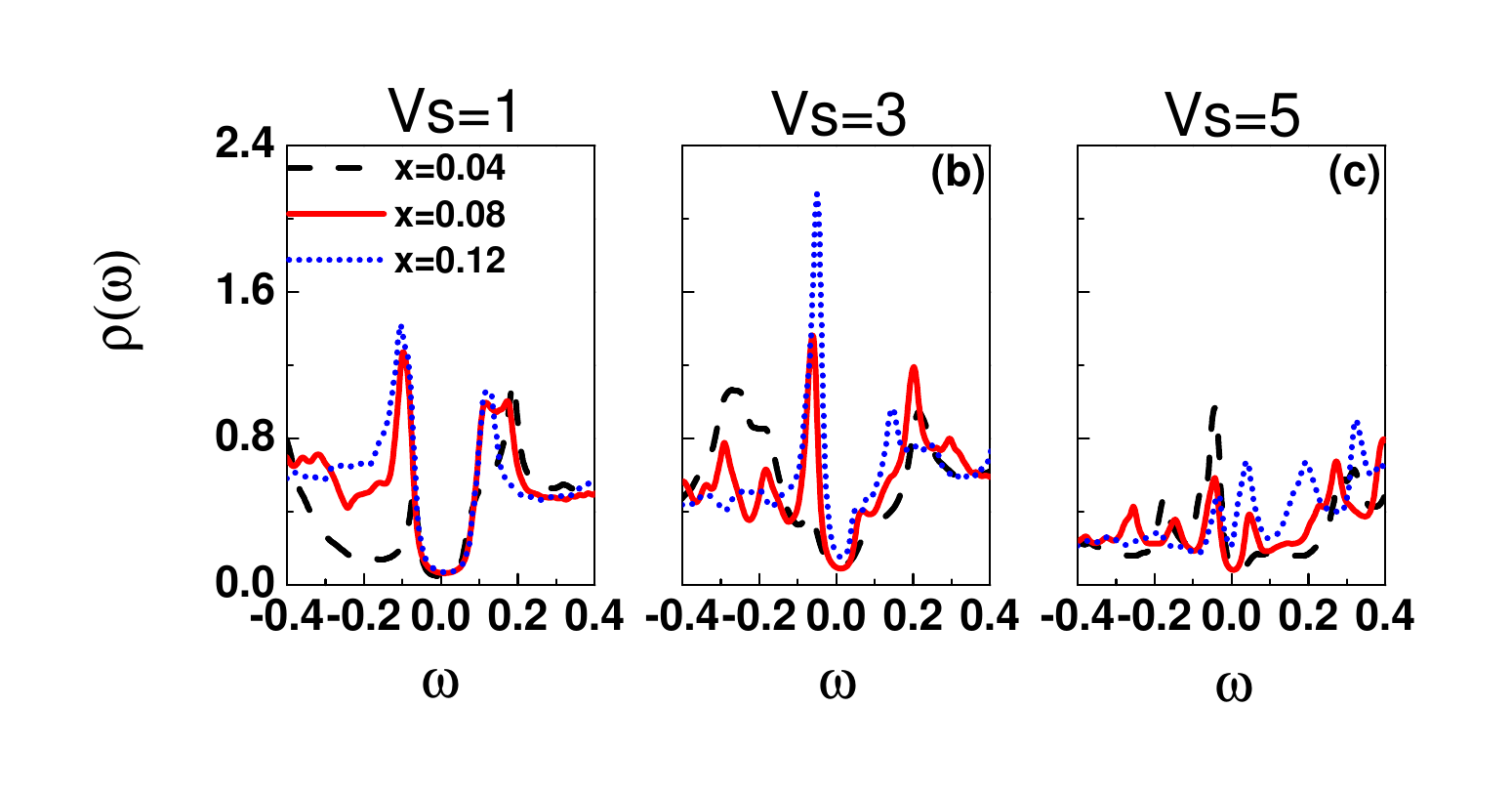}
\caption{(Color online) The LDOS at the impurity site as a function
of $\omega$ for $V_s=1,3,5$, at different doping levels
$x=0.04,0.08,0.12$. } \label{dos:zimp}
\end{figure}

The LDOS has finite values at the impurity site for small SP in all
doped samples, but unlike the undoped one, the double resonance
peaks are absent. For $V_s=1$, the effect of the impurity on the
LDOS is shown in the left panel of Fig.~\ref{dos:zimp}. In this
case, at $x=0.04$, the intensities of the SC coherence peaks at both
positive and negative energies are suppressed by the impurity. On
the other hand, at doping $x\geq0.08$, the negative SC coherence
peak is enhanced by the impurity while the positive one remains
almost unchanged. For larger SP $V_s=3$, at $x=0.04$, the
intensities of the SC coherence peaks are further suppressed. When
$x=0.08$ there is a sharp in-gap resonance peak located at negative
energy and close to the SC coherence peak [see
Fig.~\ref{dos:zimp}(b)]. As doping is increased to $x=0.12$, the
intensity of the in-gap peak becomes higher. On the other hand, for
moderate SP $V_s=5$, at $x=0.04$, there exists an in-gap bound state
at negative energy while at both $x=0.08$ and $x=0.12$, there are
two in-gap bound states, one at positive energy, the other one at
negative energy. The magnitude of the LDOS at all doping levels
becomes considerably smaller and reaches to zero for larger SP.

\begin{figure}
\centering
  \includegraphics[width=9.5cm]{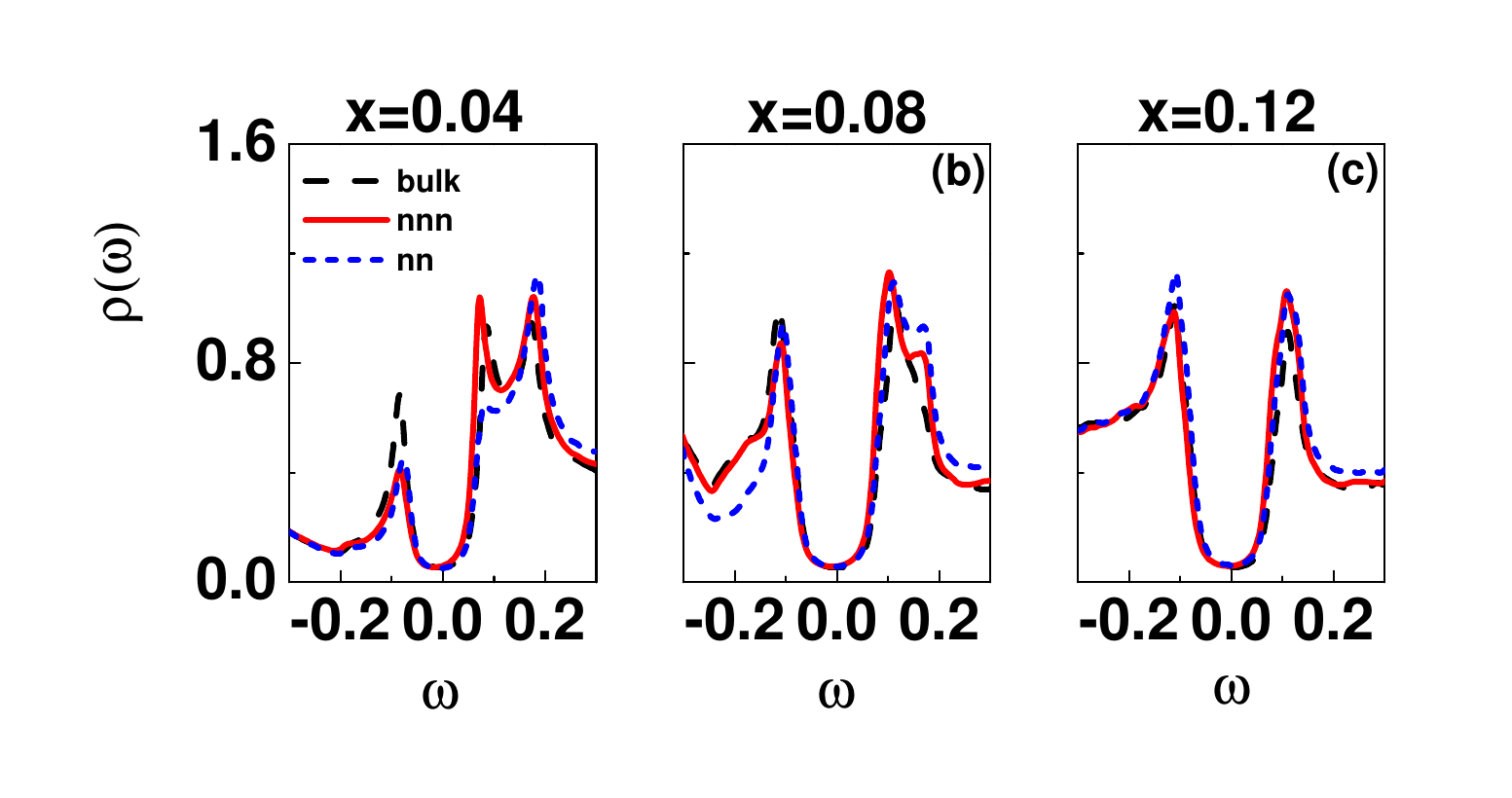}
\caption{(Color online) For $V_s=1$, the LDOS on nn and nnn  sites
of the impurity as a function of $\omega$ at various doping levels.
The black dashed line represents the bulk LDOS.} \label{dos:v1}
\end{figure}

For $V_s=1.0$, the impurity induces only minor modulations on the
LDOS around the impurity site, which is similar to the bulk LDOS at
all doping levels [see Fig.~\ref{dos:v1}]. The positive energy peak
at nn and nnn sites is higher than the negative one at low doping
$x=0.04$. As $x$ increases, the intensity of the negative resonance
peak gradually becomes higher than that of the positive one, similar
to the evolution of the bulk LDOS with doping.~\cite{panli} Although
the system does not have $C_4$ symmetry, the main features of the
LDOS at the four nnn (nn) sites are similar to each other, thus in
Fig.~\ref{dos:v1}, we only plot the LDOS at one of the nnn (nn)
sites for clarity.

\begin{figure}
\centering
  \includegraphics[width=9.5cm]{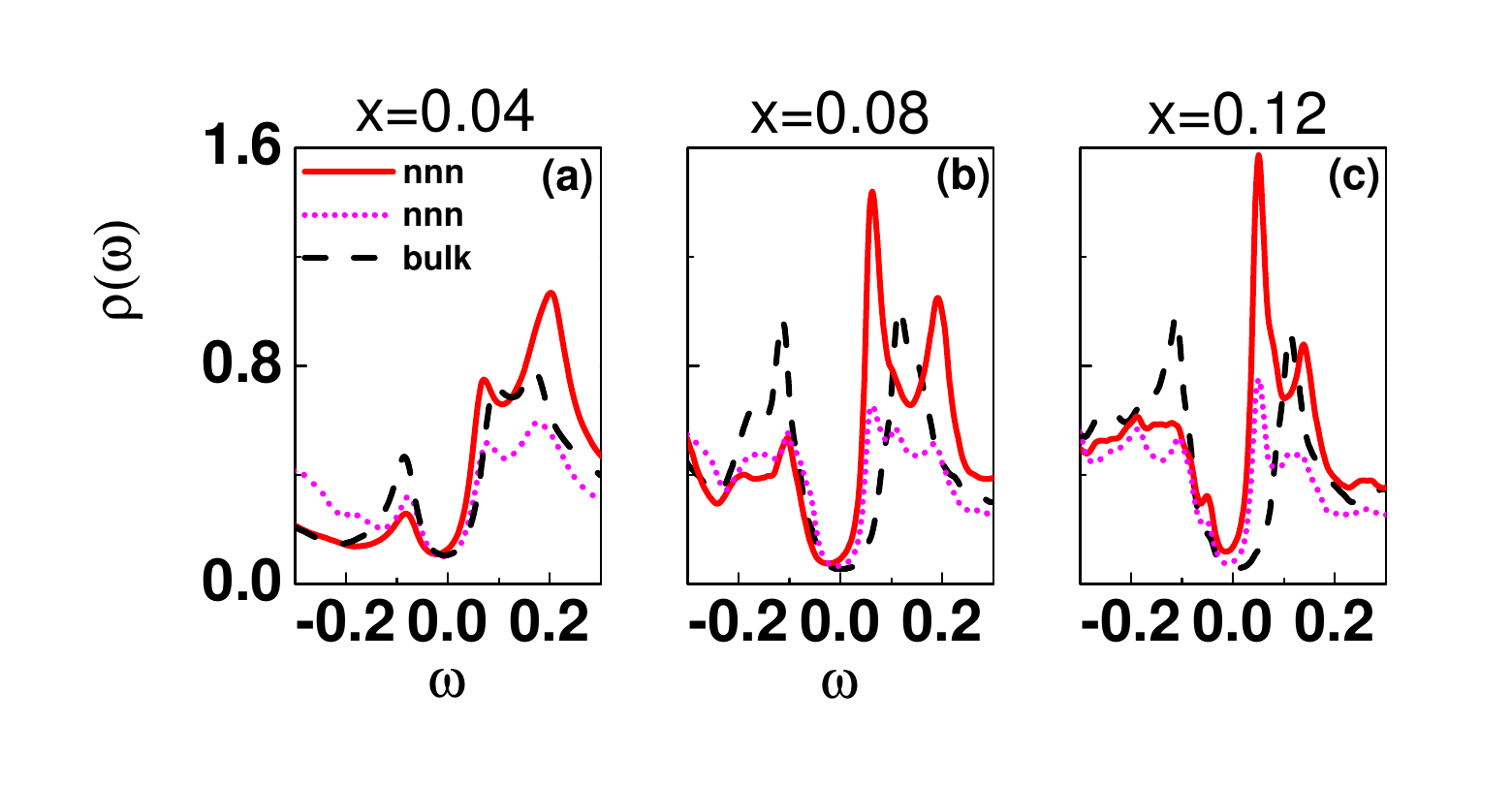}
  \includegraphics[width=9.5cm]{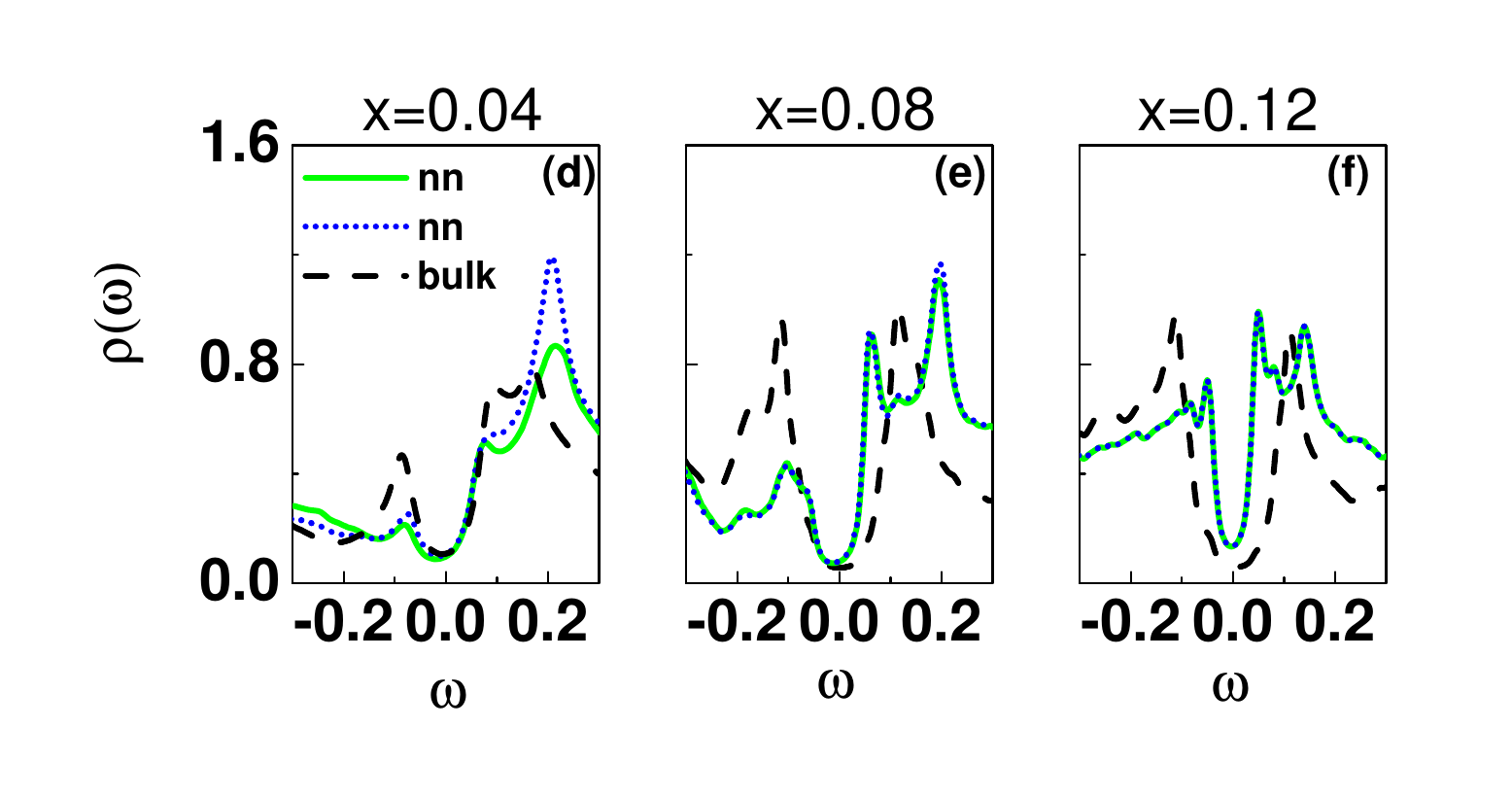}
\caption{(Color online) For $V_s=3$, the LDOS on nn and nnn sites of
the impurity as a function of $\omega$ at various doping levels. The
upper panel is for the two nnn sites $(15,15)$ (red solid) and
$(15,17)$ (pink short dot) and the lower one is for the two nn sites
$(15,16)$ (green solid) and $(16,15)$ (blue short dot). The black
dashed line represents the bulk LDOS.} \label{dos:v3}
\end{figure}

As the SP increases to $V_s=3.0$, we show the LDOS at two
inequivalent nnn (nn) sites in the upper (lower) panel of
Fig.~\ref{dos:v3}. At low doping $x=0.04$, the effect of the
impurity is weak and no in-gap bound states exist at nnn and nn
sites. At higher doping $x=0.08$, a single in-gap resonance peak
shows up at both the two nnn sites with different intensities, but
their positions are similar to each other, both are located at
positive energy and close to one of the SC coherence peaks. There is
also a single in-gap peak at the two nn sites, the LDOS of which is
identical to each other. As doping increases to $x=0.12$, the LDOS
at the nnn sites is similar to the $x=0.08$ case, except for a
higher peak intensity at positive energy and the addition of a hump
at negative energy on one of the nnn sites, which will evolve into a
resonance peak when further increasing doping (not shown here). The
LDOS at the two nn sites is also identical to each other and clearly
shows two in-gap resonance peaks.

\begin{figure}
\centering
  \includegraphics[width=9.5cm]{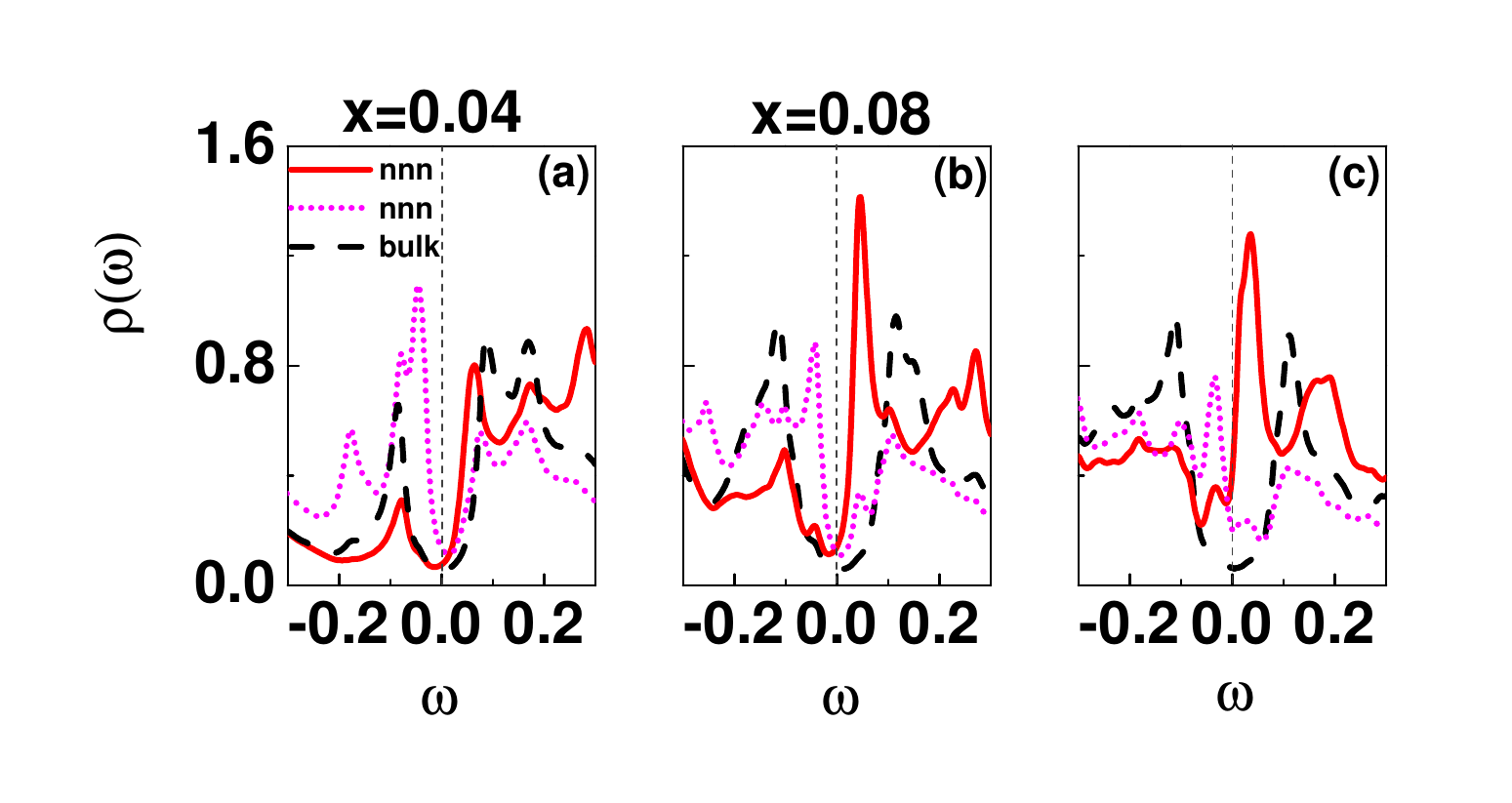}
   \includegraphics[width=9.5cm]{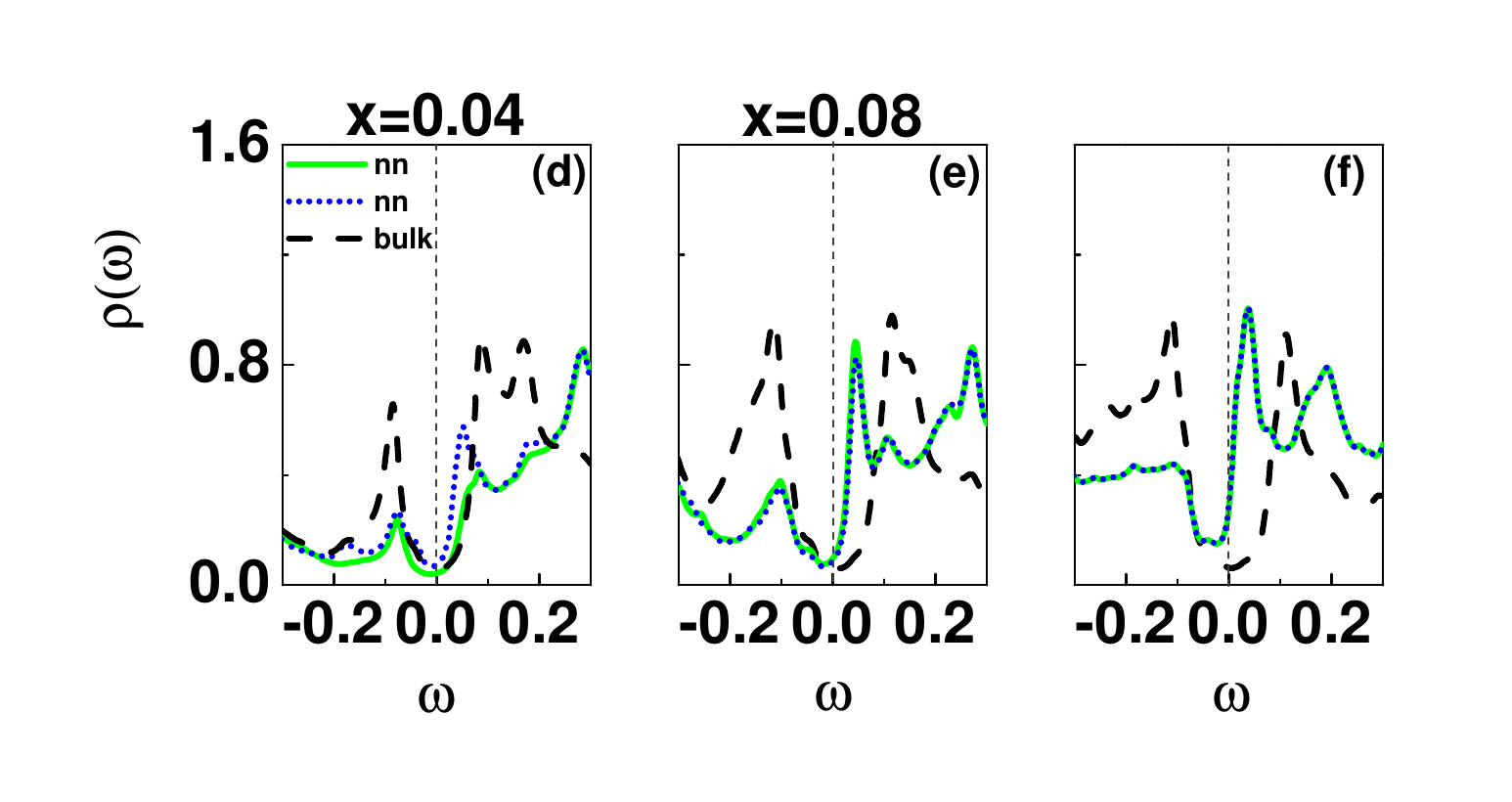}
\caption{(Color online) Similar to Fig.~\ref{dos:v3}, bur for
$V_s=5$.} \label{dos:v5}
\end{figure}

For moderate SP $V_s=5$, the difference of the LDOS between the two
inequivalent nnn sites $(15,15)$ and $(15,17)$ becomes remarkable
[see the upper panel of Fig.~\ref{dos:v5}]. At one of the nnn sites
the single in-gap resonance peak is located at positive energy while
at the other one it is located at negative energy. As doping
increases, the peaks at both the two nnn sites move closer to the
Fermi energy. On the other hand, at the two nn sites $(15,16)$ and
$(16,15)$, there is a single in-gap resonance peak located above the
Fermi energy at all dopings. At low doping the LDOS spectra at the two nn site are different. As doping increases
the peaks shift to the Fermi energy and the LDOS at the two nn
sites will become identical.

\begin{figure}
\centering
  \includegraphics[width=1.65in]{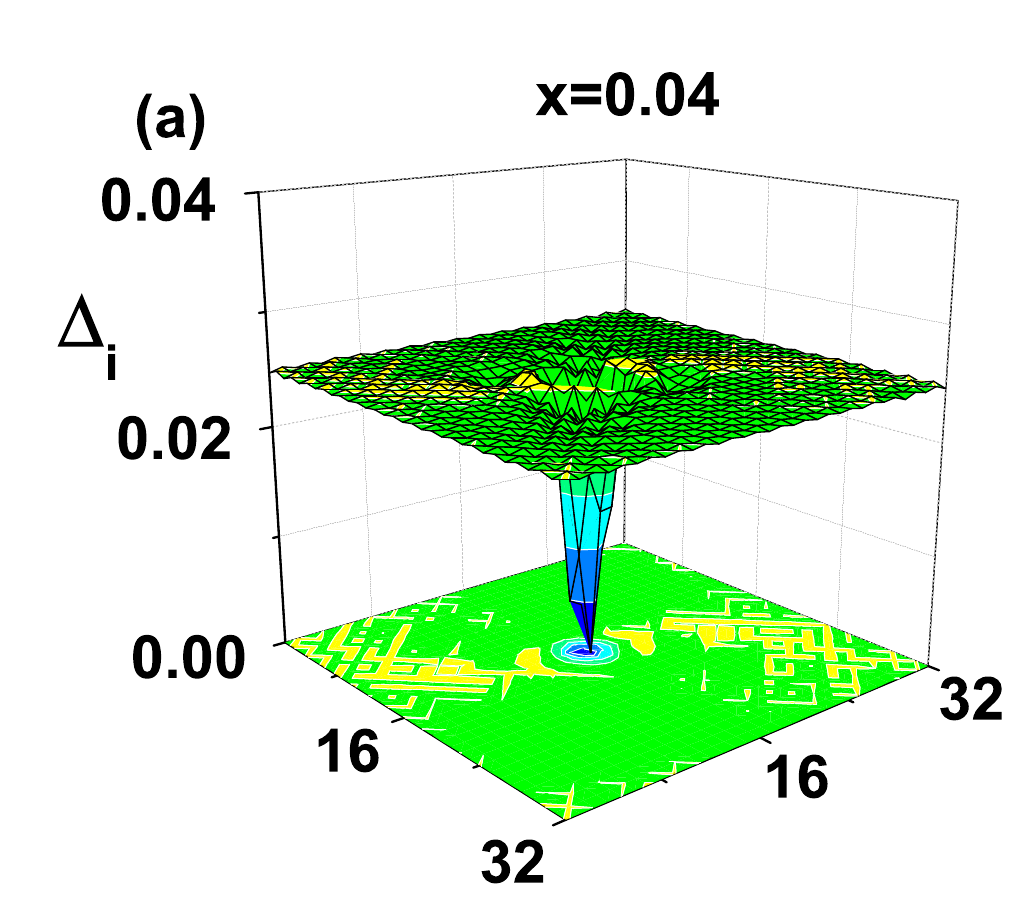}
    \includegraphics[width=1.65in]{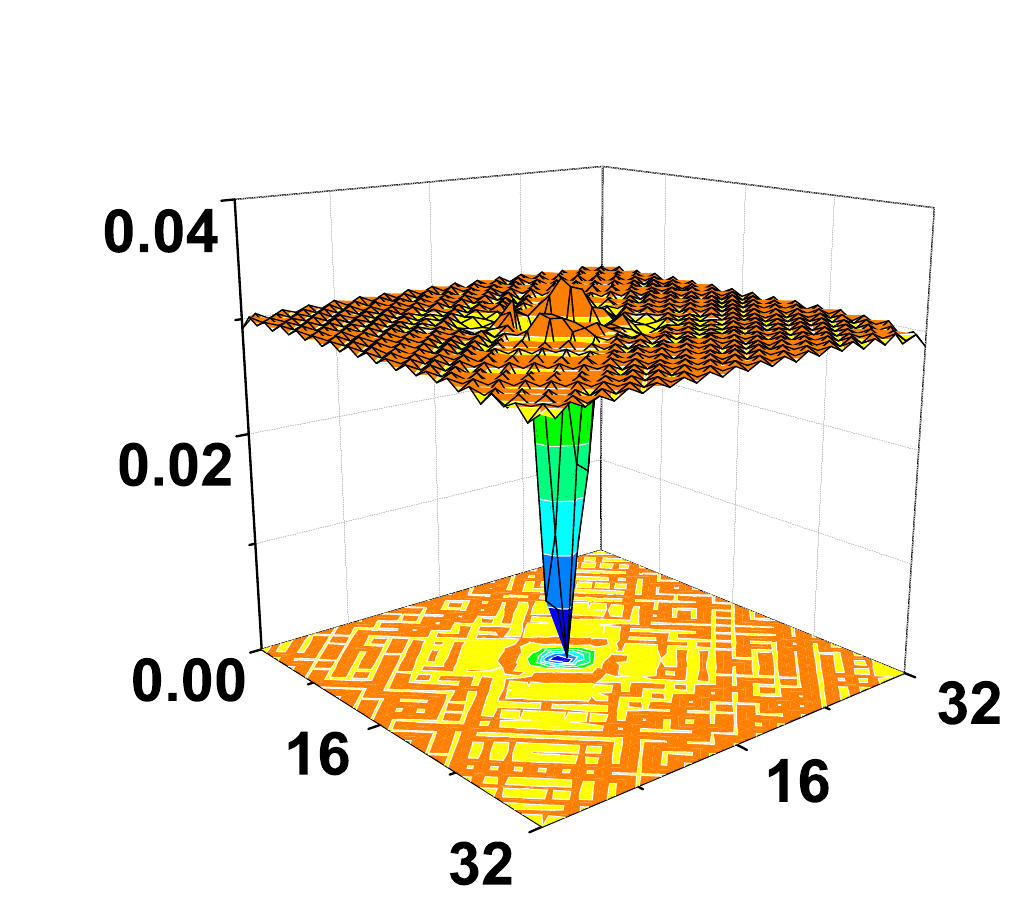}
      \includegraphics[width=1.65in]{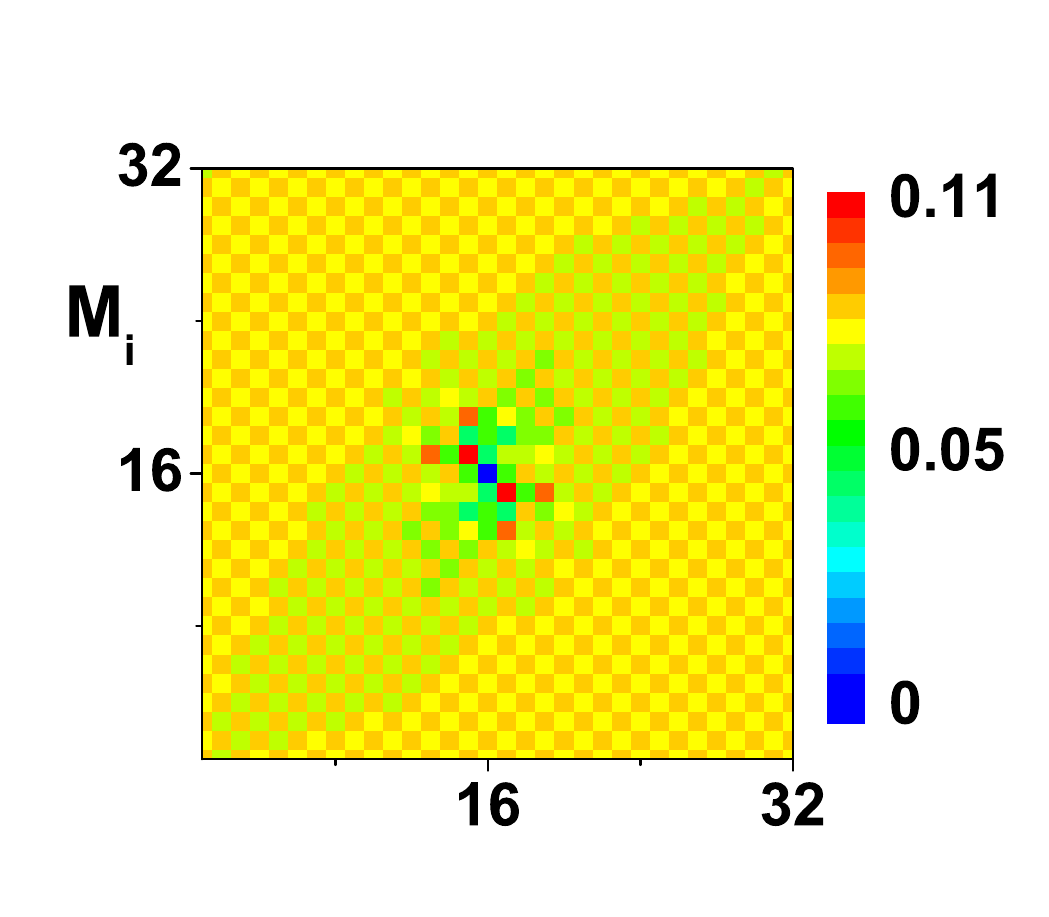}
        \includegraphics[width=1.65in]{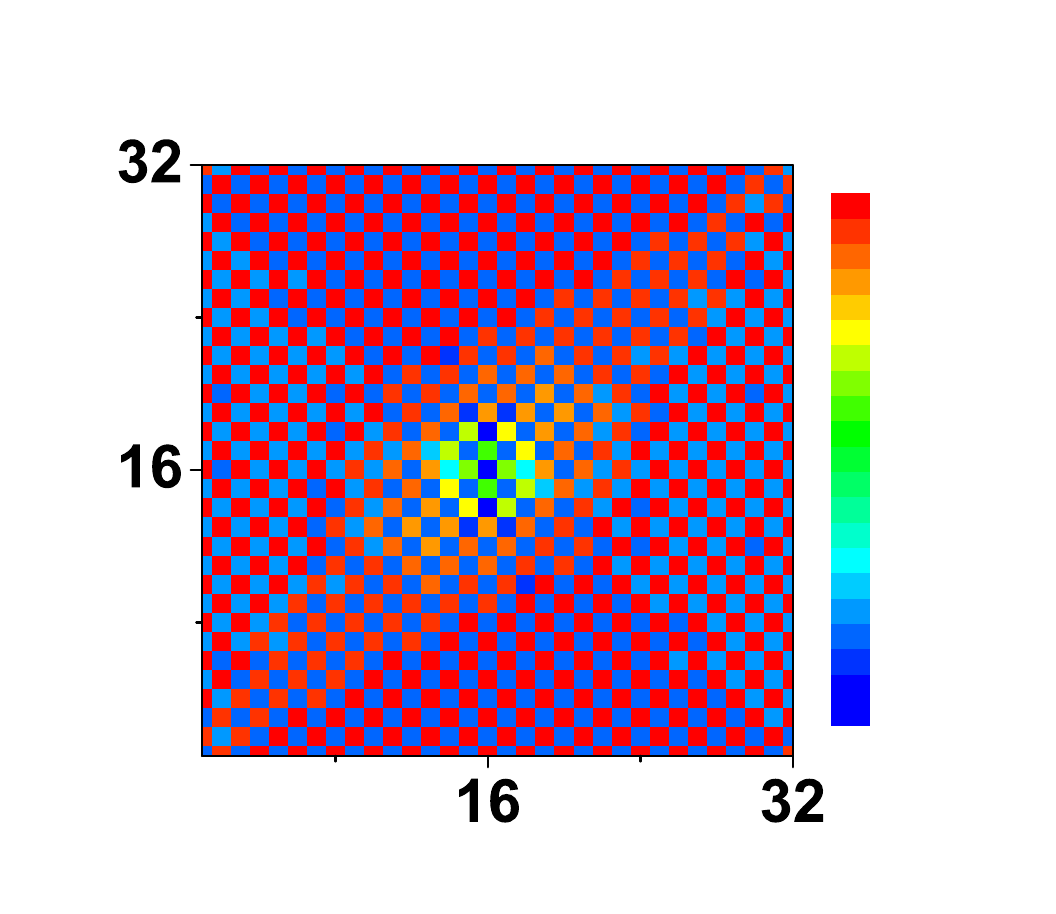}
\caption{(Color online) Similar to Fig.~\ref{v1:OP1}, but for
$V_s=100$.} \label{v100:OP1}
\end{figure}

For nearly unitary positive SP $V_s=100$, the SC and magnetic orders are
both suppressed and oscillate in the vicinity of the impurity site
and their magnitudes reach the minimum exactly at the impurity site.
The suppressed order parameters recover to their bulk value at
about $3$ lattice constants away from the impurity.  We note that when $x\geq0.04$
the system will separate into two sublattices. The difference of
$M_i$ between the two sublattices is larger at $x=0.08$ than that at
$x=0.04$ while the magnitude of $M_i$ decreases with doping and will
vanish at $x>0.1$.

\begin{figure}
\centering
  \includegraphics[width=9.5cm]{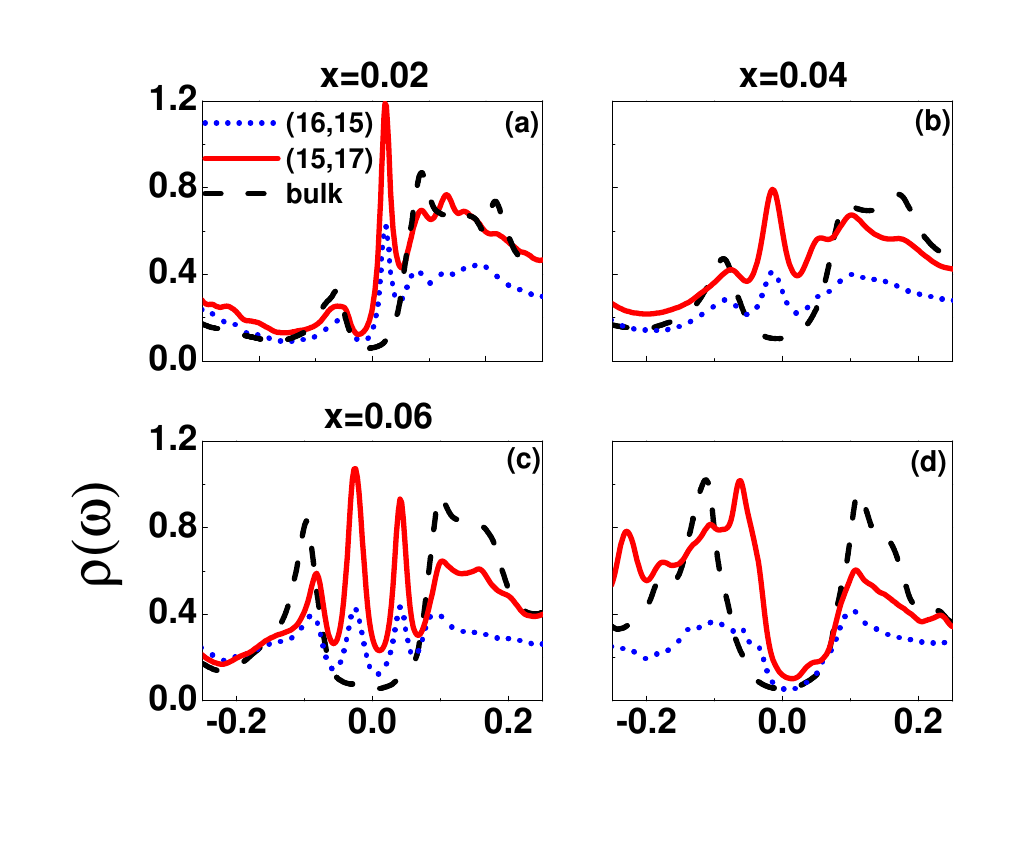}
\caption{(Color online) The LDOS on nn and nnn sites of the impurity
as a function of $\omega$, for $V_s=100$. Different panels
correspond to different doping levels. The red solid line denotes
the nnn site $(15,17)$ while the blue dashed line denotes the nn
site $(16,15)$, with the bulk LDOS denoted by the black dashed
line.} \label{dos:v100}
\end{figure}

Since the LDOS at the impurity site is zero for such a strong SP, we
thus plot it on nnn and nn sites of the impurity in
Fig~.\ref{dos:v100}, at various dopings. At $x=0.02$, a sharp in-gap
resonance peak appears close to zero energy on the positive side. It
shifts to negative energy with reduced intensity as $x$ increases to
$0.04$. When $x\geq0.05$ two in-gap resonance peaks show up. As
doping increases further, they are pushed away by each other and
finally merge into the SC coherence peaks of the bulk LDOS at
$x\geq0.08$. In all cases, the LDOS exhibits clear $C_2$ symmetry.
However, the LDOS on the inequivalent nn (nnn) sites is
qualitatively the same, thus we choose the sites $(16,15)$ and
$(15,17)$ as an example for convenience.

\section{negative impurity scattering in doped samples }
\label{SEC:negative} In real materials, both positive and negative
SPs are possible, and the response of the system to the impurity may
depend on the sign of the SP, thus we discuss the negative SP case
in this section.
\begin{figure}
\centering
       \includegraphics[width=1.65in]{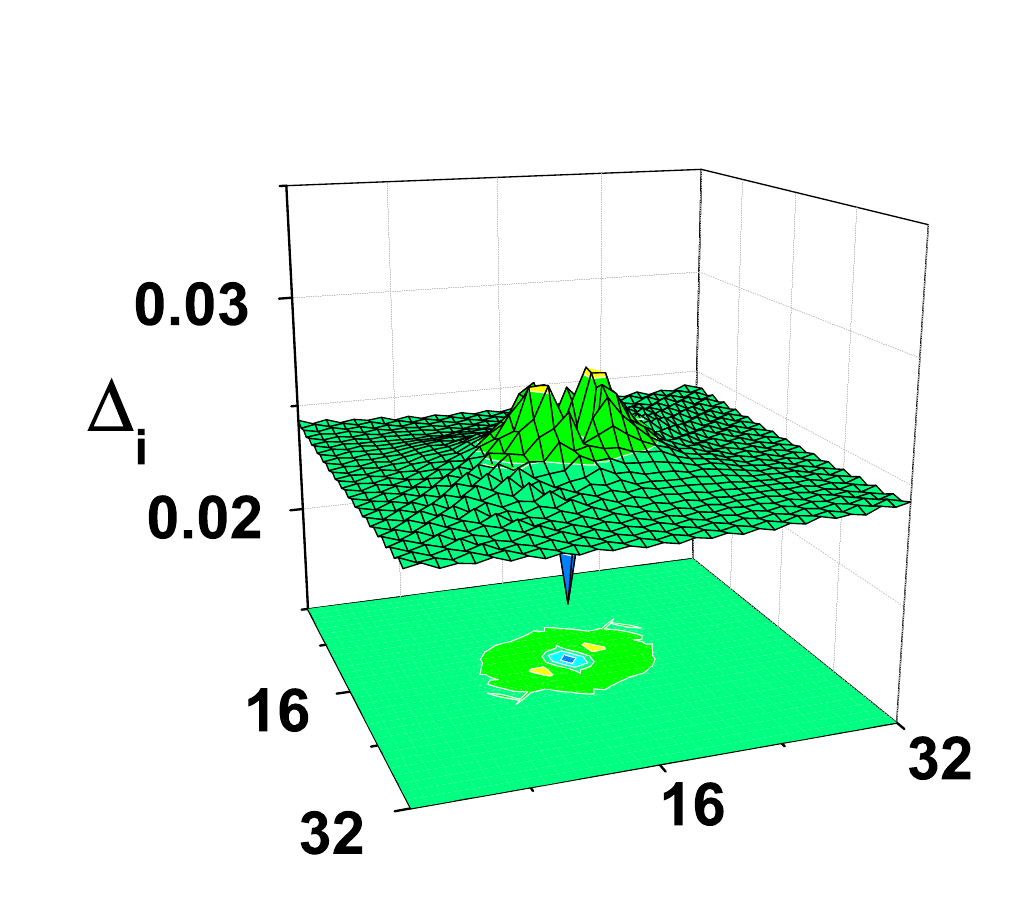}
              \includegraphics[width=1.65in]{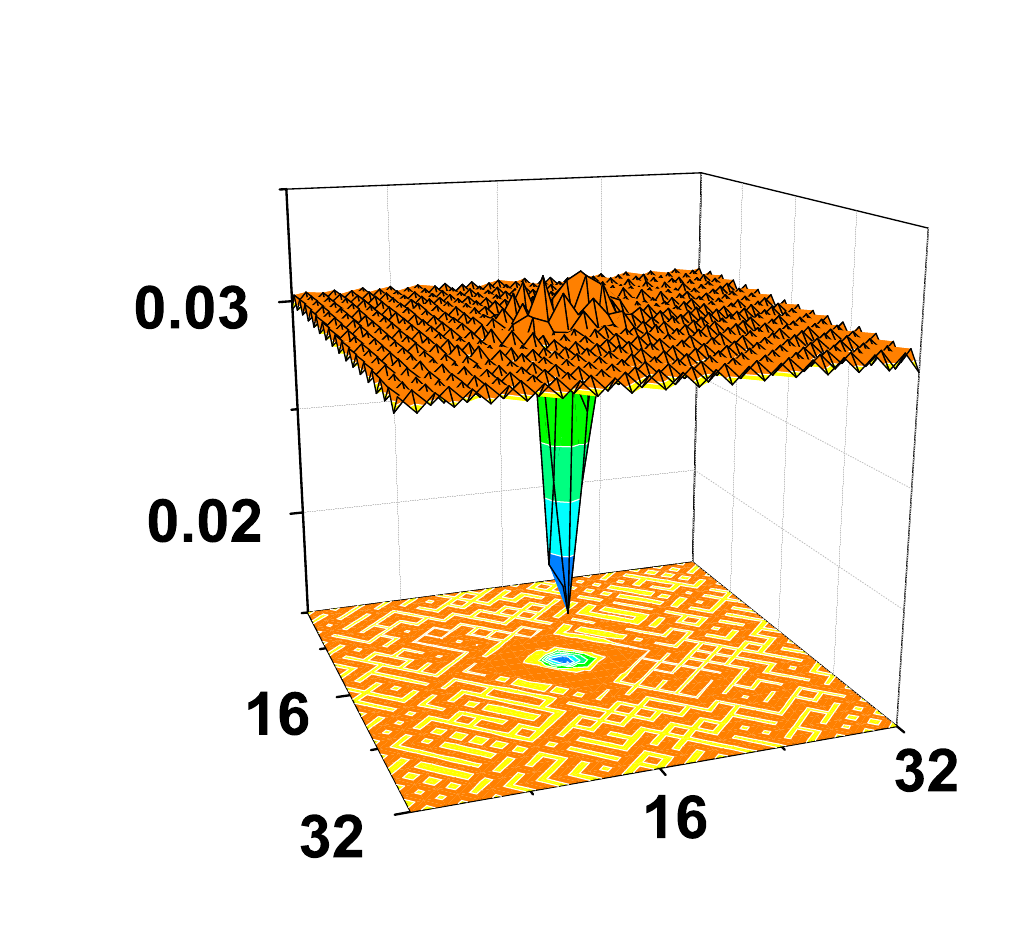}
              \includegraphics[width=1.65in]{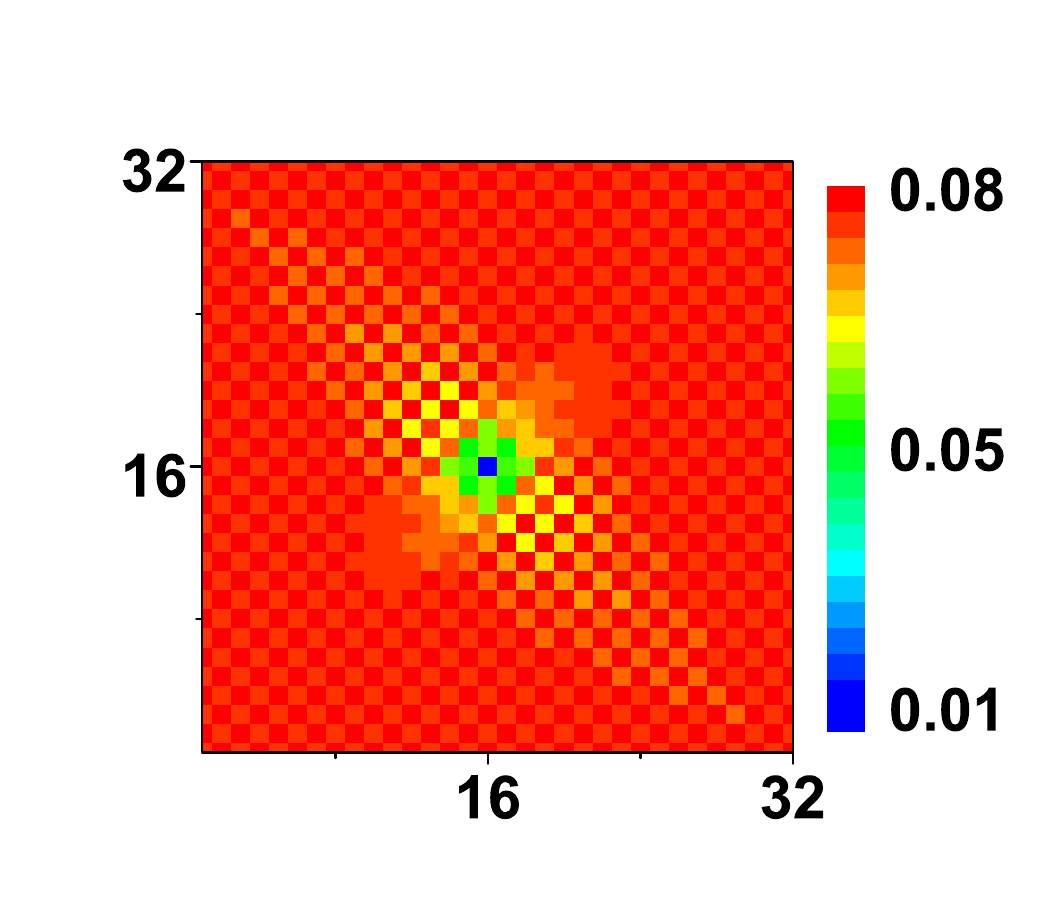}
              \includegraphics[width=1.65in]{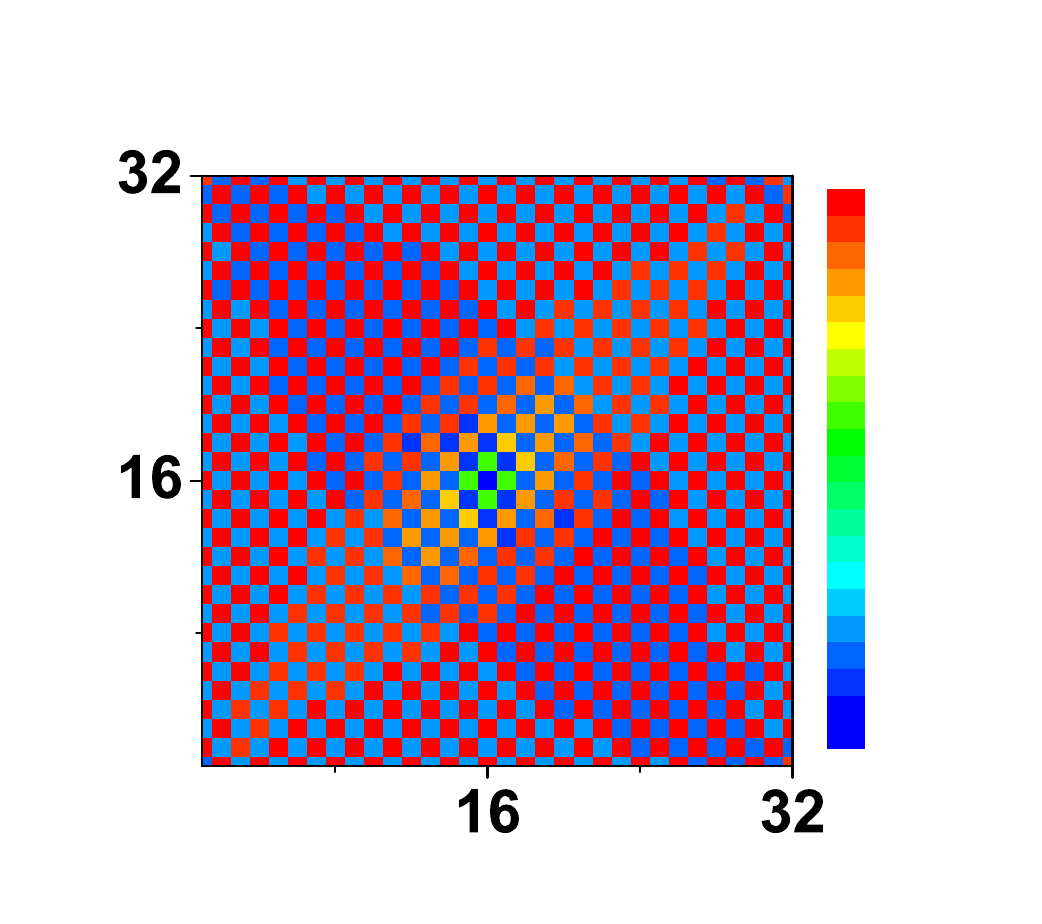}
\caption{(Color online) Similar to Fig.~\ref{v1:OP1}, but for
$V_s=-3$.} \label{vf3:order}
\end{figure}

As we can see from Fig.~\ref{vf3:order}, for $V_s=-3$, the impurity
induces oscillation of the SC order with $\Delta_i$ being suppressed
at the impurity site and enhanced on several nearby sites at
$x=0.04$. At higher doping $x=0.08$, around the impurity site
$\Delta_i$ is suppressed and the oscillation is not distinct. The
magnitude of $M_i$ is suppressed at the impurity site at all doping
levels and apparently $M_i$ will divide into two sublattices at
relatively higher doping. At $x=0.08$, at the impurity site $M_i$ is
close to zero, and is $0.05$ and $0.01$ on the two sublattices,
respectively.

\begin{figure}
\centering
  \includegraphics[width=9.5cm]{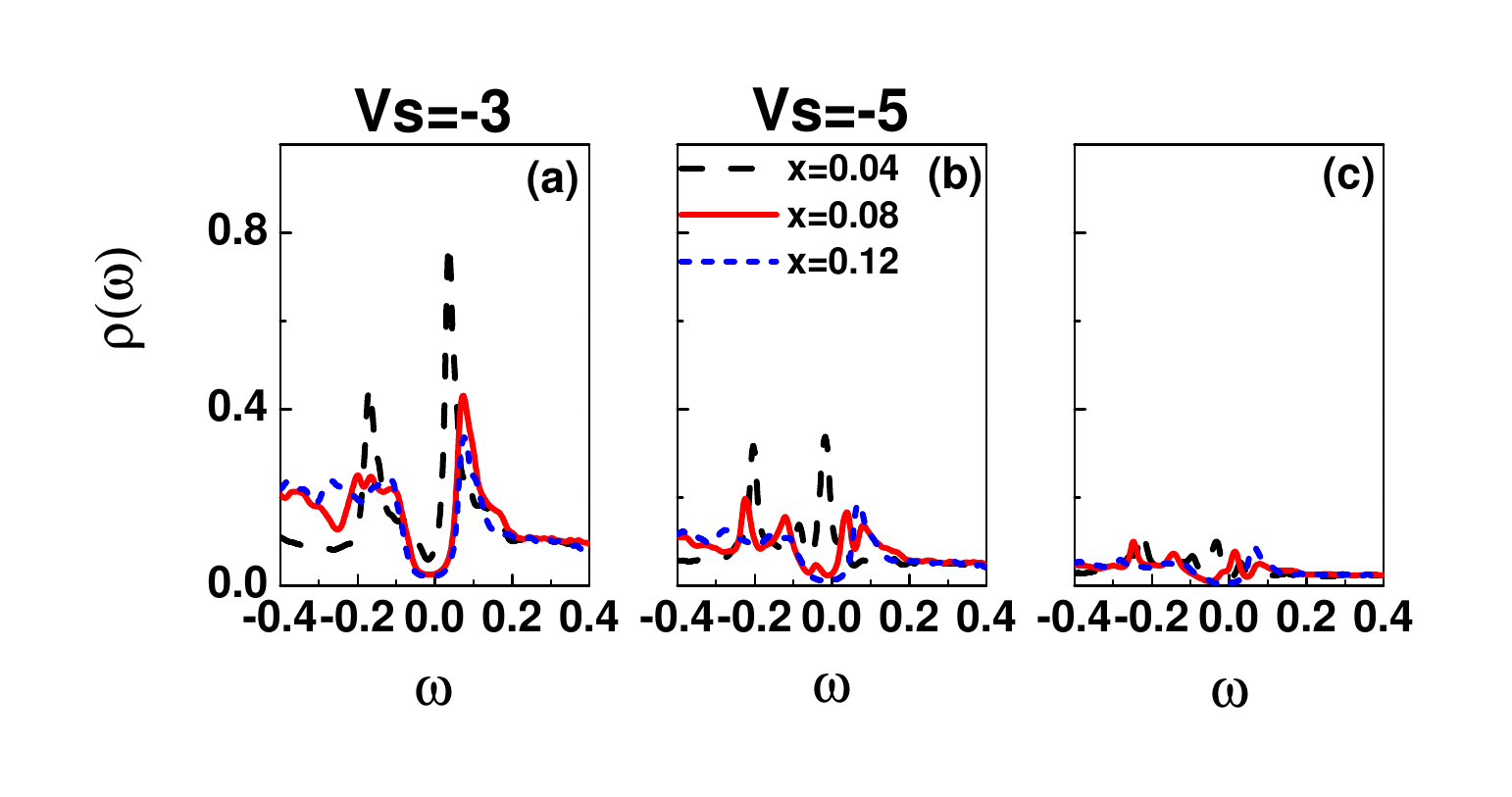}
\caption{(Color online) The LDOS at the impurity site as a function
of $\omega$ at different dopings $x=0.04,0.08,0.12$, for
$V_s=-3,-5,-8$.} \label{dos:fimp}
\end{figure}

\begin{figure}
\centering
  \includegraphics[width=9.5cm]{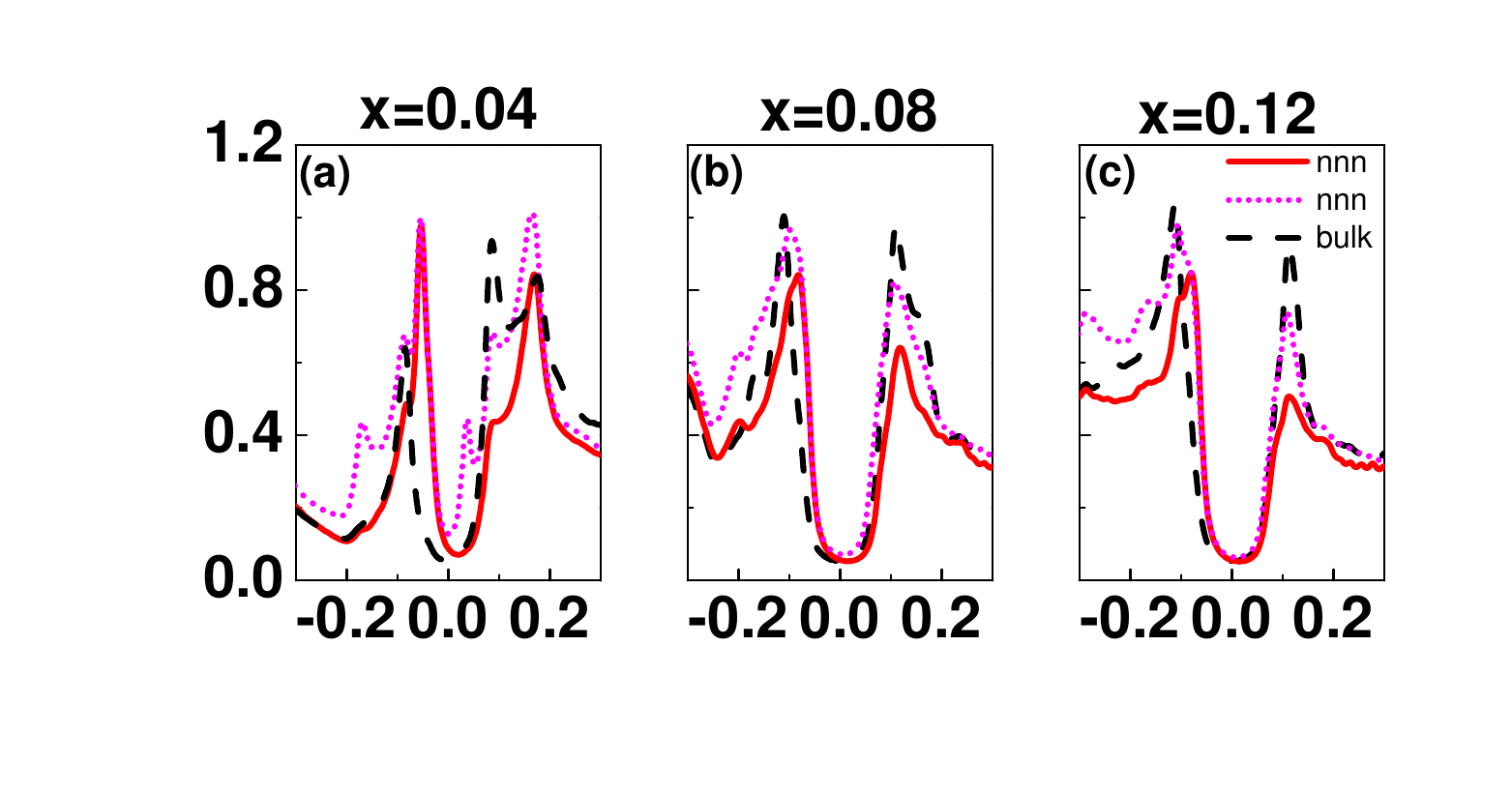}
   \includegraphics[width=9.5cm]{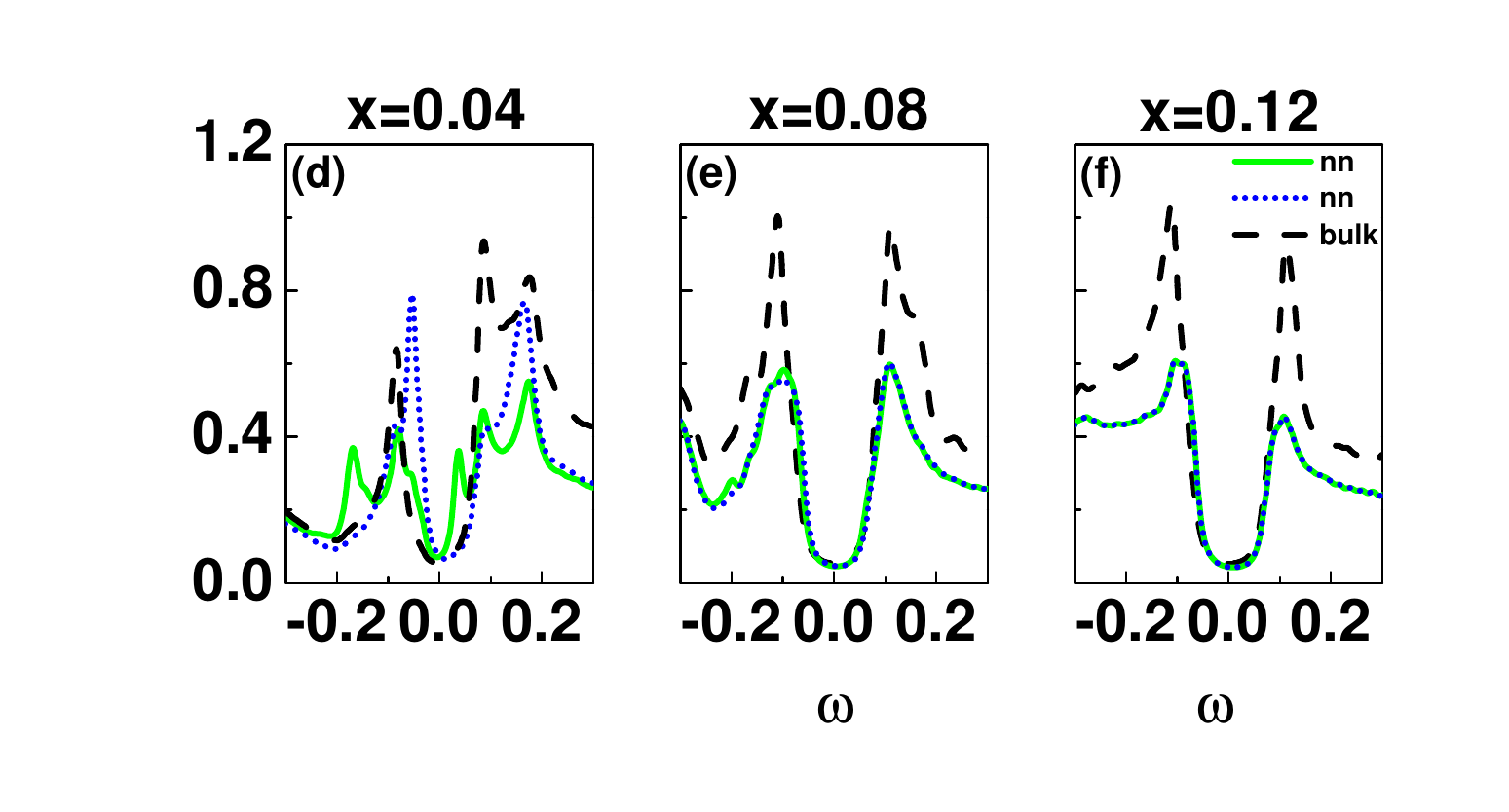}
\caption{(Color online) For $V_s=-3$, the LDOS on nnn and nn sites
of the impurity as a function of $\omega$ at various dopings. The
upper and lower panels are for nnn and nn sites, respectively. The
bulk LDOS is denoted by the black dashed line.} \label{dos:vf3}
\end{figure}

\begin{figure}
\centering
  \includegraphics[width=9.5cm]{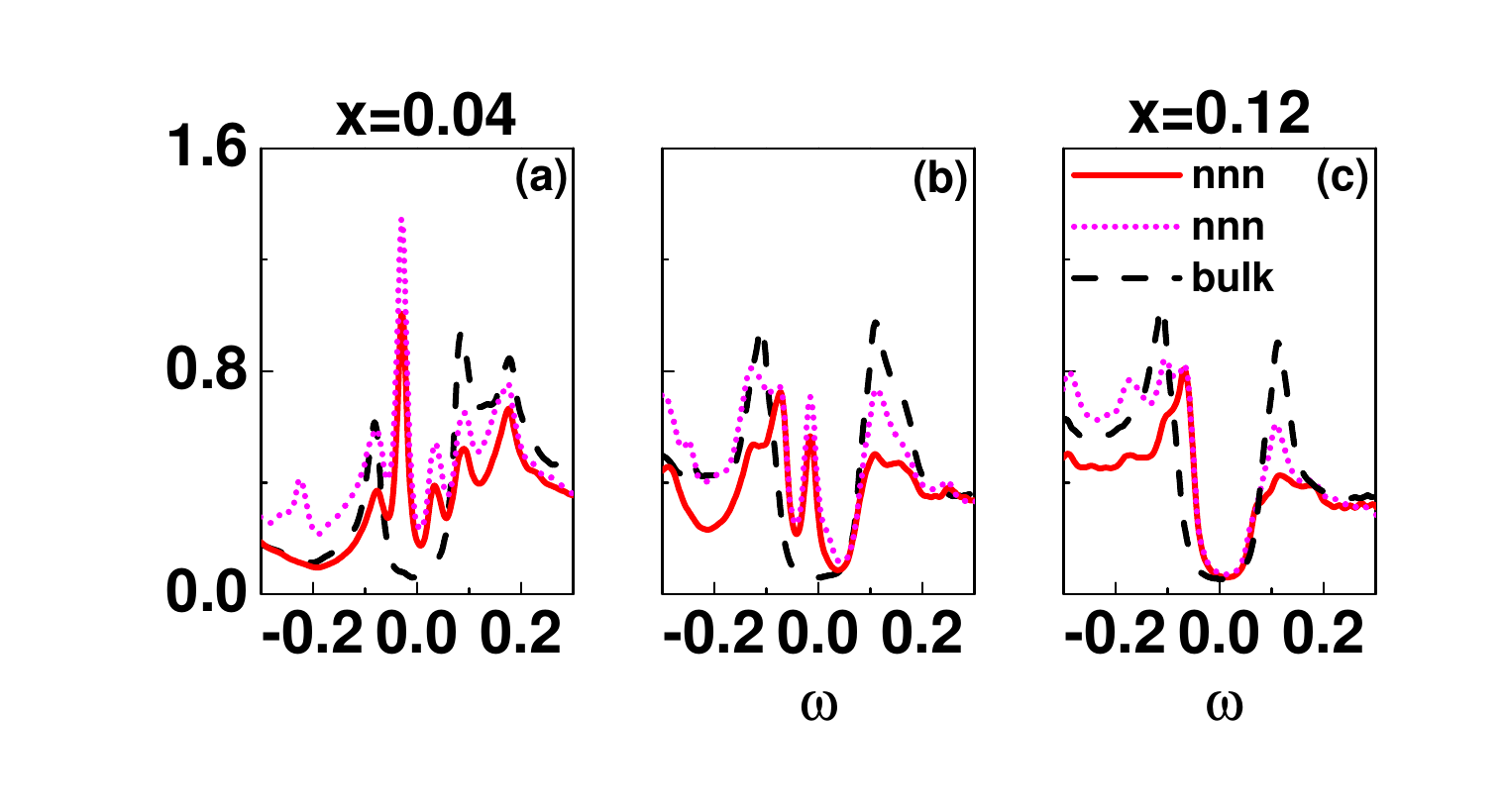}
   \includegraphics[width=9.5cm]{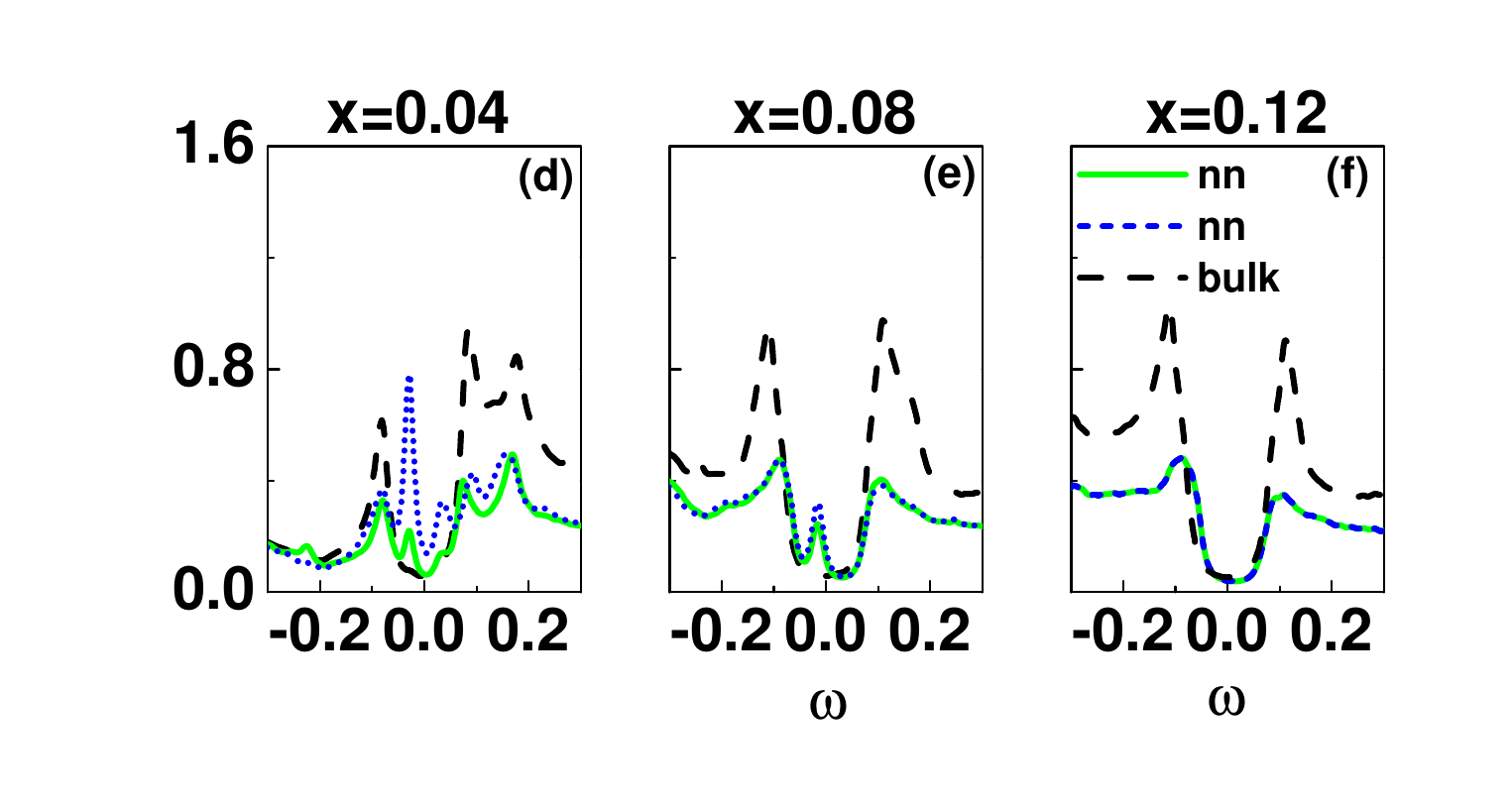}
\caption{(Color online) Similar to Fig.~\ref{dos:vf3}, but for
$V_s=-8$.} \label{dos:vf8}
\end{figure}

Contrary to the $V_s=3$ case where the impurity effect is stronger
at relatively higher doping, the in-gap resonance peak for negative
SP is clearer at lower doping. From Fig.~\ref{dos:fimp} we can see,
for $V_s=-3$, at the impurity site there is a sharp in-gap resonance
peak at positive energy at low doping $x=0.04$. At higher doping,
the LDOS at the impurity site has two peaks at the edge of the SC
coherence peaks with the right peak being higher than the left one.
Here we do not show the SC coherence peaks in Fig.~\ref{dos:fimp},
but in Figs.~\ref{dos:vf3} and\ref{dos:vf8} we plot the
corresponding bulk LDOS in which the SC coherence peaks are
explicitly shown. As the SP strength increases to $V_s=-5$, the
corresponding peaks are all suppressed and in the $x=0.04$ case, the
in-gap resonance peak shifts towards the Fermi energy. From the
 Fig.~\ref{dos:fimp}(b) we can see that the intensity of
the LDOS at the impurity site nearly vanishes for $V_s=-8$ .

Then we plot the LDOS on nnn and nn sites for $V_s=-3$ in
Fig.~\ref{dos:vf3}, at different dopings. At $x=0.04$, the LDOS on
both the nnn and nn sites shows the existence of in-gap resonance
peaks, which gradually merge into the SC coherence peaks as doping
increases. Again, the breaking of four-fold symmetry is more obvious
on the nnn sites than it is on the nn sites and as doping increases,
this asymmetry tends to diminish.

\begin{figure}
\centering
  \includegraphics[width=1.65in]{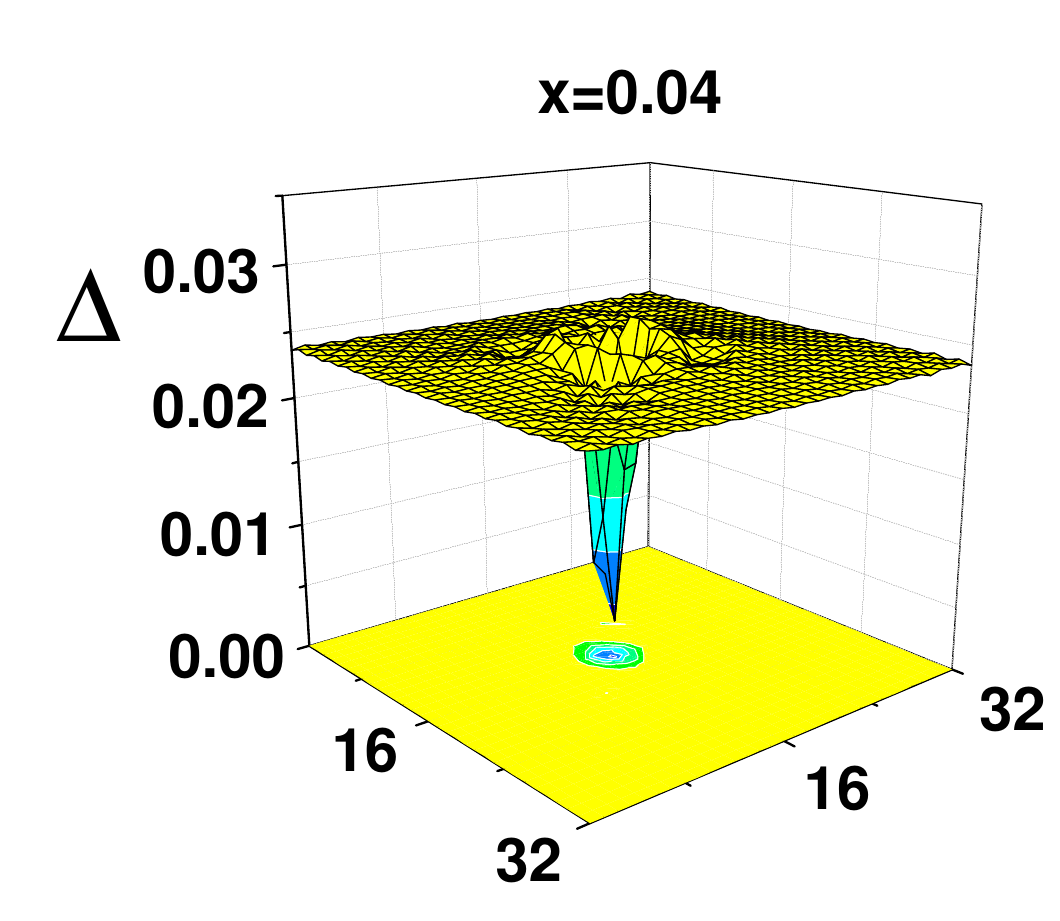}
      \includegraphics[width=1.65in]{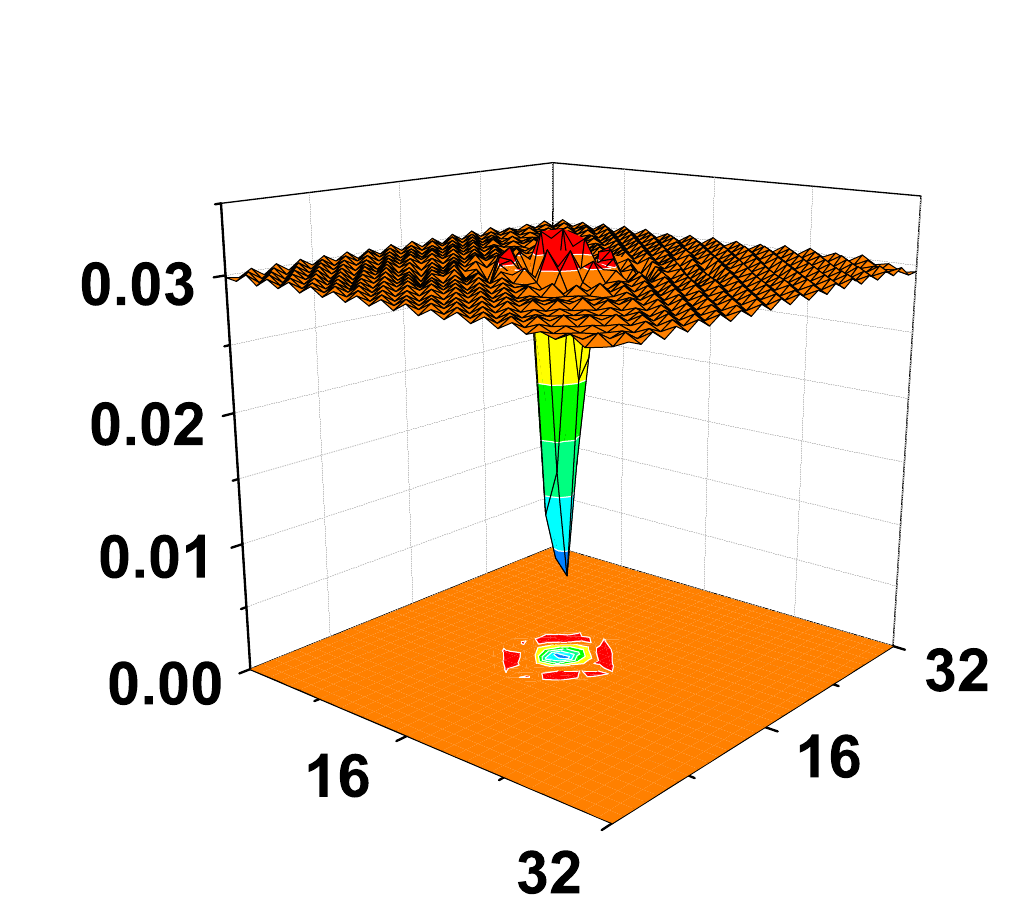}
      \includegraphics[width=1.65in]{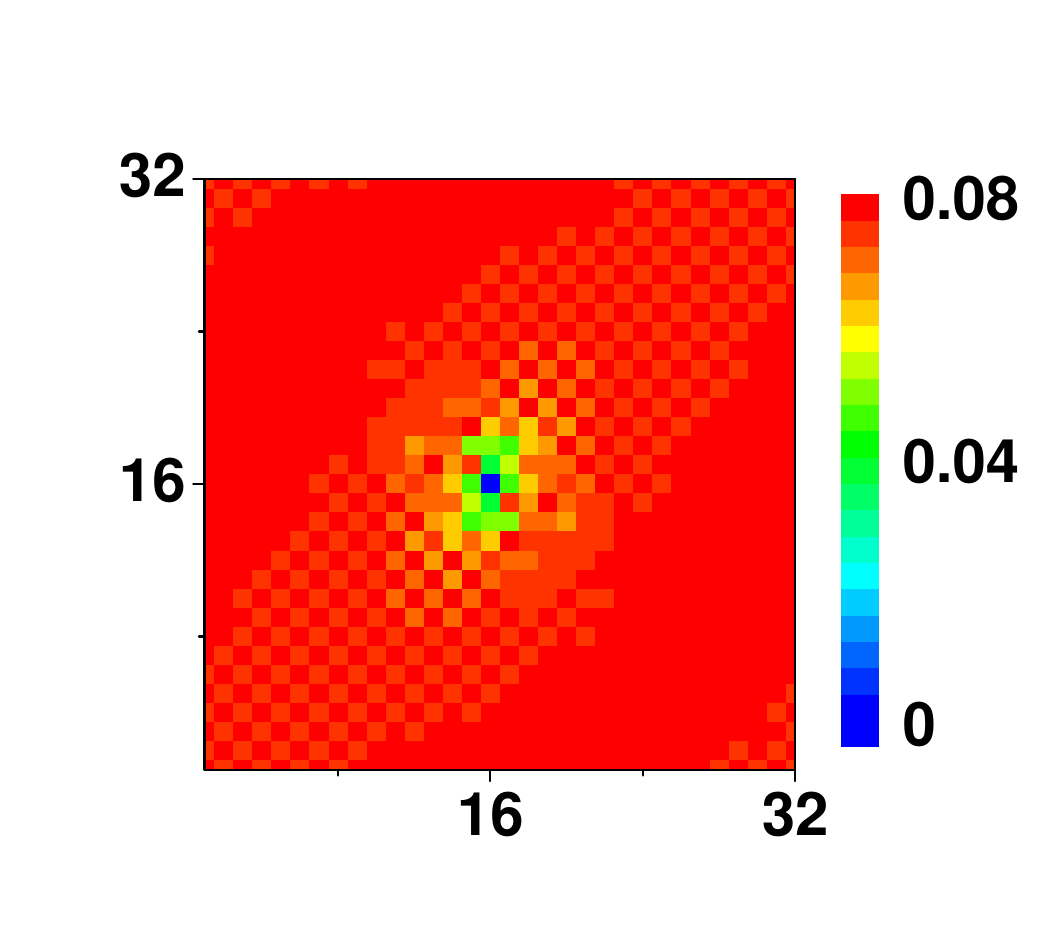}
\includegraphics[width=1.65in]{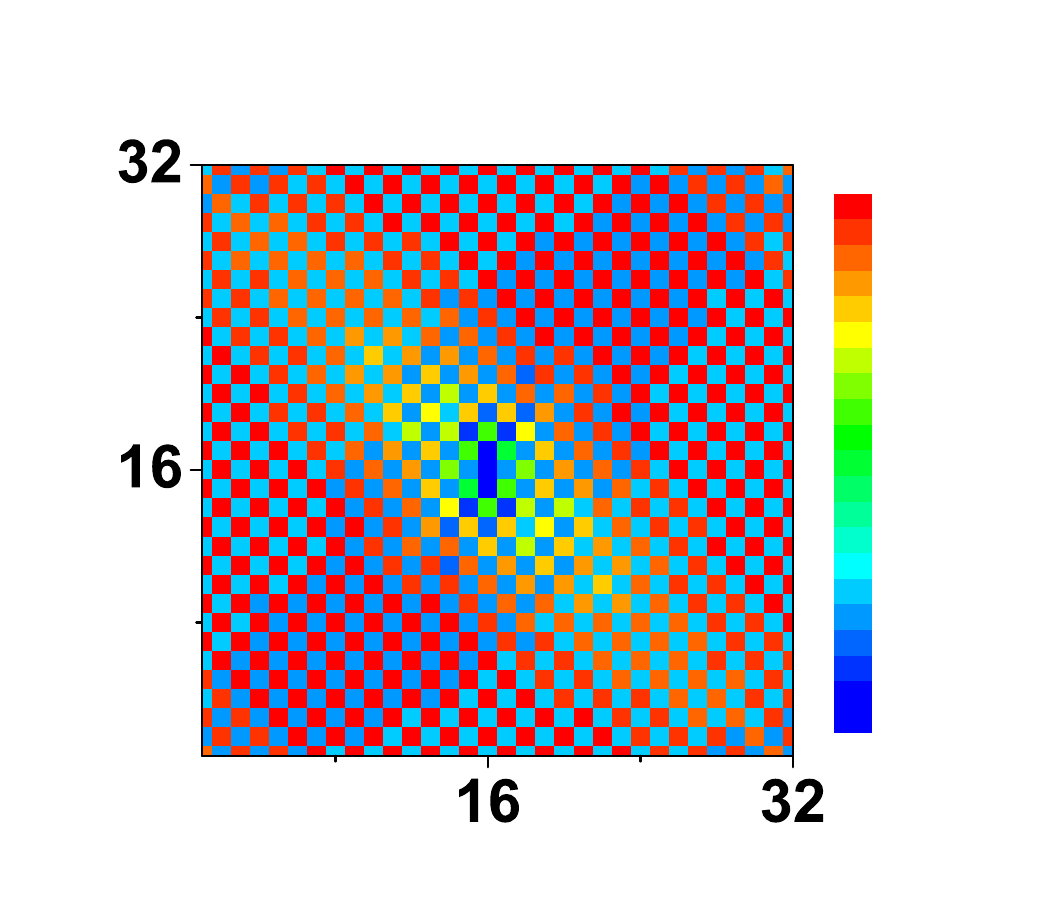}
\caption{(Color online) Similar to Fig.~\ref{v1:OP1}, but for
$V_s=-8$.} \label{vf8:order}
\end{figure}

For strong negative potential $V_s=-8$, both the SC and magnetic
orders are suppressed at the impurity site and oscillate around it
[see Fig.~\ref{vf8:order}]. At about $4$ lattice constants away from
the impurity, the SC and magnetic orders recover to their bulk
values. At doping $x=0.08$, $M_i$ also separates into two
sublattices as can be seen from Fig.~\ref{vf8:order} (d).

As shown in Fig.~\ref{dos:vf8}, there are in-gap resonance peaks on
all the nnn and nn sites at $x=0.04$. The breaking of the four-fold
symmetry is minor, although still visible. The intensities of the
in-gap peaks are suppressed with increased doping. At $x=0.12$ no
in-gap peaks exist. The profile of the LDOS on nn sites is similar
to that on nnn sites, except that the corresponding peaks are lower
and the asymmetry is weaker.

\begin{figure}
\centering
\includegraphics[width=9.5cm]{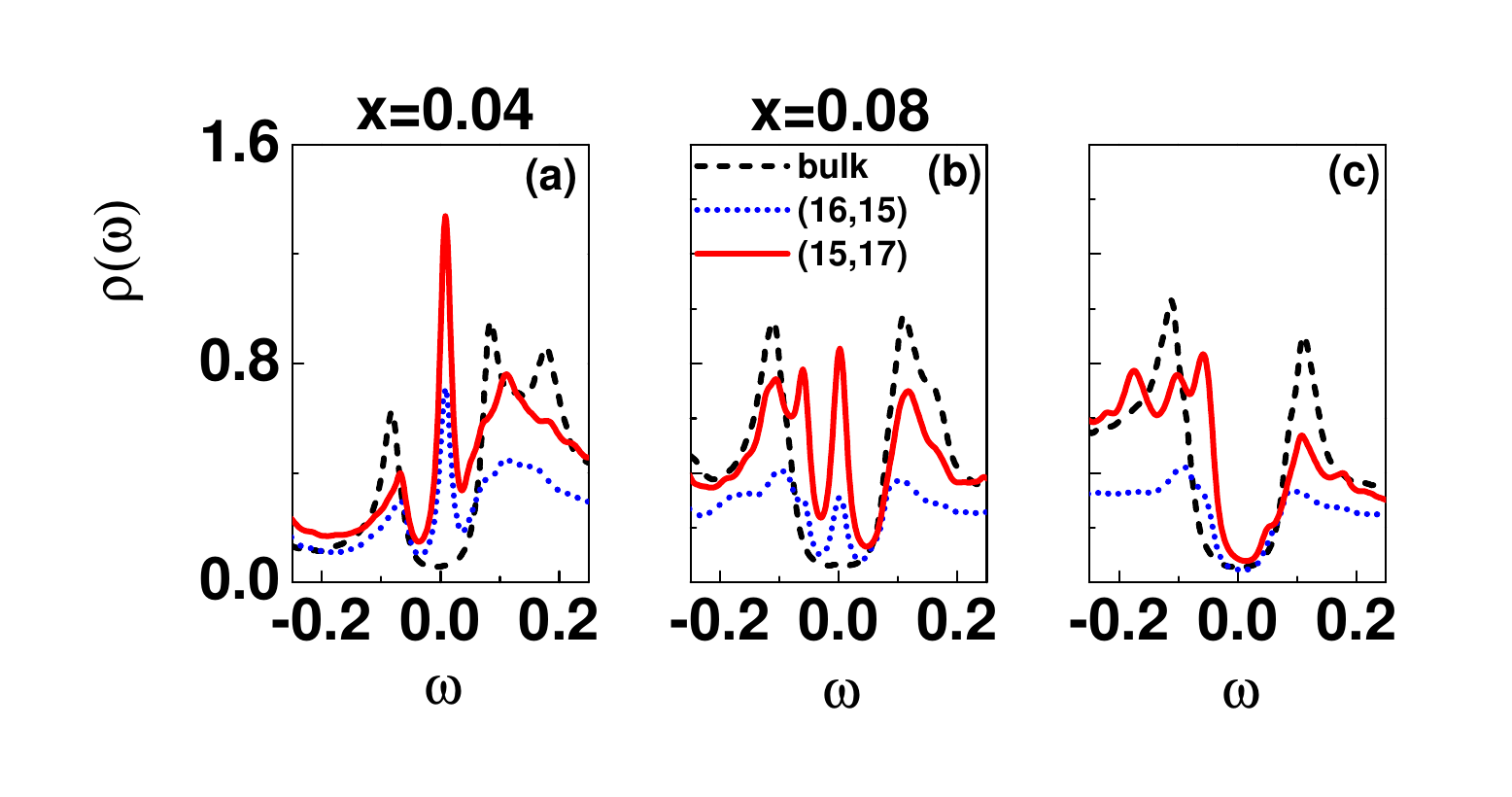}
\caption{(Color online) For $V_s=-100$, the LDOS on nnn and nn sites
as a function of $\omega$, at different dopings. The red solid and
blue dotted lines denote the LDOS on nnn and nn sites, respectively.
The black dashed line represents the bulk LDOS. } \label{dos:vf100}
\end{figure}

For nearly unitary negative potential $V_s=-100$, the real space
distributions of the SC and magnetic orders are similar to the
positive unitary potential case. Fig.~\ref{dos:vf100} shows the LDOS
on nnn and nn sites at various dopings. Similar to the strong
positive SP case, there is a sharp in-gap peak at low doping
$x=0.04$, the position of which is almost at the Fermi energy. We
believe that this sharp peak is due to the existence of the SDW
order. As doping increases to $x=0.08$, the height of the in-gap
peak drops and becomes lower than that of the SC coherence peak. As
doping increases to $x=0.12$, the in-gap peak disappears with two
resonance peaks being very close to the SC coherence peaks.

\section{summary}
\label{SEC:summary}

By solving the BdG equations self-consistently, it is shown that
without impurity, at zero doping and in the SDW state, there exist
equal-sized electron- and hole-like FS pockets along the $\Gamma-M$
line of the BZ, inside which the Dirac cones form. The electron- and
hole-like Dirac cones appear in-pairs near the Fermi energy and are
located very close to each other. The effect of
doping is mainly to reduce the size of the hole pockets while
increase that of the electron ones, consistent with the increased
electron density.

When impurity is introduced into the system, we find that in the
parent compound, strong SP, no matter repulsive or attractive, could
induce considerably large oscillation of the magnetic order around
the impurity site. In addition, for all the SP strength we
investigated, there exists one-dimensional modulation of the LDOS,
similar to the experimentally observed nematic electronic structure,
thus supporting the impurity effect as a possible candidate for the
formation of nematic order. Furthermore, two impurity-induced
resonance peaks are observed around the impurity site and they are
shifted to higher (lower) energies as the strength of the positive
(negative) SP is increased.

In doped samples, generally speaking, the SC and magnetic orders are
suppressed at and around the impurity site, with more complicated
variations compared to those in cuprates. However, for positive SP
at higher doping or negative SP at lower doping, the SC order may
even be enhanced at or around the impurity site, suggesting that the
impurity is not a pair breaker in this case. In addition, impurity could separate the system
into two sublattices denoted by two different values of magnetic
order which can be seen more clearly at relatively higher doping.
Furthermore, there exist impurity-induced bound states at and around
the impurity site, whose positions and numbers depend on the
strength and sign of the SP, as well as on the doping concentration.
For weak and moderate SPs, a distinct bound state exists explicitly
at the nnn sites of the impurity. For unitary impurity, there is a
sharp in-gap peak at low doping, while at high doping, the impurity
induced bound state is close to the SC coherence peaks. On the other
hand, in a small range of moderate doping there are two in-gap peaks
only for positive SP.

In all cases, impurity breaks the four-fold symmetry of the system
and has a stronger effect on nnn sites than it does on nn sites as
can be seen from the LDOS. This symmetry breaking is induced not
only by the SDW order, but also by the intrinsic asymmetry in our
model pinned by the impurity. All the above features could be used
to detect the presence of the SDW order and to probe the coexistence
of the SDW and SC orders.

\begin{acknowledgements}
This work was supported by the Texas Center for Superconductivity at
the University of Houston and by the Robert A. Welch Foundation
under the Grant No. E-1146 (T. Z., H. X. H. and C. S. T.) and E-1070
(Y. G.), the NSFC under Grant No. 11004105 (T. Z.), the U.S. DOE at
LANL under Contract No. DE-AC52-06NA25396, the U.S. DOE Office of
Basic Energy Science, and the LANL LDRD Program (J. X. Z.).

\end{acknowledgements}

\end{document}